\documentclass{article} 
\usepackage{iclr2027_conference,times}

\renewcommand{\iclrruler}[1]{}
\usepackage{amsmath,amssymb,mathtools}
\usepackage{graphicx,booktabs,array,multirow,tabularx,longtable,makecell}
\usepackage{placeins,flafter}
\usepackage{wrapfig}
\usepackage{listings}
\usepackage{multicol}
\usepackage{fvextra}
\usepackage{xcolor,colortbl,pifont}
\usepackage{pgfplots}
\pgfplotsset{compat=1.18}
\usepgfplotslibrary{groupplots}
\usepackage{hyperref}
\usepackage{url}
\usepackage{enumitem}
\usepackage[most]{tcolorbox}
\newcommand{\bench}{APTInvestBench}
\newcommand{\ind}{\mathbf{1}}

\newcolumntype{L}[1]{>{\raggedright\arraybackslash}p{#1}}
\definecolor{ocblue}{HTML}{0072B2}
\definecolor{octeal}{HTML}{009E73}
\definecolor{ocorange}{HTML}{E69F00}
\definecolor{ocred}{HTML}{D55E00}
\lstdefinestyle{ocprompt}{basicstyle=\ttfamily\footnotesize,breaklines=true,columns=fullflexible,keepspaces=true,showstringspaces=false,frame=single,rulecolor=\color{gray},aboveskip=8pt,belowskip=8pt}
\DefineVerbatimEnvironment{tracecall}{Verbatim}{fontsize=\scriptsize,breaklines=true,breakanywhere=true,breaksymbolleft={},breaksymbolright={}}

\title{APTInvestBench: Evaluating Autonomous APT Investigation under Varying Telemetry}
\author{Anonymous authors}

\begin{document}
\maketitle

\begin{abstract}
Large language model (LLM) agents could help security operations centers (SOCs) investigate advanced persistent threats (APTs) by turning weak leads into evidence for intrusion scoping and response. Yet success under one telemetry setting does not establish robustness to changes in log collection, retention, or sampling. We introduce \bench{}, a benchmark for evaluating \emph{cross-telemetry robustness} in autonomous APT investigation. It comprises 370 cases across seven SOC-inspired conditions, derived from 56 report-informed attack reconstructions with 16.4 million log records. Agents investigate unverified leads and submit reports with record-level citations. Fixed action-level support requirements track sufficient evidence across available logs, query returns, and formal citations, separating telemetry limitations from acquisition and reporting gaps. Across eleven LLMs, agents acquire sufficient evidence for 44.3\% of recoverable attack actions on average, while formal citations support only 25.0\%. More importantly, aggregate coverage can conceal substantial instability: from Full to endpoint-only telemetry, coverage declines by only 1.6 percentage points, yet 35.5\% of previously covered actions lose sufficient citation support despite remaining recoverable. Across four frameworks, such losses persist even when registered supporting records remain unchanged. \bench{} provides reusable investigation environments and diagnostic evaluation for identifying these gaps and developing more reliable defensive agents.
\end{abstract}
\section{Introduction}

An advanced persistent threat (APT) investigation begins with weak leads amid ordinary activity. Recent work has advanced APT investigation through graph representation learning, few-shot learning, and threat intelligence \citep{magic2024,dupin2026,knowhow2026}, with growing emphasis on the practical value of these systems and their outputs in security operations centers (SOCs) \citep{simpler2025,orthrus2025,provx2026}. 
In practice, these systems should help analysts determine what the attacker did, which assets were affected, and how the activities are connected. Grounded in specific records, these findings form an \emph{evidence-backed account of intrusion scope} \citep{nist_ir2025}.

Conducting such investigations across large volumes of SOC logs and alerts is time-consuming and difficult to scale manually \citep{yang2024alerts,siem_soar2025}. This workload motivates the use of large language model (LLM) agents to automate log searches and pursue investigation leads. Real-world deployment requires direct evaluation of their ability to conduct such investigations. Benchmarks such as CyberGym and Cybench evaluate agents through vulnerability reproduction and capture-the-flag challenges \citep{cybergym2026,cybench2025}. These tasks assess vulnerability-focused reasoning and exploitation capabilities, but do not directly test whether agents can reconstruct attack sequences from audit logs and support their findings with sufficient evidence.

\begin{figure}[t]
\centering
\input{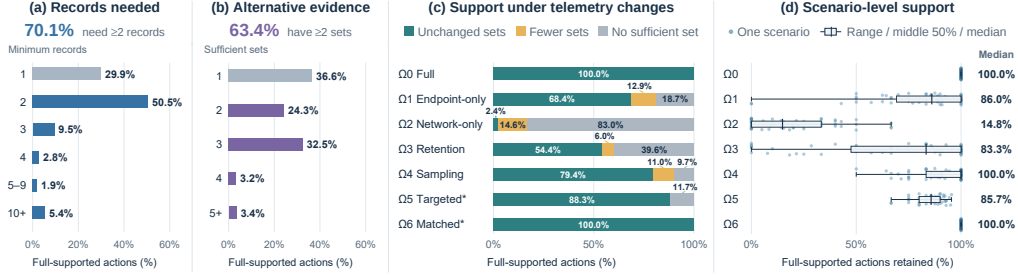}
\caption{Evidence requirements and support under telemetry changes in \bench{}.}

\label{fig:motivation}
\end{figure}

Recent benchmarks partially address this gap by evaluating query-driven investigation, attack-chain reconstruction, and the effects of limited telemetry \citep{excytin2026,diagchain2026,synthchain2026}, as summarized in Table~\ref{tab:aptinvestbench-comparison}. However, they do not establish whether investigation performance generalizes across SOCs with different log collection, retention, and sampling practices \citep{logging2024,gcp_flow_sampling2026}. These differences can alter both which attack activities remain supportable and how agents find and use the available evidence, making performance in one telemetry setting an insufficient indicator of performance in another. Therefore, this raises a central question: when the same actions remain supportable across telemetry conditions, can agents still acquire sufficient evidence and preserve it in their reports? We refer to this capability as \emph{cross-telemetry robustness}.

Evaluating this capability requires resolving this ambiguity by checking for sufficient support for the same attack actions, rather than merely matching individual records. For example, a file-creation record may identify the creator only by a process identifier, while a separate process record links that identifier to a program. Together, these records identify the program that created the file, although neither is sufficient alone. Alternative record combinations that establish the same action should receive equal credit. These patterns are common in our benchmark: among actions supported under full telemetry, 70.1\% require at least two records and 63.4\% admit multiple sufficient record sets (Figure~\ref{fig:motivation}). Applying fixed action-level evidence requirements consistently to available logs, query returns, and formal citations then reveals whether sufficient support is unavailable, available but not acquired, or acquired but not preserved in the report's citations.

We introduce \bench{} to evaluate \emph{cross-telemetry robustness} in autonomous APT investigation. The benchmark comprises 370 cases derived from 56 report-informed attack reconstructions with 16.4 million log records, exposed under seven SOC-inspired telemetry conditions. Each incident is reused across views that vary available evidence while keeping \textbf{reference actions and evidence requirements fixed}. Agents investigate unverified leads and submit reports with record-level citations. By applying the same action-level requirements to \textbf{available logs, query returns, and formal citations}, \bench{} \emph{distinguishes evidence availability from evidence acquisition and delivery}, while report interpretation is assessed separately.

We evaluate eleven LLMs and four agent frameworks on \bench{}. Across models, agents acquire sufficient evidence for 44.3\% of recoverable actions on average, while formal citations support only 25.0\%. More importantly, \textbf{similar aggregate coverage can conceal different action-level outcomes}: from Full to endpoint-only telemetry, citation coverage declines by only 1.6 percentage points, yet 35.5\% of previously covered attack actions lose sufficient citation support despite remaining recoverable. A stricter comparison also shows lower citation coverage for all frameworks despite unchanged registered evidence across views. Query audits further reveal that some acquired support is not fully preserved in formal citations. These findings show that assessing cross-telemetry robustness requires action-level evidence tracking beyond aggregate coverage to determine whether gaps call for additional logs, further investigation, or more complete citations. In summary, we make the following key contributions:

\setlength{\topsep}{0pt}
\setlength{\partopsep}{0pt}
\begin{itemize}\setlength{\itemsep}{2pt}\setlength{\parskip}{0pt}
    \item \textbf{A large-scale benchmark for autonomous APT investigation.}
    We provide report-informed multi-step attacks embedded in normal activity, with a unified environment for agents to investigate unverified leads and submit reports with record-level citations.

    \item \textbf{Controlled evaluation of cross-telemetry robustness.}
    We automatically derive investigation cases under seven SOC-inspired conditions while fixing incident identities, reference actions, and support rules. This enables comparisons of the same recoverable actions, including stricter tests with unchanged registered witnesses.

    \item \textbf{Sufficient-evidence evaluation and process-level diagnosis.}
    Action-specific rules combine complementary records and accept alternative witnesses. Applying the same rules to available logs, query results, and formal citations distinguishes telemetry limitations from incomplete evidence acquisition and reporting omissions.

    \item \textbf{Empirical insights into investigation completeness.}
    We evaluate agents' investigative capabilities. Similar overall coverage can conceal substantial losses of citation support for attack actions, exposing hidden gaps in the evidence delivered for intrusion scoping.
\end{itemize}

\section{Autonomous multi-step APT investigation}
\label{sec:task}

The agent investigates a multi-step incident amid normal host and network activity. Starting from unverified leads, it queries available records, distinguishes incident actions from background events, and submits a structured report with record-level citations.

\begin{table}[t]
\centering
\begingroup
\hypersetup{pdfborder={0 0 0}}
\setlength{\abovecaptionskip}{0pt}
\setlength{\belowcaptionskip}{5pt}
\caption{Investigation benchmarks and cross-telemetry evaluation.}
\label{tab:aptinvestbench-comparison}
\fontsize{7.6}{9.0}\selectfont
\setlength{\tabcolsep}{1.5pt}
\renewcommand{\arraystretch}{1.12}
\setlength{\heavyrulewidth}{0.7pt}
\setlength{\lightrulewidth}{0.4pt}
\newcommand{\aptYes}{\ding{51}}
\newcommand{\aptNo}{\ding{55}}
\begin{tabular*}{\linewidth}{@{\extracolsep{\fill}}c l c c c c c c@{}}
\toprule
\textbf{Benchmark}
& \textbf{Task}
& \textbf{Scenarios}
& \textbf{Data}
& \textbf{Construction}
& \makecell[c]{\textbf{Telemetry}\\\textbf{variation}}
& \makecell[c]{\textbf{Gap}\\\textbf{diagnosis}}
& \makecell[c]{\textbf{Support-matched}\\\textbf{comparisons}} \\
\midrule
\makecell[c]{ExCyTIn-Bench\\(\citealp{excytin2026})}
& Investigation QA
& 8
& \makecell[c]{923.2K\\rows}
& \makecell[c]{Attack\\reenactment}
& Sources / time
& \aptNo
& \aptNo \\
\addlinespace[1.8pt]
\makecell[c]{DiagChain\\(\citealp{diagchain2026})}
& \makecell[l]{Attack-chain\\reconstruction}
& 23
& \makecell[c]{12.8K\\cards}
& \makecell[c]{Public-log\\integration}
& Source profiles
& \aptYes
& \aptNo \\
\addlinespace[1.8pt]
\makecell[c]{SynthChain\\(\citealp{synthchain2026})}
& \makecell[l]{Supply-chain\\reconstruction}
& 7
& \makecell[c]{593.7K\\records}
& \makecell[c]{Attack\\reenactment}
& \makecell[c]{Sources /\\sampling}
& \aptNo
& \aptNo \\
\addlinespace[1.8pt]
\makecell[c]{RAG-SIA\\(\citealp{ragsia2026})}
& \makecell[l]{Forensic QA /\\reconstruction}
& $17+1$
& \makecell[c]{94.9K\\logs}
& \makecell[c]{Public data +\\reenactment}
& Sources / context
& \aptYes
& \aptNo \\
\midrule
\textbf{APTInvestBench}
& \makecell[l]{\textbf{Multi-step APT}\\\textbf{investigation}}
& \textbf{56}
& \makecell[c]{\textbf{16.4M}\\\textbf{records}}
& \makecell[c]{\textbf{Report-informed}\\\textbf{reconstruction}}
& \makecell[c]{\textbf{Seven SOC views}}
& \aptYes
& \makecell[c]{\textbf{Common actions +}\\\textbf{unchanged witnesses}} \\
\bottomrule
\end{tabular*}
\endgroup
\end{table}

\textbf{Inputs and investigation.}
An \emph{observation view} contains the records exposed under one telemetry condition. In the controlled study, the agent receives one view, a manifest of its log sources, read-only query tools, and three unverified SOC leads. One lead originates from the incident and two from background activity. Their roles and the historical incident name remain hidden. After executing the prescribed lead queries, the agent chooses follow-up queries, requests additional result pages, inspects complete records, and pivots across host, process, file, and network identifiers. The reference actions, support rules, and scoring feedback remain hidden from the agent. 

\textbf{Structured report.}
The report represents each claimed activity as a formal stage with entities, time, confidence, and record citations. Relations specify their endpoints and supporting records. A stage marked \emph{likely} or assigned low confidence remains a formal claim. Unsupported interpretations belong in the hypothesis or uncertainty fields. The primary coverage measure uses only citations attached to activity stages, termed \emph{formal-stage citations}. Citations appearing only in entities, relations, hypotheses, or uncertainties are excluded from this measure. 

\textbf{Concrete evidence example.}
An investigation of the 2019 APT27 reconstruction under Full telemetry illustrates this process. From an unverified web-request lead, the agent follows the client address to a later POST request for \texttt{error2.aspx}. It examines endpoint records to identify \texttt{w3wp.exe} as the file's creator. Process identifiers and parent-child links guide investigation into subsequent activity, including \texttt{s.exe} and \texttt{thinprobe.exe}. An outbound flow from \texttt{thinprobe.exe} leads to a matching Zeek TLS session. The agent distinguishes a nearby certificate-status query from the incident and organizes the file, process, and network observations into a cited report.

\textbf{Task formulation.}
Let $s$ denote a reconstructed incident, $v$ an observation condition, and $E_{sv}$ the available records, represented by their stable public identifiers. An investigation case is the pair $(s,v)$. Let $\pi$ denote the investigation policy implemented by a model and agent framework. A run executes this configuration on one case.
Let $L_s$ denote the public leads, which remain fixed across views, and $m_{sv}$ the source manifest for the current view. The initial history $h_0$ contains $L_s$, $m_{sv}$, and the tool and reporting instructions. The policy follows the prescribed lead-query procedure and otherwise selects operations adaptively. 
The interaction is
\begin{equation}
u_t \sim \pi(\cdot \mid h_{t-1}), \qquad
o_t = \mathcal{T}_{sv}(u_t), \qquad
h_t = h_{t-1}\oplus(u_t,o_t),
\label{eq:investigation-process}
\end{equation}
where $u_t$ is a tool operation, $\mathcal{T}_{sv}$ is the read-only interface over $E_{sv}$, $o_t$ is its response, and $\oplus$ appends the operation and response to the history.

After $T$ tool operations, including the prescribed lead queries, the agent produces a structured report from the accumulated history. Let $P_{sv}$ be its formal activity stages and $C_p$ the record identifiers cited by stage $p$. The acquired identifiers and the formal citation union are
\begin{equation}
D_{sv}
=
\bigcup\nolimits_{t=1}^{T}\operatorname{IDs}(o_t),
\qquad
C^\cup_{sv}
=
\bigcup\nolimits_{p\in P_{sv}} C_p,
\label{eq:task-record-sets}
\end{equation}

where $\operatorname{IDs}(o_t)$ extracts citable record identifiers from a response, including those accompanying previews. Valid online submissions satisfy $C^\cup_{sv}\subseteq D_{sv}\subseteq E_{sv}$, so formal stages can cite only records returned through the investigation. These identifiers resolve to the corresponding record contents during evaluation. Inclusion in $D_{sv}$ does not imply that the complete record was read or understood. Section~\ref{sec:evaluation} applies the support rules to $E_{sv}$, $D_{sv}$, and $C^\cup_{sv}$ to determine which actions have sufficient support in the available records, query results, and formal citations. Table~\ref{tab:aptinvestbench-comparison} compares this task with related benchmarks.


\section{Benchmark resource and controlled telemetry}
\label{sec:benchmark}

\begin{figure}[t]
\centering
\input{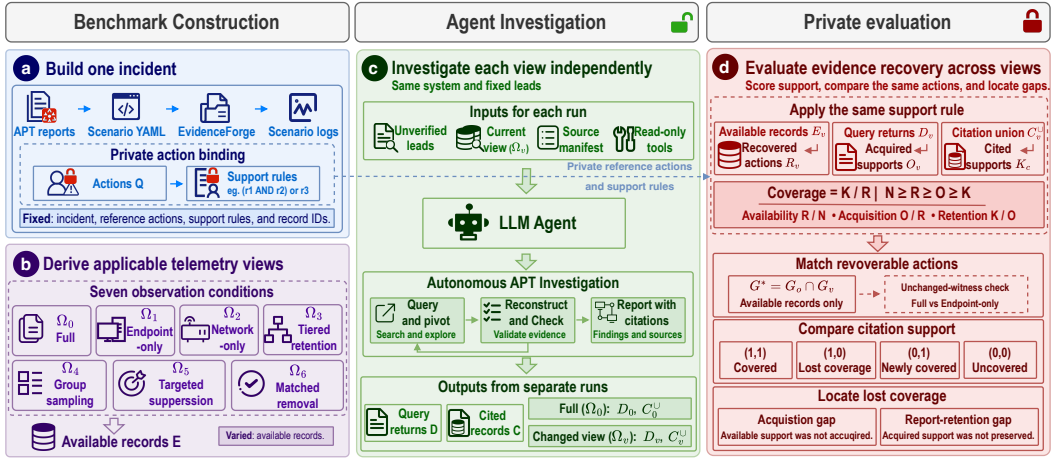}
\caption{Construction and cross-view evaluation in \bench{}.}
\label{fig:framework}
\end{figure}

We instantiate the task in Section~\ref{sec:task} through report selection, scenario compilation, evidence binding, and controlled view construction. Figure~\ref{fig:framework} connects this construction to public investigation and private cross-view evaluation.

\textbf{Report-informed scenario specification.}
We screen a collection of 6,217 APT reports and construct 56 multi-step investigation scenarios from selected reports. Each YAML scenario distinguishes reported attack behavior, configured environment details, and generated execution context. Four resource families support authoring and instantiation: scenario templates, command-line matching rules, parameter pools and configuration templates, and benign-activity patterns. Together, they specify attack activities, supply environment details, and generate normal background activity.

\textbf{Event compilation and correlated rendering.}
The pipeline converts scenario descriptions and supported command patterns into normalized events, resolves entity references, and provides host, process, and connection context. Rule-based expansion adds prerequisite events for the scenario. Building on EvidenceForge's event model and renderers \citep{evidenceforge2026}, the pipeline renders these events into correlated endpoint, network, and application records. Shared event context and consistent entity identifiers preserve links across sources. Background activity is generated in the same environment. Generated details make the reconstruction concrete without adding historical facts. Appendix~\ref{app:generation_lineage} describes construction and provides a worked evidence-binding example.

\textbf{Reference actions and evidence bindings.}
\bench{} binds each reference action instance to required facts, entity links, and temporal constraints. Each binding is checked automatically and reviewed by annotation owners, retaining record locations and rule inputs for inspection. Repeated uses of a technique can form separate action instances, and registered temporal relations require evidence for both endpoints. Support rules distinguish individual events from recurring activity (Table~\ref{tab:support_rule_examples}), while Figure~\ref{fig:motivation}(a--b) summarizes record requirements and alternative sufficient sets under Full telemetry. The full resource comprises 56 scenarios, 523 annotated action instances, 16 log formats, and six environment types. Attack-labelled records account for less than 0.07\% of the corpus. Sparse attack signals are also reported in a real-world SOC, where only 0.01\% of network alerts were associated with true attacks \citep{yang2024alerts}. The main model and framework comparisons use ten selected scenarios. Appendix~\ref{app:implementation} provides full-resource and per-scenario statistics.


\textbf{Seven observation views.}
Each applicable condition selects records from the same rendered scenario to form a view $E_{sv}$, keeping the incident, reference actions, and support rules fixed. Full telemetry ($\Omega_0$) retains all records, while endpoint-only ($\Omega_1$) and network-only ($\Omega_2$) retain their respective source families. Tiered retention ($\Omega_3$) applies source-specific windows, and group sampling ($\Omega_4$) retains or removes related records together. Inspired by anti-forensics, targeted suppression ($\Omega_5$) removes records to break every registered sufficient set for a selected action. Matched benign removal ($\Omega_6$) removes the same number of background records while preserving target support. The applicable conditions yield 370 investigation cases, with recoverability recomputed for each view. Table~\ref{tab:observation_motivation} documents the motivations, parameters, and matching criteria.



\section{Evaluating the same actions across telemetry conditions}
\label{sec:evaluation}

Using the record sets defined in Section~\ref{sec:task}, we evaluate action support within each view, compare the same recoverable actions across views, and locate gaps using query audits.

\subsection{Support at three investigation boundaries}

Let $\mathcal{B}_q$ contain the record combinations satisfying the registered support rule for action $q\in Q_s$. Each such combination is a \emph{witness}. For any record set $S$,
\begin{equation}
\phi_q(S)=\ind\!\left[\exists B\in\mathcal{B}_q:B\subseteq S\right].
\label{eq:proof}
\end{equation}


\begin{table}[t]
\centering
\footnotesize
\setlength{\tabcolsep}{3pt}
\caption{Action-level evidence metrics. Ratios are defined only for nonzero denominators.}
\label{tab:evaluation-boundaries}
\begin{tabularx}{\linewidth}{@{}l c X@{}}
\toprule
Metric & Rate & Interpretation \\
\midrule
Availability
& $R/N$
& Fraction of reference actions supportable in the view. \\

Acquisition
& $O/R$
& Fraction of recoverable actions supported by query returns. \\

Citation coverage
& $K/R$
& Fraction of recoverable actions supported by formal-stage citations. \\

Citation retention
& $K/O$
& Fraction of actions with acquired support that retain sufficient formal-stage citations. \\
\bottomrule
\end{tabularx}
\end{table}

A witness is \emph{minimal} if removing any record makes it insufficient. An action is \emph{recoverable} when the available records contain a witness. 
Let $G_{sv}=\{q\in Q_s:\phi_q(E_{sv})=1\}$ denote the recoverable actions, with $N_{sv}=|Q_s|$ and $R_{sv}=|G_{sv}|$. For one selected recorded run, we define
\begin{equation}
O_{sv}=\sum_{q\in G_{sv}}\phi_q(D_{sv}),
\qquad
K_{sv}=\sum_{q\in G_{sv}}\phi_q(C^\cup_{sv}),
\qquad
N_{sv}\ge R_{sv}\ge O_{sv}\ge K_{sv}.
\label{eq:boundaries}
\end{equation}
Here, $N$, $R$, $O$, and $K$ count all reference actions and those with sufficient support in available records, query returns, and formal-stage citations, respectively. Our primary score, citation coverage $K/R$, measures the fraction of recoverable attack actions supported by the report's formal citations. Table~\ref{tab:evaluation-boundaries} summarizes the four rates and their denominators.
For counts pooled over the same population with nonzero denominators, overall evidence delivery decomposes as
\begin{equation}
K/N = (R/N)\cdot(O/R)\cdot(K/O).
\label{eq:decomp}
\end{equation}

\subsection{Three levels of cross-telemetry comparison}
\label{sec:support_matched}

Per-view rates use $G_{sv}$ and therefore need not compare the same actions. For Full telemetry ($v=0$) and a changed view $v$, we instead compare their common-recoverable set,
\begin{equation}
G^*_{s,v}=G_{s,0}\cap G_{s,v}.
\label{eq:common_support}
\end{equation}

Paired runs keep the system, reconstructed incident, action identities, and support rules fixed. We perform comparisons at three levels. First, per-view analysis evaluates each view using its own recoverable action set \(G_{sv}\). Second, common-recoverable analysis restricts comparisons to actions recoverable in both Full and changed views, \(G^{*}_{s,v}=G_{s,0}\cap G_{s,v}\), avoiding changes caused solely by different evidence availability. Third, unchanged-witness analysis further restricts this set to actions whose registered candidate records and minimal-witness sets remain identical across views. Both runs must satisfy the selection criteria and have usable evidence audits, so each condition pair has its own analysis population.




Actions recoverable in both views may still have different witnesses. We therefore conduct a stricter comparison between Full and endpoint-only telemetry, retaining only common-recoverable actions whose registered candidate records and minimal-witness sets remain unchanged. Eligibility is determined from available records and fixed support rules, independently of agent outcomes. This tests whether investigation results change despite unchanged registered support. It does not hold investigation difficulty constant or cover evidence paths outside the registered catalogue. 

\subsection{Tracking coverage changes and locating gaps}
\label{sec:telemetry_transitions}

We track which attack actions retain sufficient citation support as telemetry changes. For each system, we compare its separately executed Full and changed-view runs on $G^*_{s,v}$. We count actions supported in both reports ($n_{11}$), only in Full ($n_{10}$), only in the changed view ($n_{01}$), or in neither ($n_{00}$). Each action is counted once per system in a condition comparison.

With $M=\sum_{i,j}n_{ij}>0$, the pooled citation-coverage change is
\begin{equation}
\Delta(K/R)_{\mathrm{common}}
=
(n_{01}-n_{10})/{M}.
\label{eq:coverage_transition}
\end{equation}
Newly covered actions can therefore offset lost coverage. When the denominator is nonzero, we also report $n_{11}/(n_{11}+n_{10})$, the fraction of actions supported under Full telemetry that remain supported in the changed-view reports. This cross-run measure differs from within-run citation retention $K/O$. We compute analogous transition counts for acquired support.


\textbf{Scoring checks and scope.}
\label{sec:scoring_checks}
We check that complete reference witnesses yield 100\% coverage, empty reports yield 0\%, and citation-preserving transformations leave $K$ unchanged. Re-scoring the same reports under any-overlap and all-candidate rules tests the effect of accepting partial evidence or requiring redundant records (Figure~\ref{fig:scoring-criteria-comparison}).

\section{Experiments}
\label{sec:experiments}
We organize the experiments around four research questions: RQ1: How far does autonomous investigation progress? RQ2: How does telemetry change affect action support? RQ3: Do gaps persist with unchanged evidence? RQ4: Is acquired support preserved in citations?

\textbf{Experimental setup.}
We evaluate eleven LLMs and four agent frameworks on \bench{} under seven SOC-inspired telemetry conditions. Model comparisons use Codex, while framework comparisons hold MiniMax-M3 fixed across Codex, Claude Code, OpenCode, and Ref-ReAct \citep{codex_cli2026,claude_code2026,opencode2026,react2023}. To manage the cost of multi-round investigations across systems and telemetry conditions, we concentrate the main evaluation on ten of the 56 scenarios. This allocation prioritizes paired comparisons and query-level analysis of where agents fail to acquire or preserve attack evidence.

\subsection{RQ1: How far does autonomous investigation progress?}
\label{sec:investigation_extent}
\label{sec:rule_comparison}
\label{sec:rule_release}
\label{sec:model_results}
\label{sec:framework_results}

\begin{wrapfigure}[21]{r}{0.66\textwidth}
  \centering
  \includegraphics[width=\linewidth]{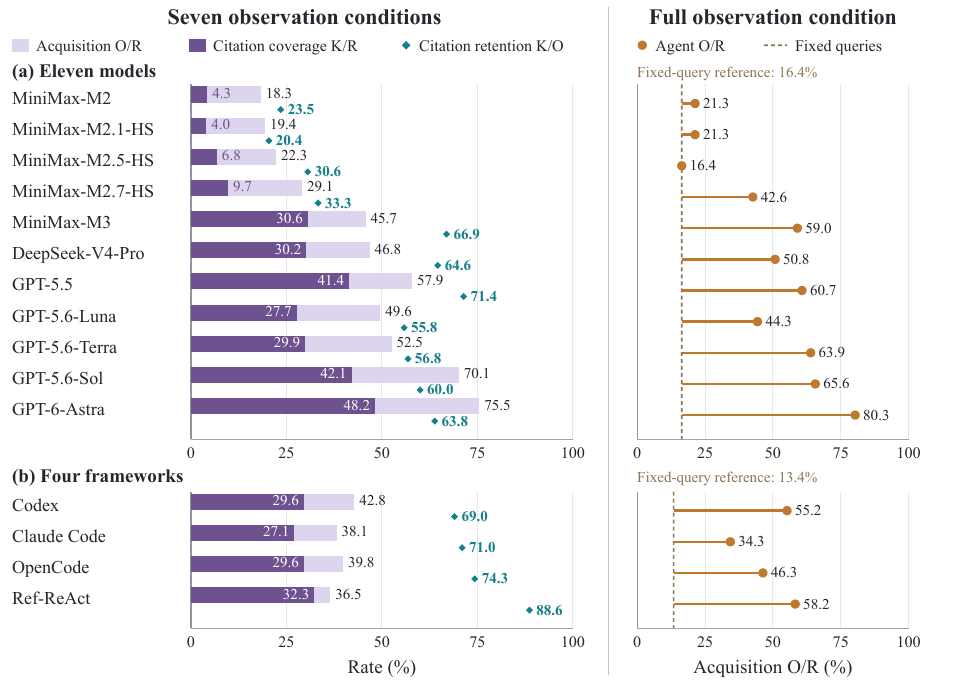}
  \caption{Investigation progress. Left: pooled acquisition, citation coverage, and retention across seven conditions. Right: Full-telemetry acquisition versus fixed lead queries.}
  \label{fig:investigation-progress-overview}
\end{wrapfigure}

Figure~\ref{fig:investigation-progress-overview} summarizes the investigation results. Across the seven telemetry conditions, the eleven models acquire sufficient evidence for 44.3\% of recoverable attack actions on average, but their formal citations support only 25.0\%. Frameworks also differ in how much acquired support reaches 

their reports. Codex acquires more support than Ref-ReAct, but Ref-ReAct preserves a larger share in citations and achieves higher citation coverage.

To assess progress beyond the supplied leads, we use fixed lead queries to distinguish the evidence they already provide from evidence obtained through further investigation. Across the shared Full-telemetry cases, these queries alone return sufficient evidence for 16.4\% of recoverable actions, compared with 59.0\% for MiniMax-M3 and 80.3\% for GPT-6-Astra after autonomous investigation (Figure~\ref{fig:investigation-progress-overview}). Agents therefore extend the evidence-supported scope of the investigation beyond supplied queries.

These results show useful investigative progress based on LLM-based agents but incomplete evidence delivery for intrusion scoping. Obtaining support for more attack activities does not ensure that the final report preserves it. Appendix~\ref{app:investigation_profiles} examines investigation depth, behavior coverage, and explicit temporal links, while Appendix~\ref{app:paired_estimates} checks ranking sensitivity. 

\subsection{RQ2: How does telemetry change affect action support?}
\label{sec:observation_effects}
\label{sec:observation_results}

\begin{wrapfigure}[21]{r}{0.69\textwidth}
\centering
\includegraphics[width=\linewidth]{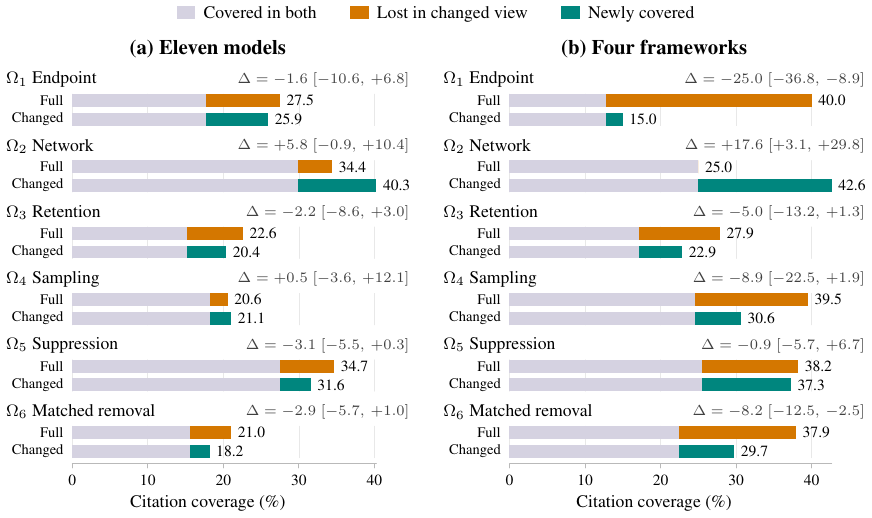}
\caption{Citation transitions for common-recoverable actions. Bars show retained, lost, and new coverage. $\Delta$ is changed--Full difference. Brackets indicate exploratory 95\% source-group bootstrap intervals.}
\label{fig:telemetry-paired-summary}
\end{wrapfigure}

\textbf{Determining what remains supportable.}
In the controlled study, Full telemetry supports 97.3\% of eligible actions, compared with 85.3\% under endpoint-only and 37.3\% under network-only telemetry. Either source family alone suffices for 25.3\% of eligible actions. Retention, sampling, and targeted suppression leave 65.3\%, 94.7\%, and 85.3\% recoverable, respectively, while matched benign removal preserves the Full level. To separate this change in available support from investigation performance, we restrict the following comparisons to actions recoverable in both views. Appendix~\ref{app:observation_details} reports results for every condition.

Figure~\ref{fig:telemetry-paired-summary} compares each changed condition with Full telemetry on common-recoverable actions. In the Full-to-endpoint model comparison, pooled citation coverage falls from 27.5\% to 25.9\%, a 1.6-point decline. Yet only 64.5\% of the actions supported in Full reports remain supported in endpoint-only reports. Newly covered actions partly offset these omissions, keeping overall coverage similar while changing which attack activities the reports substantiate. The framework comparison under targeted suppression shows similar cancellation: 12.7\% of paired action observations lose citation support and 11.8\% become newly covered, yielding a net decline of 0.9 percentage points. This comparison excludes the suppressed target, which is examined separately alongside the matched-removal control in Appendix~\ref{app:paired}.

\begin{wrapfigure}[14]{r}{0.325\textwidth}
\centering
\vspace{-22pt}
\includegraphics[width=\linewidth]{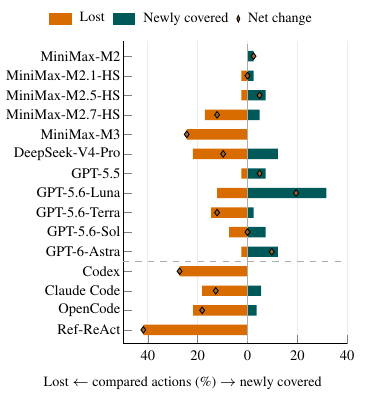}
\caption{System comparison on common-recoverable actions.}
\label{fig:telemetry-endpoint-systems}
\end{wrapfigure}

\textbf{The same view change does not affect every system alike.}
Figure~\ref{fig:telemetry-endpoint-systems} shows losses and gains by system in citation coverage. GPT-5.6-Luna's coverage rises from 24.4\% under Full telemetry to 43.9\% under endpoint-only telemetry, yet its reports preserve sufficient support for only half of the previously covered actions. MiniMax-M3 instead loses coverage without covering new actions. Thus, higher overall coverage can conceal lost support for previously covered attack activities. The net coverage change also depends on incident composition (Appendix~\ref{app:telemetry_robustness}).

For intrusion scoping, support for newly covered activities does not fill evidence gaps for other activities. Similar or higher overall coverage therefore does not imply that reports substantiate the same parts of an intrusion. These comparisons hold action recoverability fixed but allow the available witnesses to differ.

\subsection{RQ3: Do gaps persist with unchanged evidence?}

\label{sec:evidence_loss_results}

\begin{wrapfigure}[21]{r}{0.65\textwidth}
\centering
\vspace{-3pt}
\includegraphics[width=\linewidth]{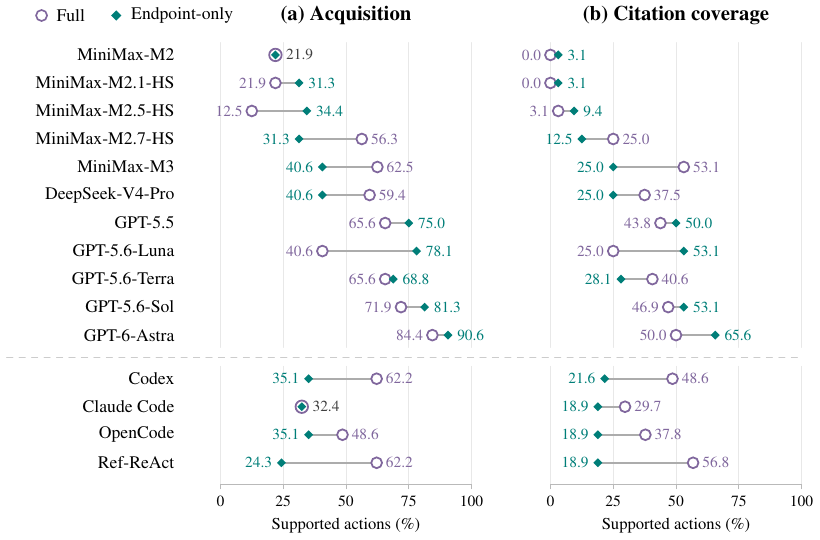}
\caption{Full-to-endpoint investigation outcomes with unchanged candidate records and registered minimal witnesses. Models and frameworks use separate matched action sets.}
\label{fig:telemetry-witness-all-qualified}
\end{wrapfigure}

We compare only actions whose registered candidate records and minimal-witness sets are identical across Full and endpoint-only telemetry. Investigation results still differ despite unchanged registered evidence (Figure~\ref{fig:telemetry-witness-all-qualified}). MiniMax-M3's acquisition rate falls from 62.5\% to 40.6\%, whereas all five GPT models acquire more support under endpoint-only telemetry. GPT-5.6-Terra further illustrates distinction between acquisition and reporting: its acquisition rate rises from 65.6\% to 68.8\%, while its citation coverage falls from 40.6\% to 28.1\%. In the framework comparison, citation coverage decreases for all four frameworks. The direction of change depends on incident composition (Appendix~\ref{app:telemetry_witnesses}). Thus, preserving registered evidence does not guarantee consistent evidence acquisition or citation coverage across telemetry conditions.

The APT27 Ref-ReAct runs provide a concrete example. Both views preserve the registered endpoint records establishing that \texttt{w3wp.exe} created \texttt{error2.aspx}. Under Full telemetry, a host query returns the complete witness before any network record is returned. Under endpoint-only telemetry, the agent queries an unavailable web-access source and never acquires that witness. This case illustrates a limitation in the agent's ability to adapt its investigation to available telemetry, even when the required supporting evidence remains intact.

\subsection{RQ4: Is acquired support preserved in citations?}
\label{sec:submission_results}
\label{sec:failure_mechanisms}

We revisit the citation-support losses in RQ2 to determine whether the changed-view investigations acquired sufficient evidence for the affected actions. In the eleven-model Full-to-endpoint comparison, the query returns lack a complete witness for 72.7\% of these losses. For the remaining 27.3\%, they contain a complete witness, but the formal-stage citations do not. Under tiered retention, this 

\begin{wrapfigure}[20]{r}{0.53\textwidth}
\centering
\vspace{-5pt}
\includegraphics[width=\linewidth]{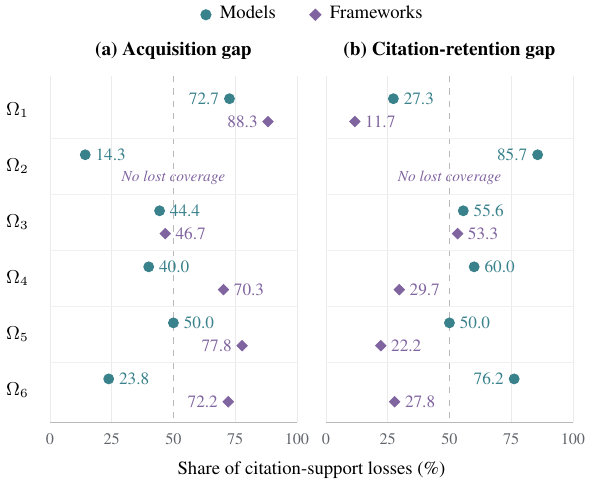}
\caption{Sources of citation-support loss. Percentages partition Full-to-changed losses into acquisition and citation-retention gaps.}
\label{fig:telemetry-loss-location}
\end{wrapfigure}

reporting gap accounts for 55.6\% of citation-support losses (Figure~\ref{fig:telemetry-loss-location}). Beyond these identifier-based checks, a content-level audit of one OpenCode/MiniMax-M3 Spyder run confirms that complete evidence was returned for five actions still lacking sufficient formal-stage citations (Appendix~\ref{app:examples}). Thus, lost report coverage cannot always be explained by insufficient evidence acquisition.

Further analyses show investigations can extend beyond the supplied leads while their reports omit support already returned by the lead queries (Appendix~\ref{app:r1_synthesis}). These results reveal a limitation in agents' ability to preserve acquired evidence in their reports. For intrusion scoping, obtaining more evidence is not enough. Agents must also retain the record combinations needed to substantiate attack activities in final reports.

\WFclear
\FloatBarrier

\section{Related work}
\label{sec:related}

\textbf{APT investigation.}
Provenance research provides intrusion-analysis datasets \citep{darpatc2020,optc2020} and methods for detecting, reconstructing, and attributing attack activity \citep{holmes2019,kairos2024,magic2024,dupin2026,knowhow2026,orthrus2025,provx2026}, supporting identification and correlation from system records. Re-evaluations show that event granularity and preprocessing can affect measured performance \citep{simpler2025}, motivating consistent evaluation targets. \bench{} complements this work by evaluating autonomous LLM-agent investigation from unverified leads to cited reports, examining whether evidence is acquired and preserved as telemetry changes.

\textbf{Security benchmarks.}
Cybench and CyberGym evaluate agents through capture-the-flag tasks and executable vulnerability reproduction, respectively \citep{cybench2025,cybergym2026}. Closer to intrusion investigation, ExCyTIn-Bench evaluates threat-investigation QA through interactive log queries \citep{excytin2026}, DiagChain evaluates ordered attack-chain reconstruction across noise profiles \citep{diagchain2026}, and SynthChain examines software supply-chain attack reconstruction under different source combinations \citep{synthchain2026}. \bench{} instead evaluates cross-telemetry robustness by comparing the same recoverable actions under fixed support rules, including stricter comparisons with unchanged registered evidence. Applying the same sufficiency criteria to available records, query returns, and formal citations separates telemetry limitations from failures to acquire or preserve support. Table~\ref{tab:aptinvestbench-comparison} summarizes these differences.
\section{Conclusion and Future Work}
\label{sec:conclusion}

We introduce \bench{}, a benchmark for autonomous multi-step APT investigation with 370 cases derived from 56 report-informed reconstructions under seven SOC-inspired telemetry conditions. Our ten-scenario study of eleven LLMs and four agent frameworks shows that agents advance investigations beyond the supplied leads, but evidence delivery remains incomplete. Across the eleven models, investigations acquire sufficient support for 44.3\% of recoverable actions on average, while formal citations support only 25.0\%. Similar overall coverage can conceal lost support for previously covered activities, and unchanged registered evidence does not guarantee consistent investigation outcomes. By distinguishing telemetry limitations from failures to acquire and preserve evidence, \bench{} identifies concrete targets for improving autonomous investigation and the evidence delivered for intrusion scoping.
We will extend \bench{} beyond LLM agents to provenance-based investigation methods and expand the resource with newly reported APT activity and defensive knowledge under versioned development and held-out splits. These extensions aim to support defensive blind-spot analysis and the development of stronger security-focused models.

\label{main-text-end}

\bibliography{references}
\bibliographystyle{iclr2027_conference}

\appendix
\makeatletter
\patchcmd{\LT@output}{\vss}{\vskip 0pt plus 1fil minus \dimexpr\ht\@arstrutbox+\dp\@arstrutbox\relax}
  {\typeout{APPENDIX-LT-PATCH-1 applied}}
  {\PackageWarning{appendix-longtable}{First compatibility patch not applied}}
\patchcmd{\LT@output}{\vss}{\vskip 0pt plus 1fil minus \dimexpr\ht\@arstrutbox+\dp\@arstrutbox\relax}
  {\typeout{APPENDIX-LT-PATCH-2 applied}}
  {\PackageWarning{appendix-longtable}{Second compatibility patch not applied}}
\makeatother

\FloatBarrier
\clearpage
\section*{Appendix overview}

The appendices provide additional details on task construction, evaluation methodology, expanded results, and case studies, organized as follows:

\begin{itemize}
\setlength{\itemsep}{7pt}
\setlength{\parsep}{0pt}

\item \textbf{Appendix~\ref{app:implementation}: Investigation task construction and execution.}
We describe resource scopes, scenario attribution, report-informed construction, evidence binding, observation conditions, and the investigation interface and execution contract.

\item \textbf{Appendix~\ref{app:rule_comparison}: Evidence requirements and scoring checks.}
We specify scoring conventions and witness requirements, distinguish returned identifiers from complete content, and verify implementation behavior under alternative scoring criteria.

\item \textbf{Appendix~\ref{app:results}: Experimental configurations and analysis populations.}
We document execution settings, resource use, latency and
query-level acquisition progress, and the denominators used
in complementary analyses.

\item \textbf{Appendix~\ref{app:investigation_profiles}: Supplementary investigation diagnostics.}
We expand baseline investigation results through lead-query references, investigation depth, behavior coverage, temporal-relation diagnostics, and sensitivity to incident composition.

\item \textbf{Appendix~\ref{app:telemetry_robustness}: Robustness across telemetry conditions.}
We report condition-specific results and paired action transitions, assess unchanged-witness comparisons, and distinguish telemetry-induced support changes from investigation failures.

\item \textbf{Appendix~\ref{app:r1_synthesis}: Acquisition gaps and citation retention.}
We analyze acquisition gaps and citation omissions through lead-query comparisons, matched-support controls, witness requirements, and citation-selection analyses.

\item \textbf{Appendix~\ref{app:examples}: Investigation examples and limits of citation-based evaluation.}
We present APT27 and Spyder investigations, verified cases where acquired support was omitted from formal citations, and assertion reviews illustrating the semantic limits of citation-based evaluation.

\end{itemize}
\section{Investigation task construction and execution}
\label{app:implementation}


\subsection{Resource scopes and scoring units}
Table~\ref{tab:corpus_scopes} summarizes the complete resource and controlled study, with exact counts and duration ranges. A \emph{scenario} is one report-informed attack reconstruction with normal background activity. A \emph{case} pairs a scenario with an applicable observation condition, and a \emph{run} executes one configuration on that case. An \emph{action--view pair} is one reference action evaluated under one observation condition. A \emph{source-incident group} contains all reconstructions derived from the same reported incident. Resampling keeps each group together. Physical-record totals count each scenario once, regardless of how many cases or runs reuse its records.

\begin{table}[htbp]
\caption{Physical-scenario inventory of APTInvestBench and its controlled-study subset. Attack spans and observation windows describe the reconstructed episode.}
\label{tab:corpus_scopes}
\centering\small
\setlength{\tabcolsep}{3pt}
\renewcommand{\arraystretch}{1.08}
\begin{tabularx}{\linewidth}{@{}Xrr@{}}
\toprule
Inventory item & \shortstack{Complete resource\\(56 scenarios)} & \shortstack{Controlled study\\(10 scenarios)} \\
\midrule
Scenarios & 56 & 10 \\
Physical records & 16,431,768 & 1,831,162 \\
Malicious records & 10,507 & 1,473 \\
Benign background records & 16,421,261 & 1,829,689 \\
Malicious share (\%) & 0.0639 & 0.0804 \\
Records per scenario (k), range & 19.6--527.6 & 109.0--231.1 \\
\midrule
Annotated action instances & 523 & 76 \\
Raw threat-family labels & 28 & 9 \\
Named actor labels / other labels & 23 / 3 & 7 / 2 \\
Log formats / environment types & 16 / 6 & 16 / 3 \\
\midrule
Attack span (min), median & 173.25 & 282.45 \\
Attack span (min), range & 63.9--1422.8 & 90.7--1422.8 \\
Observation window (h), median & 54.5 & 32.5 \\
Observation window (h), range & 16--90 & 16--44 \\
\bottomrule
\end{tabularx}
\par\vspace{3pt}\begin{minipage}{\linewidth}\footnotesize Actions count source event instances. Benign records are normal background within attack scenarios.\end{minipage}
\end{table}

\subsection{Scenario catalogue, attribution}
Table~\ref{tab:scenes} lists all 56 scenarios, with the ten controlled-study scenarios first. Stable identifiers link each reconstruction to its attribution and duration. Threat labels follow the source reports and may name an actor, a campaign, or malware. 

The complete resource has 28 raw threat-family labels. Verified alias normalization merges UNC3524 with APT29 and APT36 with Transparent Tribe \citep{unc3524_mandiant2022,transparenttribe_mitre2026}. It contains 23 named-actor label clusters and three separately retained malware or unresolved-attribution labels: COATHANGER, Purple Fox, and Zardoor. The controlled study retains seven named-actor clusters, with Purple Fox and Zardoor as additional labels \citep{purplefox_guardicore2021,zardoor_talos2024}. Spyder retains the source report's tentative Patchwork attribution \citep{spyder_qianxin2023}. These labels can overlap across vendor naming systems.

\paragraph{Attack span and observation window.}
Attack span runs from the first to the last annotated event, in minutes. The observation window is the declared collection interval, in hours, including surrounding background activity. Scenario identifiers link each saved reconstruction to the supplied duration workbook. All entries agree with intervals calculated from the reference timestamps at the workbook's displayed precision. 

\begingroup
\footnotesize
\setlength{\tabcolsep}{2pt}
\renewcommand{\arraystretch}{1.00}
\setlength{\LTleft}{0pt}
\setlength{\LTright}{0pt}
\setlength{\LTcapwidth}{\linewidth}
\setlength{\LTpost}{6pt}
\begin{longtable}{@{}
  >{\raggedright\arraybackslash}p{\dimexpr.39\linewidth-\tabcolsep\relax}
  >{\raggedright\arraybackslash}p{\dimexpr.175\linewidth-2\tabcolsep\relax}
  >{\raggedleft\arraybackslash}p{\dimexpr.09\linewidth-2\tabcolsep\relax}
  >{\raggedleft\arraybackslash}p{\dimexpr.10\linewidth-2\tabcolsep\relax}
  >{\raggedleft\arraybackslash}p{\dimexpr.06\linewidth-2\tabcolsep\relax}
  >{\raggedleft\arraybackslash}p{\dimexpr.095\linewidth-2\tabcolsep\relax}
  >{\raggedleft\arraybackslash}p{\dimexpr.09\linewidth-\tabcolsep\relax}@{}}
\caption{All 56 APTInvestBench scenarios. The ten controlled-study scenarios appear first, followed by the remaining 46 scenarios in the complete resource. Stable IDs identify source reconstructions; embedded dates are not collection dates.}\label{tab:scenes}\\
\toprule
Scenario ID & Threat label & Logs (k) & Mal. (\%) & Acts & \shortstack{Attack\\(min)} & \shortstack{Window\\(h)} \\
\midrule
\endfirsthead
\multicolumn{7}{@{}l}{\tablename\ \thetable\ (continued)}\\
\toprule
Scenario ID & Threat label & Logs (k) & Mal. (\%) & Acts & \shortstack{Attack\\(min)} & \shortstack{Window\\(h)} \\
\midrule
\endhead
\midrule
\multicolumn{7}{r@{}}{\footnotesize Continued on next page}\\
\endfoot
\bottomrule
\endlastfoot
\multicolumn{7}{@{}l}{\textit{Controlled-study subset: ten scenarios}}\\*
\texttt{2014-\allowbreak 06-\allowbreak 30-\allowbreak dragonfly-\allowbreak energetic-\allowbreak bear} & Dragonfly & 109.0 & 0.0670 & 7 & 90.7 & 16 \\
\texttt{2018-\allowbreak 04-\allowbreak 25-\allowbreak apt-\allowbreak c-\allowbreak 01} & APT-C-01 & 231.1 & 0.1653 & 5 & 268.0 & 34 \\
\texttt{2019-\allowbreak 05-\allowbreak 28-\allowbreak apt27} & APT27 & 166.3 & 0.0505 & 13 & 318.5 & 44 \\
\texttt{2021-\allowbreak 12-\allowbreak 23-\allowbreak apt-\allowbreak q-\allowbreak 38-\allowbreak donot} & Donot & 212.5 & 0.0339 & 11 & 1,422.8 & 34 \\
\texttt{2023-\allowbreak 04-\allowbreak 07-\allowbreak muddywater} & MuddyWater & 200.4 & 0.0200 & 5 & 304.4 & 30 \\
\texttt{2023-\allowbreak 07-\allowbreak 04-\allowbreak spyder} & Patchwork$^\ddagger$ & 118.2 & 0.0973 & 10 & 399.7 & 18 \\
\texttt{apt-\allowbreak c-\allowbreak 44-\allowbreak purplefox-\allowbreak 2fe8} & Purple Fox$^\dagger$ & 224.1 & 0.0165 & 6 & 151.6 & 39 \\
\texttt{apt27} & APT27 & 196.5 & 0.1517 & 4 & 173.2 & 33 \\
\texttt{desert-\allowbreak falcons} & Desert Falcons & 181.3 & 0.0463 & 7 & 93.8 & 31 \\
\texttt{unknown-\allowbreak zaardoor} & Zardoor$^\dagger$ & 191.7 & 0.1502 & 8 & 296.9 & 32 \\
\addlinespace[2pt]
\multicolumn{2}{@{}l}{Controlled-study subtotal} & 1,831.2 & 0.0804 & 76 & --- & --- \\
\midrule
\multicolumn{7}{@{}l}{\textit{Complete resource: remaining 46 scenarios}}\\*
\texttt{2016-\allowbreak 11-\allowbreak 17-\allowbreak apt23} & APT23 & 454.4 & 0.0139 & 11 & 449.9 & 69 \\
\texttt{2016-\allowbreak 11-\allowbreak 22-\allowbreak apt23} & APT23 & 441.3 & 0.0145 & 10 & 753.8 & 76 \\
\texttt{2018-\allowbreak 03-\allowbreak 13-\allowbreak muddywater} & MuddyWater & 322.5 & 0.0301 & 9 & 442.5 & 76 \\
\texttt{2021-\allowbreak 03-\allowbreak 15-\allowbreak bitter} & Bitter & 364.3 & 0.0804 & 11 & 320.1 & 68 \\
\texttt{2022-\allowbreak 09-\allowbreak 08-\allowbreak lazarus} & Lazarus & 527.6 & 0.0398 & 18 & 145.3 & 90 \\
\texttt{2023-\allowbreak 05-\allowbreak 22-\allowbreak bluenoroff} & BlueNoroff & 215.0 & 0.1330 & 6 & 101.7 & 38 \\
\texttt{2024-\allowbreak 02-\allowbreak 19-\allowbreak lazarus-\allowbreak defense-\allowbreak supply-\allowbreak chain} & Lazarus & 342.7 & 0.0161 & 8 & 120.0 & 57 \\
\texttt{2024-\allowbreak 08-\allowbreak 26-\allowbreak apt-\allowbreak q-\allowbreak 12} & APT-Q-12 & 317.4 & 0.0495 & 7 & 144.9 & 59 \\
\texttt{2025-\allowbreak 07-\allowbreak 11-\allowbreak apt-\allowbreak q-\allowbreak 36-\allowbreak patchwork} & Patchwork & 372.7 & 0.0910 & 15 & 291.6 & 57 \\
\texttt{apt10} & APT10 & 480.7 & 0.0212 & 8 & 108.5 & 72 \\
\texttt{apt10-\allowbreak apt10} & APT10 & 329.8 & 0.0246 & 6 & 136.2 & 62 \\
\texttt{apt27-\allowbreak bf71} & APT27 & 293.1 & 0.1075 & 4 & 124.5 & 53 \\
\texttt{apt27-\allowbreak e617} & APT27 & 442.7 & 0.0147 & 4 & 223.5 & 79 \\
\texttt{apt29-\allowbreak 72e3} & APT29 & 315.3 & 0.0323 & 10 & 479.9 & 56 \\
\texttt{apt29-\allowbreak 9568} & APT29 & 462.7 & 0.0140 & 10 & 480.5 & 82 \\
\texttt{apt29-\allowbreak b113} & APT29 / UNC3524 & 346.3 & 0.0494 & 14 & 425.6 & 70 \\
\texttt{lazarus-\allowbreak de46} & Lazarus & 305.0 & 0.0646 & 13 & 112.1 & 48 \\
\texttt{lazarus-\allowbreak f951} & Lazarus & 184.9 & 0.2883 & 12 & 351.3 & 30 \\
\texttt{muddywater-\allowbreak dev} & MuddyWater & 374.2 & 0.0110 & 5 & 275.2 & 71 \\
\texttt{uat4356} & UAT4356 & 409.5 & 0.0051 & 6 & 97.5 & 75 \\
\texttt{2014-\allowbreak 07-\allowbreak 02-\allowbreak dragonfly-\allowbreak energetic-\allowbreak bear} & Dragonfly & 266.3 & 0.0601 & 5 & 94.5 & 40 \\
\texttt{2018-\allowbreak 09-\allowbreak 20-\allowbreak apt-\allowbreak c-\allowbreak 01} & APT-C-01 & 19.6 & 2.1879 & 21 & 170.9 & 18 \\
\texttt{2018-\allowbreak 11-\allowbreak 19-\allowbreak apt29-\allowbreak not-\allowbreak so-\allowbreak cozy} & APT29 & 177.1 & 0.0779 & 9 & 71.9 & 27 \\
\texttt{2019-\allowbreak 03-\allowbreak 05-\allowbreak transparenttribe} & APT36 & 188.0 & 0.1702 & 18 & 731.6 & 31 \\
\texttt{2021-\allowbreak 05-\allowbreak 11-\allowbreak lazarus} & Lazarus & 343.1 & 0.0443 & 7 & 156.9 & 57 \\
\texttt{2021-\allowbreak 06-\allowbreak 08-\allowbreak sidewinder} & SideWinder & 300.7 & 0.1240 & 9 & 231.7 & 49 \\
\texttt{2021-\allowbreak 11-\allowbreak 23-\allowbreak donot-\allowbreak apt-\allowbreak q-\allowbreak 38} & Donot & 129.1 & 0.0992 & 29 & 63.9 & 22 \\
\texttt{2022-\allowbreak 02-\allowbreak 24-\allowbreak muddywater} & MuddyWater & 325.0 & 0.0757 & 7 & 206.8 & 49 \\
\texttt{2022-\allowbreak 12-\allowbreak 27-\allowbreak bluenoroff-\allowbreak stardust-\allowbreak chollima} & BlueNoroff & 334.2 & 0.0185 & 6 & 115.2 & 53 \\
\texttt{2024-\allowbreak 10-\allowbreak 12-\allowbreak bitter-\allowbreak apt-\allowbreak q-\allowbreak 37} & Bitter & 23.9 & 0.2013 & 13 & 110.8 & 26 \\
\texttt{2024-\allowbreak 11-\allowbreak 26-\allowbreak apt-\allowbreak c-\allowbreak 48-\allowbreak cnc-\allowbreak v2} & APT-C-48 & 284.1 & 0.0148 & 6 & 135.3 & 45 \\
\texttt{2025-\allowbreak 02-\allowbreak 28-\allowbreak apt-\allowbreak q-\allowbreak 38-\allowbreak donot} & Donot & 485.9 & 0.0338 & 14 & 688.5 & 77 \\
\texttt{2025-\allowbreak 06-\allowbreak 24-\allowbreak apt-\allowbreak q-\allowbreak 14} & APT-Q-14 & 341.6 & 0.0817 & 6 & 173.3 & 56 \\
\texttt{2025-\allowbreak 08-\allowbreak 28-\allowbreak lazarus} & Lazarus & 437.4 & 0.1216 & 8 & 116.0 & 68 \\
\texttt{apt36} & APT36 & 466.8 & 0.0962 & 15 & 465.2 & 82 \\
\texttt{apt36-\allowbreak 83d4} & APT36 & 338.5 & 0.0251 & 11 & 76.5 & 65 \\
\texttt{bluenoroff-\allowbreak lazarus} & BlueNoroff & 345.9 & 0.0755 & 8 & 150.6 & 64 \\
\texttt{bluenoroff-\allowbreak stardust-\allowbreak chollima} & BlueNoroff & 256.9 & 0.0315 & 6 & 108.8 & 48 \\
\texttt{chimera} & Chimera & 132.1 & 0.0197 & 10 & 92.5 & 20 \\
\texttt{chinese-\allowbreak state-\allowbreak sponsored} & COATHANGER$^\dagger$ & 225.9 & 0.1107 & 6 & 153.9 & 39 \\
\texttt{dragonfly-\allowbreak energetic-\allowbreak bear} & Dragonfly & 335.3 & 0.0438 & 5 & 98.7 & 63 \\
\texttt{earth-\allowbreak yako} & Earth Yako & 370.1 & 0.0838 & 8 & 283.7 & 68 \\
\texttt{earth-\allowbreak yako-\allowbreak yako} & Earth Yako & 314.8 & 0.1223 & 8 & 191.0 & 59 \\
\texttt{molerats} & Molerats & 303.0 & 0.0211 & 7 & 217.8 & 57 \\
\texttt{muddywater} & MuddyWater & 398.6 & 0.0753 & 10 & 423.7 & 74 \\
\texttt{oceanlotus} & OceanLotus & 154.7 & 0.2237 & 8 & 187.4 & 25 \\
\addlinespace[2pt]
\multicolumn{2}{@{}l}{Remaining-resource subtotal} & 14,600.6 & 0.0619 & 447 & --- & --- \\*
\midrule
\multicolumn{2}{@{}l}{\textbf{Complete resource total}} & 16,431.8 & 0.0639 & 523 & --- & --- \\
\end{longtable}
\noindent\begin{minipage}{\linewidth}\footnotesize Acts: source action instances. Mal.: malicious share of physical records. Attack: annotated event span (min). Window: collection duration (h). $^\dagger$Malware/campaign or unresolved attribution; $^\ddagger$tentative source-report attribution. Labels follow frozen metadata and may overlap across naming systems. Exact values, workbook cells, environments, and source hashes accompany the machine-readable catalogue.\end{minipage}
\endgroup

\FloatBarrier

\subsection{Controlled observation conditions}
\label{app:observation_conditions}
The seven observation conditions vary the available telemetry while preserving the incident and reference actions. Table~\ref{tab:observation_motivation} separates each benchmark definition from its operational motivation and sources. The references document coverage gaps, unequal retention, sampling, and selective log removal. Window lengths, sampling probabilities, and matching rules are controlled benchmark settings.

\begin{table}[t]
\caption{Definitions and real-world motivations for the seven observation conditions. Linked source titles open the original documents. The benchmark applies each condition to the same incident.}
\label{tab:observation_motivation}
\centering\footnotesize
\setlength{\tabcolsep}{3pt}
\renewcommand{\arraystretch}{1.13}
\begin{tabularx}{\linewidth}{@{}L{0.14\linewidth}L{0.265\linewidth}>{\raggedright\arraybackslash}XL{0.25\linewidth}@{}}
\toprule
Condition & Benchmark definition & Operational motivation & Sources \\
\midrule
$\Omega_0$ Full telemetry & Retain all generated telemetry as the reference record set. & SOC investigations correlate host, identity, network, DNS, and other records to establish incident scope. & \href{https://www.cyber.gov.au/business-government/detecting-responding-to-threats/event-logging/best-practices-for-event-logging-and-threat-detection}{Joint logging guidance}~\citep{logging2024} \\
\addlinespace[4pt]
$\Omega_1$ Endpoint-only & Retain endpoint-origin records, including network facts recorded by endpoint sensors. Remove independent network sources. & Host monitoring and network-flow logging are configured separately, allowing asymmetric source coverage. & \href{https://learn.microsoft.com/en-us/sysinternals/downloads/sysmon}{Sysmon}~\citep{sysmon_docs2026}\par \href{https://docs.cloud.google.com/vpc/docs/using-flow-logs}{VPC Flow Logs setup}~\citep{gcp_flow_config2026} \\
\addlinespace[4pt]
$\Omega_2$ Network-only & Retain network-origin records and remove endpoint sources. & Devices can lack endpoint agents or usable local logs. Network monitoring supplies visibility into such gaps, also documented in BRICKSTORM intrusions. & \href{https://www.cyber.gov.au/business-government/detecting-responding-to-threats/event-logging/best-practices-for-event-logging-and-threat-detection}{Joint logging guidance}~\citep{logging2024}\par \href{https://cloud.google.com/blog/topics/threat-intelligence/brickstorm-espionage-campaign}{BRICKSTORM report}~\citep{brickstorm2025} \\
\addlinespace[4pt]
$\Omega_3$ Tiered retention & Use source-specific history windows: endpoint 8\,h, network/application 6\,h, and perimeter 4\,h. & Log types and licenses have different retention periods. Long intrusions can outlast retained evidence of initial access. & \href{https://learn.microsoft.com/en-us/entra/identity/monitoring-health/reference-reports-data-retention}{Entra retention}~\citep{entra_retention2026}\par \href{https://cloud.google.com/blog/topics/threat-intelligence/brickstorm-espionage-campaign}{BRICKSTORM report}~\citep{brickstorm2025} \\
\addlinespace[4pt]
$\Omega_4$ Group sampling & Retain network/perimeter event groups with seeded probability $p=0.5$, keeping records within a group together. & Flow selection and production flow-log sampling reduce collection and processing demands. Event groups preserve within-group associations in this controlled instance. & \href{https://www.rfc-editor.org/rfc/rfc7014.html}{RFC 7014}~\citep{rfc7014}\par \href{https://docs.cloud.google.com/vpc/docs/flow-logs}{VPC Flow Logs sampling}~\citep{gcp_flow_sampling2026} \\
\addlinespace[4pt]
$\Omega_5$ Targeted suppression & Remove records intersecting every registered sufficient witness of an initially supported target action. & Selective log deletion conceals attacker activity. The Volt Typhoon advisory documents this behavior in confirmed intrusions. & \href{https://www.cyber.gov.au/about-us/view-all-content/alerts-and-advisories/prc-state-sponsored-actors-compromise-and-maintain-persistent-access-us-critical-infrastructure}{Volt Typhoon advisory}~\citep{volt_typhoon2024} \\
\addlinespace[4pt]
$\Omega_6$ Matched benign removal & Remove an equal number of benign records, matching format and instance and approximating time and size. Preserve registered target support. & Routine filtering also changes searchable records. We construct a matched control to compare targeted support removal with general data reduction. & \href{https://www.cyber.gov.au/business-government/detecting-responding-to-threats/event-logging/implementing-siem-soar-platforms/implementing-siem-and-soar-platforms-practitioner-guidance}{SIEM/SOAR guidance}~\citep{siem_soar2025}\par \href{https://www.itl.nist.gov/div898/handbook/pri/section3/pri332.htm}{NIST control principles}~\citep{nist_blocking} \\
\bottomrule
\end{tabularx}
\par\vspace{4pt}\begin{minipage}{\linewidth}\footnotesize
Sources motivate the mechanisms. Retention windows, sampling probabilities, and removal rules are reproducible benchmark settings. Full telemetry refers to the scenario's generated records. Suppression is defined over registered support. The NIST reference motivates control of nuisance factors, while the matched log-removal procedure is our design.
\end{minipage}
\end{table}

\subsection{Report-informed construction and evidence binding}
\label{app:generation_lineage}
Construction turns incident reports into structured YAML scenarios, revised action descriptions, and environment configurations. EvidenceForge supplies the structured-event model, shared environment state, and correlated sensor records \citep{evidenceforge2026}. \bench{} binds the generated records to reviewed reference actions and sufficient-support rules, then selects the records exposed in each case.

\begin{figure}[t]
\centering 
\includegraphics[width=0.95\textwidth]{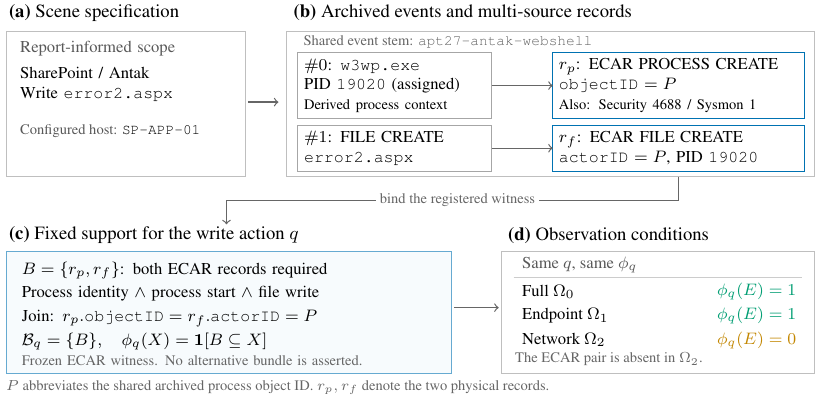} 
\caption{From a reconstructed action to a registered witness in the APT27 case. The file-creation action is supported by linked ECAR process and file records that jointly establish the process identity and file write. Additional startup records observe the same process but are not part of the registered witness. The left panel summarizes archived scenario metadata, and reconstruction-specific identifiers are assigned during instantiation. Cross-telemetry comparisons evaluate whether this registered support remains available and is recovered under different observation conditions.}
\label{fig:generation_support_example}
\end{figure}

\paragraph{From report to instantiated scenario.}
Authors encode three kinds of information. Reported behavior sets the incident scope and action sequence. Configured environment details specify hosts, users, addresses, paths, and timing. Protocol behavior and execution prerequisites add the process or connection context needed to render those actions. Scenario descriptions and selected event declarations document these roles, although field-level provenance tags remain incomplete. For example, the archived APT27 description identifies the configured values used in place of redacted victim details, while the IIS worker event identifies platform-derived process context.

The specification assigns identities to logical events and links them before rendering sensor records. A process start can appear in ECAR, Windows Security, and Sysmon, while one connection can produce endpoint-flow, network-monitor, and application records. Event templates specify the activities, environment configuration supplies their context, and format emitters render the observations. Background activity uses the same environment. Each rendering preserves its event's shared identity.

\paragraph{Archived lineage and benchmark adaptation.}
Generation receipts identify source YAML files by location and digest. The lineage manifest records these digests, generated-event counts, annotation-review receipts, and per-view bindings. The benchmark extracts fields and linking identifiers from existing raw records, then binds reviewed actions to record identities. Action definitions can change without altering physical records, as in the 2019 APT27 reconstruction's revision from nine to thirteen actions. Observation conditions select records from the fixed scenario. The same rule then tests support in available records, query returns, and formal-stage citations.

\paragraph{A concrete binding.}
Figure~\ref{fig:generation_support_example} illustrates APT27 ASPX file creation. The registered witness links process context to a file-creation record through the process object and actor identity. This witness survives full and endpoint-only collection but is absent under network-only observation. Additional startup records describe the same event. The provenance audit verifies these records at their archived byte locations, together with records of a subsequent connection.

These assets serve distinct purposes: review receipts document annotation decisions, byte checks verify record bindings, and action rules specify the evidence required for scoring.
\FloatBarrier

\subsection{Autonomous investigation interface and execution}
\label{app:online_protocol}
\paragraph{Task instructions.}
The investigator receives unverified leads, a query-client guide,
and a prediction schema. It queries the audited telemetry interface
and submits an evidence graph. Private labels, support rules,
alternate raw stores, and scenario identities remain inaccessible.
The system prompt opens with the following instructions.

\begin{tcolorbox}[
  enhanced,
  breakable,
  title={System prompt: instructions},
  colback=white,
  colframe=black!75,
  colbacktitle=black!85,
  coltitle=white,
  fonttitle=\bfseries\small,
  fontupper=\footnotesize,
  sharp corners,
  boxrule=0.5pt,
  left=7pt,
  right=7pt,
  top=6pt,
  bottom=6pt,
  before skip=7pt,
  after skip=7pt
]
\raggedright
You are a security-telemetry investigation analyst.

\medskip
Using the read-only endpoint, network, and perimeter telemetry
available through the audited interface, complete the authorized
SOC triage within a cumulative budget of 50,000 model-output tokens.

\medskip
Determine whether a security incident is supported, verify supplied
leads first, identify observable actions, and reconstruct their
timing and relationships.

\medskip
Submit an evidence graph with record-ID citations.
Take no action against real systems.
\end{tcolorbox}


The following excerpt reproduces an executed fixed-budget contract. Commands and substantive instructions are unchanged. The archived artifact preserves the verbatim executed text and original release identities.

\begin{tcolorbox}[
  enhanced,
  breakable,
  title={APTInvestBench investigation instructions},
  colback=white,
  colframe=black!75,
  colbacktitle=black!85,
  coltitle=white,
  fonttitle=\bfseries\small,
  fontupper=\footnotesize,
  sharp corners,
  boxrule=0.5pt,
  left=7pt,
  right=7pt,
  top=6pt,
  bottom=6pt,
  before skip=7pt,
  after skip=7pt
]
\raggedright
\textbf{Initial SOC handoff}
\begin{itemize}[leftmargin=1.2em,labelsep=0.4em,itemsep=3pt,topsep=3pt,parsep=0pt]
  \item \textbf{Inputs.} The clue condition is \texttt{alert\_queue}. Read the machine-readable handoff at \path{/workspace/observations/initial_clues.json}.

  \item \textbf{Unverified alerts.} The queue contains three alerts, which may include unrelated false positives. The benchmark does not reveal which alert, if any, belongs to the incident. Treat every alert as an unverified triage lead and base conclusions on case-visible records.

  \item \textbf{Required pivots.} Alert IDs identify queue entries, not telemetry records; they cannot be passed to \texttt{record\_lookup.py show}. Execute the exact \path{pivot.record_lookup_argv} supplied for each alert before interpreting it. A pivot that returns no records does not establish a false positive: the current view may not retain the originating record.
\end{itemize}

\medskip
\textbf{Core investigation contract}
\begin{itemize}[leftmargin=1.2em,labelsep=0.4em,itemsep=3pt,topsep=3pt,parsep=0pt]
  \item \textbf{Telemetry access.} Use \path{/workspace/observations/record_lookup.py} for all telemetry inventory, search, and read operations, as well as prediction validation. Do not inspect or probe alternative raw telemetry stores. Query syntax and pagination are documented in \path{/workspace/observations/README.md}.

  \item \textbf{Output specification.} Consult \path{/workspace/prediction-guide.md} and \path{/workspace/prediction.examples.json} for field definitions, safe-write instructions, and a minimal no-answer example. The machine-readable schema is \path{/workspace/prediction.schema.json}.

  \item \textbf{Submission.} Only \path{/workspace/prediction.json} is scored. First validate \path{/workspace/prediction.candidate.json}. Replace the scored file atomically only after validation returns \texttt{valid: true}.

  \item \textbf{Execution limits.} The evaluation budget is $B=50{,}000$ cumulative model-output tokens. Separate safety limits cap execution at 200 model requests and audited-query exposure at 100,000 records and 250,000,000 bytes. Query and tool-call counts do not define the evaluation budget.
\end{itemize}

\medskip
\textbf{Formal claims and uncertainty}
\begin{itemize}[leftmargin=1.2em,labelsep=0.4em,itemsep=3pt,topsep=3pt,parsep=0pt]
  \item \textbf{Supported actions.} Add an action to \texttt{stages} only when case-visible records directly support it. A stage remains subject to false-positive and overclaim checks even if it is marked \texttt{likely} or assigned low \texttt{confidence}.

  \item \textbf{Unresolved interpretations.} Place plausible but unproved interpretations in top-level \texttt{hypotheses}, and evidence gaps or competing explanations in top-level \texttt{uncertainties}. Do not encode hypotheses or unknown actions as formal stages.

  \item \textbf{Action granularity.} Merge records from different sensors when they describe the same logical event instance. Keep actions separate when they have an independent actor, object, time, or security effect. A shared host, temporal proximity, or common storyline alone does not justify merging.

  \item \textbf{Incident attribution.} Attribute suspicious background activity to the incident only when cited evidence establishes the connection. Temporal ordering alone establishes, at most, a temporal relation.

  \item \textbf{Output scope.} Include only benchmark scoring data in \texttt{prediction.json}. Exclude containment, remediation, monitoring, and other operational recommendations.
\end{itemize}
\end{tcolorbox}

The action-coverage endpoint checks formal-stage citations. Appendix~\ref{sec:relations} separately examines reported relations, and the assertion audit reviews selected claims against the prompt's semantic requirements. One preserved README sentence says that leads originate only from visible records, whereas executed lead bindings can refer to records absent from the current view. The archive retains both artifacts and documents the discrepancy.

\paragraph{Query interface.}
The query interface records what the investigator receives. The commands below use public placeholders. \texttt{summary} returns aggregate source counts and time ranges. Searches return citable records within a maximum 32-KiB response, with \texttt{has\_more} and an opaque cursor for pagination. Batch \texttt{show} accepts one to eight unique IDs and returns full records. The exposure ledger records returned IDs and bytes.

\begin{tcolorbox}[
  enhanced,
  breakable,
  title={Telemetry query interface},
  colback=white,
  colframe=black!75,
  colbacktitle=black!85,
  coltitle=white,
  fonttitle=\bfseries\small,
  sharp corners,
  boxrule=0.5pt,
  left=7pt,
  right=7pt,
  top=6pt,
  bottom=6pt,
  before skip=7pt,
  after skip=7pt
]
\begin{lstlisting}[
  style=ocprompt,
  frame=none,
  aboveskip=0pt,
  belowskip=0pt
]
python3 record_lookup.py summary
python3 record_lookup.py summary --offset <emitted_offset>

python3 record_lookup.py search '<observed_pattern>' \
  --source-format <format> --source-instance <instance> \
  --start <inclusive_time_with_timezone> \
  --end <inclusive_time_with_timezone>

python3 record_lookup.py continue <opaque_next_cursor>

python3 record_lookup.py show <record_id> [<record_id> ...]

python3 record_lookup.py validate-prediction \
  /workspace/prediction.candidate.json
\end{lstlisting}
\end{tcolorbox}

\paragraph{Submission and execution workflow.}
The schema separates supported observations from open interpretations. Formal \texttt{stages} contribute citations, including stages marked \emph{likely}. Unverified interpretations belong in \texttt{hypotheses}, gaps in \texttt{uncertainties}, and unsupported roles in null fields. Temporal edges record order, while stronger relations require a direct evidential link. An empty stage set with explicit gaps expresses insufficient support without resolving incident status. Automated validation checks structure and exposed IDs. The assertion audit separately inspects how selected reports interpret their citations.

Before execution, the builder fixes the visible records, leads, agent-visible files, and private rules. For each configuration and case, the runner creates the environment and broker and launches the investigator. It saves the available interaction stream, exposure, validation, network-audit, and finalization receipts.

In fixed-budget arms, the gateway counts cumulative model-output tokens and separately limits the number of requests. A final response that begins before the output allowance is crossed may finish beyond that allowance. A valid candidate is then published atomically for private scoring. 

\FloatBarrier
\section{Evidence requirements and scoring checks}
\label{app:rule_comparison}
This appendix defines sufficient action support and checks the scorer's behavior. It distinguishes returned record identifiers from complete response content and citation support from claim correctness. Table~\ref{tab:verification_guide} summarizes these checks. Current temporal-relation diagnostics appear with their results in Appendix~\ref{app:current_relations}.

\subsection{Current rules, denominators, and supplementary scores}
\label{app:metrics}
\label{app:current_rules}

\paragraph{Units and undefined ratios.}
Unless stated otherwise, rates pool eligible action--condition pairs over each table's specified cases. Equation~\ref{eq:decomp} applies to these pooled rates when their denominators are nonzero. Averaging per-case ratios instead gives a case-averaged rate, which need not satisfy this factorization.

Acquisition and citation coverage are undefined when $R=0$, and retention is undefined when $O=0$. We report these ratios as N/A and omit them from case-averaged rates. An accepted empty report has $K=0$ and receives zero coverage when $R>0$. Runs with missing or unusable query-return audits are excluded from acquisition and retention estimates. The exported results identify the included cases and actions for each denominator.

\paragraph{Citation scope and record-level measures.}
The primary set $C^\cup$ deduplicates citations attached to formal stages. It excludes citations found only in top-level hypotheses, uncertainties, entities, or edges. For visible source-labelled malicious records $M$, incident-record citation share and malicious-log recall describe the cited records:
\begin{equation}
\mathrm{Prec}_{m}=\frac{|C^\cup\cap M|}{|C^\cup|},\qquad
\mathrm{Rec}_{m}=\frac{|C^\cup\cap M|}{|M|}.
\label{eq:malicious_logs}
\end{equation}
Both measures pool record counts over the same eligible cases. Historical support-log precision and recall instead use citations from all report fields and compare them with visible registered support records. 

\paragraph{Claim-local evidence assignment.}
Claim-local coverage checks whether an individual claim cites a witness. The fixed matching procedure assigns each prediction object to at most one reference action and each reference to at most one prediction. With $m$ formal objects and $H$ matched pairs,
\begin{equation}
\mathrm{Precision}_{H}=H/m,\qquad
\mathrm{Recall}_{H}=H/R,\qquad
F_{1,H}=\frac{2H}{m+R}.
\label{eq:f1}
\end{equation}
Zero-denominator cases follow the preceding N/A convention. These scores depend on how a report divides its claims. Matching checks each claim's evidence without independently judging whether its text expresses the reference behavior. The same distinction applies to relation endpoints (Appendix~\ref{sec:relations}).

\subsection{An exact threshold witness}
\label{app:witness_rules}
Table~\ref{tab:support_rule_examples} contrasts complementary records for one event with repeated observations needed to establish a temporal pattern. The file-creation binding is illustrated in Appendix~\ref{app:generation_lineage}. The recurring-network rule is worked through below.

\begin{table}[tb]
\caption{Two registered support rules. Each rule checks observable facts and their links. Section~\ref{sec:task} introduces the file-creation example, and Section~\ref{sec:evaluation} follows it through a saved investigation.}
\label{tab:support_rule_examples}
\centering\small
\begin{tabularx}{\linewidth}{@{}L{0.20\linewidth}XX@{}}
\toprule
Reference action & Required observations & Binding and compatibility \\
\midrule
APT27 process creates \texttt{error2.aspx} & ECAR process creation and file creation. & Process \texttt{objectID} equals file \texttt{actorID}, linking process identity and start to the write. \\
\addlinespace
APT-C-01 recurring HTTP communication & At least three distinct HTTP observations spanning at least 60 seconds. & Compatible source and host, the required action binding, and a largest-to-smallest gap ratio of at most two. \\
\bottomrule
\end{tabularx}
\par\vspace{2pt}\begin{minipage}{\linewidth}\footnotesize
The rules establish the recorded write or temporal pattern. Webshell functionality and command-and-control attribution require additional evidence. Exact records and timing appear in Appendices~\ref{app:generation_lineage} and~\ref{app:witness_rules}.
\end{minipage}
\end{table}

The APT-C-01 recurring-network rule requires at least three distinct HTTP observations, a span of at least 60 seconds, bounded inter-arrival variation, and a binding to the reference action. Table~\ref{tab:threshold} gives one satisfying subset. The records span 410.786146 seconds, with a largest-to-smallest gap ratio of 1.4004, below the registered bound of two. Each record carries the required destination binding, so the three together form a witness.

\begin{table}[!htb]
\centering
\caption{Three records forming one sufficient recurring-network witness in APT-C-01. All are distinct Zeek HTTP events on April 25, 2018: \texttt{GET date.ocry.com:80/api/v1/status}, status 302, from the same sensor and source host in this instance. Symbols abbreviate exact physical record IDs.}
\label{tab:threshold}
\small
\begin{tabular*}{\linewidth}{@{\extracolsep{\fill}}llr@{}}
\toprule
Record & UTC timestamp & Gap from previous (s)\\
\midrule
$r_1$ & 10:48:09.820287 & --\\
$r_2$ & 10:52:09.473867 & 239.653580\\
$r_3$ & 10:55:00.606433 & 171.132566\\
\bottomrule
\end{tabular*}
\end{table}

The examples require different kinds of evidence. Process context and a file write supply complementary facts about one event. Periodicity requires repeated observations, with source and host compatibility as well as timing constraints. Counting related records alone cannot establish either rule.

\subsection{Returned identifiers versus complete content}
\label{app:contract_boundary}
Acquisition counts actions with a witness among audited returned record identifiers, including those accompanied by shortened previews. It does not measure attention or understanding. The full-content audit checks whether a witness appears in saved tool output. It matches the output to source records at verified byte locations.

The OpenCode Spyder case verifies complete response content for five omitted witnesses (Appendix~\ref{app:examples}). The two GPT traces instead reconstruct content from source records because their original wire-format responses are unavailable. Appendix~\ref{app:complete_content_omissions} reports the separate historical content audit. Neither source reconstruction nor the returned-ID ledger alone certifies complete-content exposure.

\begin{table}[ht]
\caption{Verification evidence and its scope. These checks support distinct parts of the benchmark.}
\label{tab:verification_guide}
\centering\small
\begin{tabularx}{\linewidth}{@{}L{0.24\linewidth}XX@{}}
\toprule
Check & What is established & Scope of the evidence \\
\midrule
Source-byte and binding checks & A public record maps to archived content and the inspected event identity. & Traceability of inspected bindings. Rule sufficiency also depends on the registered fact requirements. \\
\addlinespace
Oracle, empty, and report transformations & Scoring reaches expected extremes and preserves $K$ under citation-preserving transformations. & Implementation behavior under the declared rules. \\
\addlinespace
Complete-content checks & The inspected witness appears in complete persisted tool output. & Certified records and saved outputs. The main acquisition endpoint uses returned IDs. \\
\addlinespace
Bounded assertion review & Selected assertions are inspected against their own citations for activity, linkage, timing, and effects. & A case study with unresolved judgments retained. Whole-report accuracy remains unevaluated. \\
\bottomrule
\end{tabularx}
\end{table}

\subsection{Implementation stress checks and semantic limits}
\label{app:stress}
\label{sec:metric_stress}
Table~\ref{tab:verification_guide} separates implementation checks from record-exposure and assertion audits. The stress checks below test which report changes affect citation coverage. For reports $P$ and $P'$ with the same formal citation union,
\begin{equation}
\bigcup_{p\in P}C_p=\bigcup_{p'\in P'}C_{p'}\quad\Longrightarrow\quad K(P)=K(P').
\label{eq:invariance}
\end{equation}
The evaluator applies the same action rules to the same record set. Splitting, merging, or duplicating claims therefore leaves $K$ unchanged when the union is preserved. A deterministic audit verifies this property on 192 transformations of saved OpenCode reports. Duplicate objects can change the auxiliary one-to-one score $H$.




\begin{wrapfigure}{r}{0.58\textwidth}
\centering
\includegraphics[width=\linewidth]{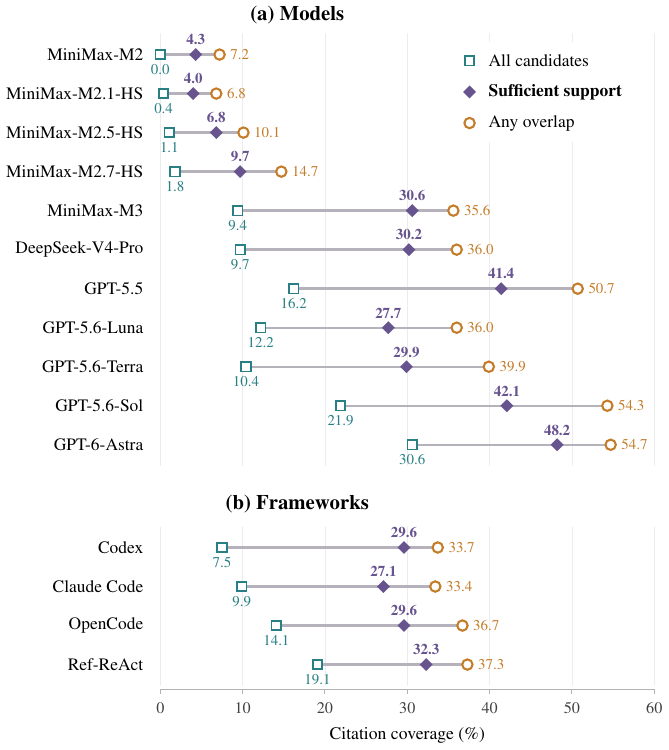}
\caption{Citation coverage under three scoring criteria on unchanged reports.}
\label{fig:scoring-criteria-comparison}
\end{wrapfigure}

\phantomsection\label{app:pooling}
\paragraph{Historical sensitivity check.}
An archived check on 64 OpenCode/MiniMax-M3 reports lets each existing formal object use the report's pooled formal-stage citations while retaining one-to-one matching. This changes $H$ without changing the reports. The primary endpoint $K$ already pools these citations and does not impose one-to-one assignment. All segmentation checks keep the reference action definitions fixed, because changing them would change which actions receive weight.

\paragraph{Effect of the evidence requirement.}
Figure~\ref{fig:scoring-criteria-comparison} compares three scoring criteria on the same saved reports. Any-overlap scoring accepts one related candidate record. Sufficient-support scoring requires a complete witness. All-candidates scoring requires every visible candidate record, including redundant alternatives.

The four earlier MiniMax variants remain at or below 14.7\% even when a single related record earns credit. Their low coverage is therefore not explained solely by the requirement to assemble a complete witness. For GPT-5.5, coverage rises from 41.4\% under sufficient support to 50.7\% under any overlap, showing that some cited fragments fail to establish complete actions. These changes isolate the effect of the scoring criterion.

\paragraph{Effect of citation placement.}
The primary score uses formal-stage citations only. A complete witness elsewhere in the report does not satisfy this submission boundary. Appendix~\ref{app:current_citation_controls} measures the effect of expanding citation extraction, while Appendix~\ref{app:examples} shows how citation placement affects individual activities.

\FloatBarrier
\section{Experimental configurations and analysis populations}
\label{app:results}

\subsection{Execution settings and resource use}
\label{sec:budget_results}
Configured limits and actual consumption are different quantities. Table~\ref{tab:runtime_config} records the executed client settings. Table~\ref{tab:resources_standard} gives consumption on each configuration's selected accepted runs.

\begin{table}[t]
\caption{Recorded client settings for the retained online configurations. Rows group configurations with shared settings. Exact per-run identities remain in the archived receipts.}
\label{tab:runtime_config}
\centering\small
\renewcommand{\arraystretch}{1.12}
\setlength{\tabcolsep}{3pt}
\begin{tabular*}{\linewidth}{@{}L{0.50\linewidth}@{\extracolsep{\fill}}llr@{}}
\toprule
Framework / models & Version & Effort & \shortstack{Response\\(K)} \\
\midrule
Codex / MiniMax models & 0.145.0 & High & 32.768 \\
Claude Code / MiniMax-M3 & 2.1.228 & High & 32.000 \\
OpenCode / MiniMax-M3 & 1.18.4 & Not set & 32.768 \\
Ref-ReAct / MiniMax-M3 & 0.2.0 & Not set & 32.768 \\
Codex / DeepSeek-V4-Pro & 0.145.0 & High & 16.384 \\
Codex / GPT-5.5, GPT-5.6-Luna, GPT-5.6-Terra, GPT-5.6-Sol, GPT-6-Astra & 0.145.0 & High & 32.768 \\
\bottomrule
\end{tabular*}
\par\vspace{3pt}\begin{minipage}{\linewidth}\footnotesize
K denotes 1,000 tokens. Entries describe configured client settings. Codex compaction thresholds are 900K for MiniMax and 80K for DeepSeek-V4-Pro. ``Not set'' refers to the effort field. The fixed-allowance MiniMax and DeepSeek-V4-Pro arms specify a 50K cumulative soft output budget and a 200-request safety limit. Higher-allowance MiniMax runs use a 1,000K/1,000-request runaway safeguard. The gateway allows 1,024 tokens of response-accounting headroom. Actual consumption appears in Table~\ref{tab:resources_standard}.
\end{minipage}
\end{table}



\paragraph{Input and output counters.}
\label{app:deep_context}
Cumulative input includes repeated processing of the conversation. It does not count distinct records read or the largest context occupied at one time. In the GPT configurations' own selected runs, cumulative input is 112.5--211.5 times output and 90.0--93.7\% is recorded as cached. Output length and formal-claim counts also depend on how a report divides its assertions and are not measures of evidence coverage.

\begin{table}[htbp]
\centering\small
\caption{Resource use on each configuration's selected accepted executions. Configured allowances and actual consumption are separate quantities.}
\label{tab:resources_standard}\label{tab:resources_natural}
\setlength{\tabcolsep}{3pt}
\renewcommand{\arraystretch}{1.08}
\begin{tabular*}{\linewidth}{@{}l@{\extracolsep{\fill}}rrrr@{}}
\toprule
System & Output (K) & P90 (K) & Over 50K (\%) & Time (s) \\
\midrule
\multicolumn{5}{@{}l}{\textit{Fixed 50K allowance}} \\
Codex / MiniMax-M2 & 8.7 & 17.2 & 0.0 & 212.9 \\
Codex / MiniMax-M2.1-HS & 9.0 & 20.5 & 0.0 & 177.8 \\
Codex / MiniMax-M2.5-HS & 8.8 & 23.8 & 0.0 & 179.2 \\
Codex / MiniMax-M2.7-HS & 9.4 & 20.4 & 0.0 & 182.7 \\
Codex / MiniMax-M3 & 26.0 & 37.5 & 0.0 & 562.0 \\
Claude Code / MiniMax-M3 & 16.2 & 25.4 & 0.0 & 262.3 \\
OpenCode / MiniMax-M3 & 11.7 & 17.8 & 0.0 & 250.3 \\
Ref-ReAct / MiniMax-M3 & 30.0 & 43.0 & 1.5 & 377.1 \\
Codex / DeepSeek-V4-Pro & 42.6 & 50.0 & 10.0 & 796.0 \\
Codex / GPT-5.5 & 16.3 & 19.6 & 0.0 & 497.2 \\
Codex / GPT-5.6-Luna & 19.5 & 24.8 & 0.0 & 586.3 \\
Codex / GPT-5.6-Terra & 10.9 & 14.0 & 0.0 & 351.0 \\
Codex / GPT-5.6-Sol & 19.9 & 26.8 & 0.0 & 954.1 \\
Codex / GPT-6-Astra & 9.8 & 12.8 & 0.0 & 493.6 \\
\midrule
\multicolumn{5}{@{}l}{\textit{MiniMax higher-allowance comparison}} \\
Codex / MiniMax-M2 & 8.3 & 14.9 & 0.0 & 205.8 \\
Codex / MiniMax-M2.1-HS & 8.8 & 15.9 & 0.0 & 197.2 \\
Codex / MiniMax-M2.5-HS & 9.3 & 17.6 & 0.0 & 199.9 \\
Codex / MiniMax-M2.7-HS & 8.8 & 20.9 & 0.0 & 190.1 \\
Codex / MiniMax-M3 & 27.1 & 46.4 & 8.1 & 601.0 \\
\bottomrule
\end{tabular*}
\par\vspace{3pt}\begin{minipage}{\linewidth}\footnotesize
Output and time are medians. Time is reported in seconds from the unrounded recorded durations. P90 is the output-token 90th percentile. Each row uses its selected accepted runs.
\end{minipage}
\end{table}

\subsection{Latency and evidence-acquisition progress}
\label{app:investigation_progress}
\label{app:exploration_value}

Table~\ref{tab:runtime_progress} combines latency and output
with evidence-acquisition progress on the shared comparison cases.
The model and framework panels retain their respective
analysis populations.

\begin{table}[t]
\centering
\caption{Latency, output, and evidence-acquisition progress
on the main comparison cases.}
\label{tab:runtime_progress}
\label{tab:runtime_summary}
\label{tab:restored_progress}

\fontsize{8}{9.5}\selectfont
\setlength{\tabcolsep}{2pt}
\renewcommand{\arraystretch}{1.12}

\begin{tabular*}{\linewidth}{
  @{}l@{\extracolsep{\fill}}rrrrrrr@{}
}
\toprule
& \multicolumn{3}{c}{\textbf{Latency and output}}
& \multicolumn{4}{c}{\textbf{Acquisition progress}} \\
\cmidrule(lr){2-4}\cmidrule(l){5-8}

\textbf{System}
& \shortstack{Median [IQR]\\(min)}
& \shortstack{P90\\(min)}
& \shortstack{Output\\(K tokens)}
& \shortstack{First\\(\%)}
& \shortstack{Partial\\(\%)}
& \shortstack{Last by\\half (\%)}
& \shortstack{Queries\\after} \\
\midrule

\multicolumn{8}{@{}l}{\textit{A. Models with Codex}} \\
MiniMax-M2
& 3.6 [3.1--4.0] & 4.9 & 8.7
& 19.4 & 72.3 & 88.2 & 9 \\
MiniMax-M2.1-HS
& 3.0 [2.6--3.5] & 4.7 & 9.1
& 17.2 & 72.3 & 85.3 & 12.5 \\
MiniMax-M2.5-HS
& 3.0 [2.7--4.0] & 4.5 & 8.8
& 16.7 & 74.5 & 82.9 & 14 \\
MiniMax-M2.7-HS
& 3.1 [2.7--3.8] & 4.8 & 8.9
& 17.6 & 72.3 & 70.6 & 11 \\
MiniMax-M3
& 9.1 [7.0--12.4] & 13.7 & 25.2
& 8.3 & 85.1 & 65.0 & 27 \\
DeepSeek-V4-Pro
& 12.9 [10.2--14.7] & 16.1 & 41.8
& 8.1 & 74.5 & 74.3 & 24 \\
GPT-5.5
& 8.3 [7.0--9.4] & 10.7 & 16.4
& 8.0 & 85.1 & 60.0 & 20 \\
GPT-5.6-Luna
& 10.0 [7.9--11.6] & 15.5 & 19.5
& 5.8 & 87.2 & 56.1 & 21 \\
GPT-5.6-Terra
& 6.1 [5.3--7.0] & 8.6 & 10.9
& 12.0 & 85.1 & 40.0 & 8 \\
GPT-5.6-Sol
& 14.6 [11.7--18.3] & 23.9 & 19.6
& 7.0 & 72.3 & 50.0 & 14.5 \\
GPT-6-Astra
& 8.2 [7.5--9.3] & 10.1 & 9.8
& 9.0 & 70.2 & 45.5 & 14 \\
\midrule

\multicolumn{8}{@{}l}{
  \textit{B. Frameworks with MiniMax-M3}
} \\
Codex
& 8.4 [6.9--12.1] & 13.8 & 25.3
& 8.3 & 80.4 & 60.0 & 28 \\
Claude Code
& 4.2 [3.1--5.5] & 7.6 & 16.2
& 17.6 & 80.4 & 55.6 & 10 \\
OpenCode
& 4.1 [3.3--5.4] & 7.7 & 11.4
& 14.5 & 80.4 & 64.4 & 13 \\
Ref-ReAct
& 6.5 [4.9--8.1] & 10.0 & 30.3
& 14.7 & 57.1 & 56.2 & 19 \\
\bottomrule
\end{tabular*}

\par\vspace{3pt}
\begin{minipage}{\linewidth}
\fontsize{8}{9.5}\selectfont

\textit{Time and output.}
Latency includes waiting within one execution selected for analysis, excluding scheduling delays, earlier attempts, and subsequent analyst review. IQR gives the 25th--75th percentiles; P90 is the 90th percentile. Output is median cumulative model output in thousands of tokens.

\par\vspace{2pt}
\textit{Query progress.}
First is the median percentage of the query sequence elapsed when
the first action becomes sufficiently supported, among runs with $O>0$. Partial is the percentage of shared-case runs with $0<O<R$. Within this subset, Last by half is the percentage whose final support gain occurs by the query midpoint. Queries after is the median number of queries following that final support gain. Later queries may check hypotheses, corroborate findings, or explore unsuccessfully.

\end{minipage}
\end{table}

\paragraph{Latency on shared cases.}
\label{sec:runtime_results}
\label{sec:operational_tradeoffs}
Comparable citation coverage can entail different completion times.
With MiniMax-M3 fixed, Codex and OpenCode both achieve 29.6\%
citation coverage, but their median execution times are
8.4 and 4.1 minutes, respectively.
Latency therefore complements coverage when assessing how quickly
a system delivers supported findings.

\paragraph{Progress across recorded queries.}
The acquisition columns describe progress along the record-query
sequence, independently of elapsed time.
They distinguish obtaining the first sufficient witness from
continuing to expand the set of supported actions.
First is computed among runs with $O>0$; the final-gain statistics
are computed among partial-acquisition runs with $0<O<R$.

Queries after the final support gain may check hypotheses,
corroborate existing findings, or explore unsuccessfully.
Their value therefore depends on their role in the investigation,
not solely on whether they add another sufficiently supported action.



\subsection{Where to find each denominator}
\label{app:analysis_ledger}

\begin{table}[htbp]
\centering\small
\caption{Per-view populations for the primary model and framework panels. Counts apply to each system within its panel. Cases and recoverable actions can differ between conditions; these are not matched-action comparisons.}
\label{tab:condition_populations}
\setlength{\tabcolsep}{5pt}
\begin{tabular*}{\linewidth}{@{}l@{\extracolsep{\fill}}rrrrrr@{}}
\toprule
& \multicolumn{3}{c}{Eleven models} & \multicolumn{3}{c}{Four frameworks} \\
\cmidrule(lr){2-4}\cmidrule(l){5-7}
Condition & Cases & Groups & $N/R$ & Cases & Groups & $N/R$ \\
\midrule
Full & 8 & 8 & 63/61 & 9 & 8 & 69/67 \\
Endpoint-only & 7 & 6 & 51/44 & 8 & 7 & 64/55 \\
Network-only & 5 & 5 & 34/18 & 5 & 5 & 32/17 \\
Tiered retention & 8 & 7 & 61/37 & 8 & 8 & 52/39 \\
Group sampling & 5 & 5 & 34/31 & 9 & 8 & 70/68 \\
Targeted suppression & 8 & 7 & 58/48 & 9 & 8 & 68/58 \\
Matched benign removal & 6 & 6 & 39/39 & 8 & 7 & 59/58 \\
\bottomrule
\end{tabular*}
\end{table}

Table~\ref{tab:condition_populations} gives each condition's cases and action counts. Table~\ref{tab:telemetry-qualification} gives the paired populations, and Table~\ref{tab:telemetry-witness-qualification} records strict evidence qualification. Depth, category, matched-acquisition, and target-audit results state their own denominators beside the results.


\FloatBarrier
\section{Investigation performance, behavior coverage, and temporal relations}
\label{app:investigation_profiles}

This appendix expands the investigation results in Section~\ref{sec:investigation_extent}. It reports complete scores, defines the lead-query references, and examines coverage depth, behavior categories, and explicit temporal links. These analyses describe the saved investigations under the current support rules. They complement the cross-telemetry comparisons rather than define additional robustness scores.

\subsection{Complete results and comparison populations}
The eleven Codex models share 47 cases with $N=340$ eligible and $R=278$ recoverable action opportunities per model. The four MiniMax-M3 frameworks share a separate set of 56 cases with $N=414$ and $R=362$ per framework. Each panel spans nine source-incident groups. Codex with MiniMax-M3 appears in both panels and is evaluated on each panel's own cases. Table~\ref{tab:full_evidence_results} reports acquisition, citation retention, citation coverage, case reach, and citation volume. Appendix~\ref{app:analysis_ledger} gives the remaining analysis populations.

\begin{table}[ht]
\centering\small
\setlength{\tabcolsep}{3pt}
\renewcommand{\arraystretch}{1.08}
\caption{Evidence recovery by model and framework. Reach is the percentage of cases with $K\geq1$.}
\label{tab:full_evidence_results}
\begin{tabular*}{\linewidth}{@{}l@{\extracolsep{\fill}}rrrrr@{}}
\toprule
System & Reach (\%) & $O/R$ (\%) & $K/O$ (\%) & $K/R$ (\%) & \shortstack{Citations/\\covered} \\
\midrule
\multicolumn{6}{@{}l}{\textit{A. Codex models: $N=340$, $R=278$, $R/N=81.8\%$}} \\
MiniMax-M2 & 21.3 & 18.3 & 23.5 & 4.3 & 15.6 \\
MiniMax-M2.1-HS & 21.3 & 19.4 & 20.4 & 4.0 & 27.7 \\
MiniMax-M2.5-HS & 21.3 & 22.3 & 30.6 & 6.8 & 17.4 \\
MiniMax-M2.7-HS & 25.5 & 29.1 & 33.3 & 9.7 & 11.7 \\
MiniMax-M3 & 66.0 & 45.7 & 66.9 & 30.6 & 8.0 \\
DeepSeek-V4-Pro & 70.2 & 46.8 & 64.6 & 30.2 & 7.7 \\
GPT-5.5 & 85.1 & 57.9 & 71.4 & 41.4 & 7.8 \\
GPT-5.6-Luna & 78.7 & 49.6 & 55.8 & 27.7 & 8.9 \\
GPT-5.6-Terra & 74.5 & 52.5 & 56.8 & 29.9 & 5.0 \\
GPT-5.6-Sol & 91.5 & 70.1 & 60.0 & 42.1 & 6.8 \\
GPT-6-Astra & 89.4 & 75.5 & 63.8 & 48.2 & 7.0 \\
\midrule
\multicolumn{6}{@{}l}{\textit{B. MiniMax-M3 frameworks: $N=414$, $R=362$, $R/N=87.4\%$}} \\
Codex & 60.7 & 42.8 & 69.0 & 29.6 & 7.1 \\
Claude Code & 75.0 & 38.1 & 71.0 & 27.1 & 8.2 \\
OpenCode & 66.1 & 39.8 & 74.3 & 29.6 & 8.2 \\
Ref-ReAct & 57.1 & 36.5 & 88.6 & 32.3 & 15.0 \\
\bottomrule
\end{tabular*}
\par\vspace{3pt}\begin{minipage}{\linewidth}\footnotesize The rate columns report acquisition, retention, and coverage. Citations/covered deduplicates record IDs within each report, sums across reports including those with $K=0$, and divides by total $K$. \end{minipage}
\end{table}

\subsection{Three lead-query references}
\label{app:lead_reference}
The fixed lead-query reference measures how much support the prescribed queries provide without adaptive investigation. The reporting controls instead use records actually returned during saved runs. Table~\ref{tab:lead_reference_guide} separates the fixed procedure, first-page records shared across models, and each run's own lead-query returns.

\begin{table}[!h]
\caption{Three uses of lead-query records. Each answers a different comparison question.}
\label{tab:lead_reference_guide}
\centering\small
\begin{tabularx}{\linewidth}{@{}L{0.24\linewidth}XX@{}}
\toprule
Reference & Record set & Comparison question \\
\midrule
Fixed lead-query reference & Independently execute the supplied pivots and pagination without adaptive follow-up. & How much support does autonomous investigation acquire beyond those queries? \\
\addlinespace
Shared first-page returns & Intersect first-page IDs actually returned to every model in a case. & Is support returned to all models preserved in their reports? \\
\addlinespace
Run-specific lead returns & Exact lead queries and linked continuation pages actually consumed in a run. Their supported actions form $B_i$. & Which supported actions are added to or omitted from formal-stage citations? \\
\bottomrule
\end{tabularx}
\end{table}

The fixed procedure measures acquisition without producing a report. Shared first-page returns fix record identities across models. Run-specific returns include matched continuation pages, which can arrive later in an investigation. Ref-ReAct transforms the supplied queries and is therefore outside the exact-query matching analysis. Appendix~\ref{app:public_lead_expansion} defines that matching procedure and compares additions with omissions.

\subsection{Coverage within individual investigations}
\label{app:coverage_depth}
An overall action percentage can reflect partial coverage of many cases or extensive coverage of a few. Table~\ref{tab:deep_delivery_thresholds} fixes the 43 model cases and 53 framework cases with at least three recoverable actions. It then raises the required coverage from any supported action to all recoverable actions, retaining zero-coverage reports in every denominator. These rates describe action coverage, not the length of a connected attack chain.

\begin{table}[htb]
\centering\small
\setlength{\tabcolsep}{4pt}
\renewcommand{\arraystretch}{1.06}
\caption{Share of investigations reaching progressively deeper citation coverage.}
\label{tab:deep_delivery_thresholds}
\begin{tabular*}{\linewidth}{@{}l@{\extracolsep{\fill}}rrrr@{}}
\toprule
System & $K\geq1$ & $K/R\geq25\%$ & $K/R\geq50\%$ & $K=R$ \\
\midrule
\multicolumn{5}{@{}l}{\textit{Models with Codex. $R\geq3$}} \\
MiniMax-M2 & 23.3 & 9.3 & 0.0 & 0.0 \\
MiniMax-M2.1-HS & 23.3 & 9.3 & 0.0 & 0.0 \\
MiniMax-M2.5-HS & 23.3 & 14.0 & 2.3 & 0.0 \\
MiniMax-M2.7-HS & 27.9 & 18.6 & 2.3 & 0.0 \\
MiniMax-M3 & 72.1 & 51.2 & 25.6 & 0.0 \\
DeepSeek-V4-Pro & 74.4 & 48.8 & 23.3 & 0.0 \\
GPT-5.5 & 90.7 & 67.4 & 37.2 & 0.0 \\
GPT-5.6-Luna & 86.0 & 53.5 & 14.0 & 0.0 \\
GPT-5.6-Terra & 79.1 & 53.5 & 14.0 & 0.0 \\
GPT-5.6-Sol & 95.3 & 74.4 & 25.6 & 4.7 \\
GPT-6-Astra & 95.3 & 79.1 & 44.2 & 0.0 \\
\midrule
\multicolumn{5}{@{}l}{\textit{Frameworks with MiniMax-M3. $R\geq3$}} \\
Codex & 64.2 & 45.3 & 24.5 & 0.0 \\
Claude Code & 79.2 & 47.2 & 22.6 & 0.0 \\
OpenCode & 69.8 & 52.8 & 30.2 & 0.0 \\
Ref-ReAct & 60.4 & 47.2 & 34.0 & 0.0 \\
\bottomrule
\end{tabular*}
\par\vspace{3pt}\begin{minipage}{\linewidth}\footnotesize Entries are percentages of the 43 model cases and 53 framework cases with $R\geq3$, including reports with zero coverage. Every threshold uses this fixed denominator within its panel. Complete coverage means that formal-stage citations sufficiently support every registered recoverable action.\end{minipage}
\end{table}

GPT-5.6-Sol and GPT-6-Astra each support at least one action in 95.3\% of these cases, but reach half coverage in 25.6\% and 44.2\%, respectively. GPT-5.6-Sol reaches complete recoverable coverage in 4.7\% despite its lower pooled score. Among frameworks, Claude Code reaches more cases, whereas Ref-ReAct more often reaches half coverage. Similar case reach therefore need not mean similar coverage within each investigation.

Complete recoverable coverage, $K=R$, remains relative to the view. Full eligible-incident coverage requires $K=N$. DeepSeek-V4-Pro's network-only Purple Fox report supports both recoverable actions in a six-action incident, covering only 33.3\% of the eligible incident. This case falls outside the table's three-action threshold but illustrates the distinction. GPT-5.6-Sol's complete-coverage cases include an APT27 investigation under group sampling with $K=N$.

\subsection{Behavior categories and their interpretation}
\label{sec:stage_results}
\label{app:category_reading_key}
The category analyses use three different groupings:
\begin{itemize}\setlength{\itemsep}{1pt}\setlength{\parskip}{0pt}
\item \textbf{Source-attested categories} include only actions whose saved annotation marks the category eligible and records a review receipt or explicit reviewer attestation. This flag documents the assignment's provenance, not model performance.
\item \textbf{Descriptive behavior groups} refine the original Other group using saved action descriptions. They provide a broader breakdown of observable behaviors.
\item \textbf{Archived reference categories} retain every saved category label, including labels without source attestation, for the complete breakdown and seven-condition matrix.
\end{itemize}
The same category name can contain different actions in these groupings. Each comparison fixes membership and cases within its panel. Acquisition $O/R$ and citation coverage $K/R$ use recoverable action opportunities, while retention $K/O$ uses each system's acquired opportunities. These analyses evaluate evidence for categorized actions, not the tactic labels predicted by agents.

\begin{figure}[ht]
\centering 
\includegraphics[width=0.95\textwidth]{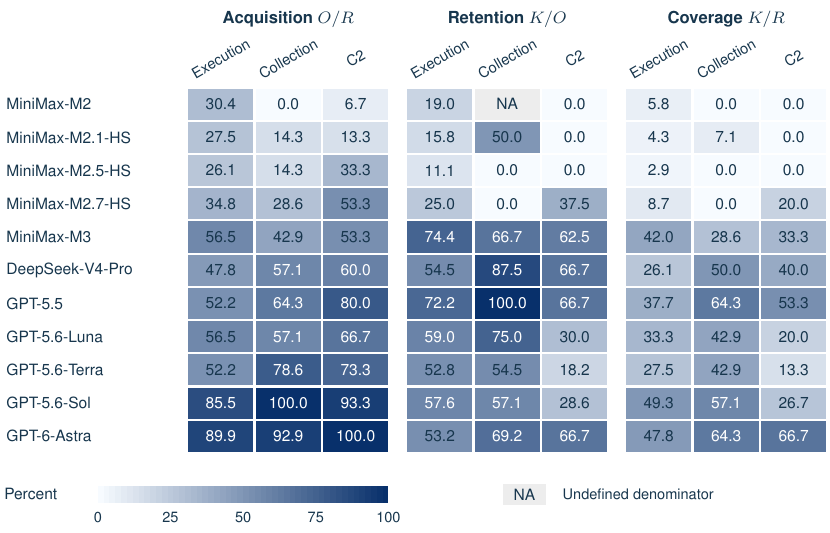} 
\caption{Acquisition, citation retention, and citation coverage within the source-attested category subset. Acquisition and citation coverage use recoverable action opportunities as the denominator. Retention uses acquired opportunities.}
\label{fig:category_attested_models}
\end{figure}

\paragraph{Source-attested behavior coverage.}
\label{app:attested_category_profiles}
Figure~\ref{fig:category_attested_models} reports the model results, and Table~\ref{tab:restored_categories_frameworks} reports the framework results.

\begin{table}[htbp]
\caption{Evidence recovery within the source-attested category subset, using four frameworks with MiniMax-M3 and the current support rules.}
\label{tab:restored_categories_frameworks}
\centering
\small
\setlength{\tabcolsep}{2.5pt}
\renewcommand{\arraystretch}{1.12}
\begin{tabular*}{\linewidth}{@{}l@{\extracolsep{\fill}}*{9}{c}@{}}
\toprule
System
& \multicolumn{3}{c}{$O/R$}
& \multicolumn{3}{c}{$K/O$}
& \multicolumn{3}{c}{$K/R$} \\
\cmidrule(lr){2-4}\cmidrule(lr){5-7}\cmidrule(l){8-10}
& Execution & Collection & C2
& Execution & Collection & C2
& Execution & Collection & C2 \\
\midrule
Codex       & 60.2 & 47.1 & 43.5 & 69.8 & 62.5  & 60.0 & 42.0 & 29.4 & 26.1 \\
Claude Code & 43.2 & 52.9 & 39.1 & 68.4 & 100.0 & 33.3 & 29.5 & 52.9 & 13.0 \\
OpenCode    & 48.9 & 58.8 & 30.4 & 74.4 & 90.0  & 57.1 & 36.4 & 52.9 & 17.4 \\
Ref-ReAct   & 34.1 & 47.1 & 39.1 & 90.0 & 87.5  & 66.7 & 30.7 & 41.2 & 26.1 \\
\bottomrule
\end{tabular*}
\par\vspace{3pt}
\begin{minipage}{\linewidth}
\footnotesize\raggedright
All rates are percentages on the shared framework-comparison cases. C2 denotes command and control.
\end{minipage}\par
\end{table}

MiniMax-M3 and DeepSeek-V4-Pro have nearly equal overall citation coverage, but MiniMax-M3 covers more source-attested Execution opportunities (42.0\% versus 26.1\%) and DeepSeek-V4-Pro covers more Collection opportunities (50.0\% versus 28.6\%). Two thirds of DeepSeek-V4-Pro's Collection advantage comes from opportunities it alone acquires. On jointly acquired opportunities, it retains 83.3\%, compared with 66.7\% for MiniMax-M3. Both acquisition and citation retention contribute to this contrast.

High acquisition also need not produce high citation coverage within a category. GPT-5.6-Sol acquires support for 85.5\% of Execution opportunities and cites support for 49.3\%. GPT-6-Astra acquires 89.9\% and cites 47.8\%. With MiniMax-M3 fixed, Codex covers more Execution opportunities than Ref-ReAct despite its lower overall coverage, whereas Claude Code and OpenCode cover more Collection opportunities than Codex. These differences concern the fixed reference subsets, not general knowledge of the corresponding tactics.

\begin{figure}[ht]
\centering 
\includegraphics[width=0.95\textwidth]{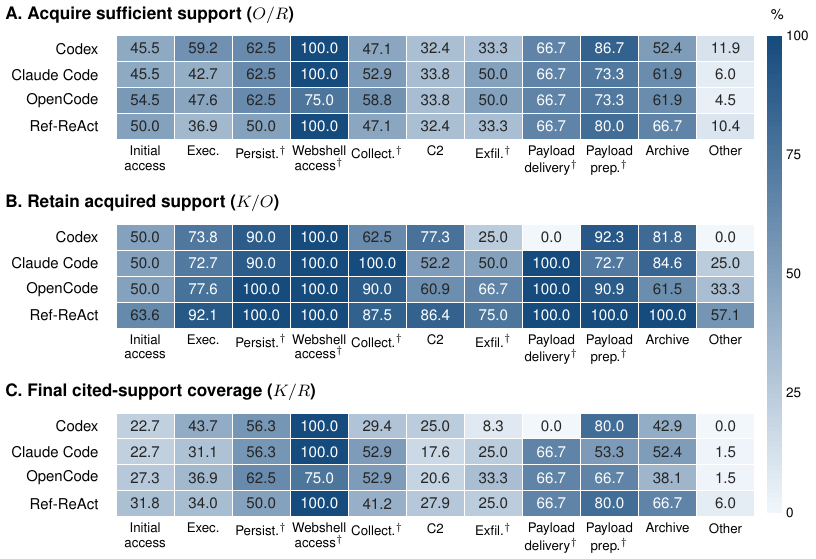} 
\caption{Acquisition, citation retention, and citation coverage by descriptive behavior group across frameworks using MiniMax-M3. Acquisition and citation coverage use recoverable action opportunities as the denominator. Retention uses acquired opportunities. Saved descriptions refine the archived categories while preserving action identities and support rules.}
\label{fig:category_behavior_diagnostic}
\end{figure}

In the descriptive Execution group, Ref-ReAct retains 92.1\% of acquired support but acquires only 36.9\%, yielding 34.0\% citation coverage. Codex acquires 59.2\% and covers 43.7\%. In descriptive C2, Codex and Claude Code acquire similar shares, 32.4\% and 33.8\%, but retain 77.3\% and 52.2\%, yielding 25.0\% and 17.6\% coverage. The first contrast emphasizes acquisition; the second emphasizes preservation of acquired support. Appendix~\ref{app:matched_acquisition} additionally compares retention after matching acquired opportunities.


\subsection{Current temporal-relation scope and correspondence}
\label{app:current_relations}
\label{sec:relations}
\label{sec:activity_relations}
The structured reports contain activity nodes, entity references, times, explicit relations, and hypotheses. This supplementary analysis checks whether reports connect supported actions through registered temporal links and whether each submitted link cites sufficient evidence. It describes how reports organize activities. The main cross-telemetry analysis instead compares action support, while a separate assertion audit examines the meaning of selected activity descriptions.

\paragraph{Reference scope and comparison population.}
The full 56-scenario collection registers 360 distinct reference relations, all expressing temporal precedence. The ten-scenario controlled collection originally registers 62. Excluding the unresolved Donot action removes its two adjacent relations without adding a replacement link across the removed action. We also update the Dragonfly and Zardoor endpoint witnesses used by adjacent relation rules, then recheck their timestamps and ordering. The current shared panels contain 208 recoverable relation--condition pairs per model and 277 per framework.

The model panel contains 517 reports and the framework panel 224. They share 39 runs of Codex with MiniMax-M3, yielding 702 unique runs. The panel counts therefore cannot be added without counting those shared runs twice. These panels use the same cases as the corresponding main action-coverage comparisons.

\paragraph{Evidence-grounded node correspondence.}
A reported activity can match a reference action only when its own citations contain a witness. The fixed one-to-one assignment first maximizes matched endpoints, then resolves ties by evidence overlap and stable ordering. Relation scores do not influence the assignment. Splitting repeated behavior across nodes or merging distinct actions into one node can reduce matched endpoints even when the report's formal-stage citation union covers those actions. This matching checks each activity's evidence without independently validating its text.

\paragraph{Actual links and their evidence.}
After mapping nodes, the evaluator records three nested counts. \emph{Endpoints} counts recoverable reference relations with both actions matched. \emph{Submitted} counts those pairs connected by an explicit directed temporal edge. \emph{Supported} additionally requires a complete ordered witness in that edge's own citations. Such a witness contains compatible evidence for the source and target actions, with the latest source observation preceding the earliest target observation. The evaluator adds no citations from endpoints or other fields.

Exact reference matching can leave legitimate alternative links unmatched. The evaluator also adds no links implied by transitivity. Unmatched links remain unjudged rather than automatically false. Different edges may use different witnesses for the same endpoint action, so sufficient evidence for each edge does not guarantee a consistent ordering of all action instances. The submitted \emph{strong} and \emph{support} relation classes remain outside the temporal reference scope.

The model panel contains 2,036 submitted edge objects, including 1,353 temporal and 683 strong or support objects. The framework panel contains 1,012, including 792 temporal and 220 strong or support objects. Thus 33.5\% and 21.7\% of submitted edges, respectively, fall outside the registered class. These percentages describe output objects, not relation correctness.

\paragraph{Offline replay checks.}
The asset audit verifies saved predictions, query-return audits, scores, reference inputs, and the archived scorer for all 959 rows in the replay index. All 3,157 distinct files match their recorded hashes. The evaluator first reproduces all 702 archived node assignments under the original rules for the current panels. After rule updates, an independent check reproduces every current citation-union $K$. Reversing input node order leaves every assignment unchanged. Empty edge citations support no relation, and direct per-edge checks agree with the archived relation-scoring procedure. No model was rerun. Table~\ref{tab:current_temporal_diagnostics} gives the resulting counts.

\begin{table}[tb]
\centering\small
\setlength{\tabcolsep}{3pt}
\caption{Current registered temporal-relation diagnostics on the 47 model cases and 56 framework cases used for the corresponding action comparisons.}
\label{tab:current_temporal_diagnostics}
\begin{tabular*}{\linewidth}{@{}l@{\extracolsep{\fill}}rrrrr@{}}
\toprule
System & Endpoints & Submitted & Supported & Time order & \shortstack{Source deletion(\%)} \\
\midrule
\multicolumn{6}{@{}l}{\textit{Codex models: 208 registered relation opportunities}} \\
MiniMax-M2 & 0 & 0 & 0 & 0/0 & 0.0--0.0 \\
MiniMax-M2.1-HS & 0 & 0 & 0 & 0/0 & 0.0--0.0 \\
MiniMax-M2.5-HS & 1 & 0 & 0 & 1/1 & 0.0--0.0 \\
MiniMax-M2.7-HS & 2 & 0 & 0 & 1/2 & 0.0--0.0 \\
MiniMax-M3 & 22 & 17 & 8 & 15/22 & 2.3--4.4 \\
DeepSeek-V4-Pro & 9 & 6 & 1 & 7/9 & 0.0--0.6 \\
GPT-5.5 & 31 & 12 & 11 & 30/31 & 0.6--6.3 \\
GPT-5.6-Luna & 16 & 8 & 5 & 15/16 & 1.7--3.0 \\
GPT-5.6-Terra & 16 & 12 & 11 & 14/16 & 0.6--6.3 \\
GPT-5.6-Sol & 36 & 22 & 22 & 35/36 & 4.4--12.5 \\
GPT-6-Astra & 41 & 24 & 24 & 40/41 & 5.5--13.6 \\
\midrule
\multicolumn{6}{@{}l}{\textit{MiniMax-M3 frameworks: 277 registered relation opportunities}} \\
Codex & 27 & 18 & 9 & 20/27 & 2.1--3.7 \\
Claude Code & 17 & 4 & 2 & 13/17 & 0.4--0.8 \\
OpenCode & 26 & 17 & 11 & 24/26 & 1.9--4.5 \\
Ref-ReAct & 25 & 9 & 6 & 19/25 & 1.4--2.5 \\
\bottomrule
\end{tabular*}
\par\vspace{3pt}\begin{minipage}{\linewidth}\footnotesize
The first three counts are nested. Time order counts registered endpoint pairs ordered by the report's own time intervals, over pairs with mapped endpoints and interpretable times. 
\end{minipage}
\end{table}

MiniMax-M3 explicitly submits 17 registered temporal relations in the model panel, but only eight have sufficient edge-level citations. GPT-5.6-Sol and GPT-6-Astra support all 22 and 24 of their exact registered submissions, yet these account for only part of their 36 and 41 matched-endpoint opportunities. The nested counts distinguish unmatched activities, missing explicit links, and links without sufficient citations.

\paragraph{Report-time check and source concentration.}
The report-time check forms all strictly non-overlapping interval pairs from the report's own start and end fields. A missing end is treated as a point at the start. Pairs are generated before matching them to reference relations, without private timestamps or reference edges. This procedure can produce more pairs than the submitted graph and supplies no additional evidence for any link. It checks whether reported times contain ordering information without controlling for the number of submitted edges. For GPT-6-Astra, report times order 40 of 41 comparable registered pairs, while 24 corresponding temporal links are explicitly submitted. A missing link can therefore coexist with usable ordering information elsewhere in the report.

Supported relations are concentrated in a few incidents. GPT-5.6-Sol and GPT-6-Astra each support 14 relation--condition pairs from Spyder. Removing that source leaves eight and ten out of 181, respectively. The table shows the range of scores after each source-incident group is removed in turn. Multiple observation conditions provide comparisons of the same incident, not additional independent incidents.

\FloatBarrier
\subsection{Sensitivity to incident composition}
\label{app:paired_estimates}
Table~\ref{tab:current_paired_system_differences} checks whether the observed model
and framework differences depend on which source incidents are included.
It combines paired source-incident bootstrap intervals with
leave-one-group-out ranges. Both analyses keep each incident group
together and report citation-coverage differences in the direction $A-B$.

\begin{table}[tb]
\centering
\small
\setlength{\tabcolsep}{2pt}
\renewcommand{\arraystretch}{1.10}
\caption{Paired citation-coverage differences and sensitivity to source-incident composition.}
\label{tab:paired_robustness}
\label{tab:current_paired_system_differences}
\label{tab:current_leave_one_incident_out}
\begin{tabular*}{\linewidth}{@{}l@{\extracolsep{\fill}}rrrc@{}}
\toprule
& & Bootstrap & \multicolumn{2}{c}{Leave-one-group-out} \\
\cmidrule(lr){3-3}\cmidrule(l){4-5}
Comparison ($A-B$)
& $\Delta$ (pp)
& 95\% interval (pp)
& Range (pp)
& $+\,/\,0\,/\,-$ (\%) \\
\midrule
\multicolumn{5}{@{}l}{\textit{Models with Codex}} \\
MiniMax-M3 $-$ MiniMax-M2 & +26.3 & $[+12.5, +36.5]$ & $[+22.8, +29.1]$ & $100.0\,/\,0.0\,/\,0.0$ \\
MiniMax-M3 $-$ MiniMax-M2.1-HS & +26.6 & $[+13.7, +36.2]$ & $[+23.2, +29.5]$ & $100.0\,/\,0.0\,/\,0.0$ \\
MiniMax-M3 $-$ MiniMax-M2.5-HS & +23.7 & $[+11.8, +33.2]$ & $[+19.8, +26.0]$ & $100.0\,/\,0.0\,/\,0.0$ \\
MiniMax-M3 $-$ MiniMax-M2.7-HS & +20.9 & $[+10.8, +30.4]$ & $[+16.5, +22.9]$ & $100.0\,/\,0.0\,/\,0.0$ \\
MiniMax-M3 $-$ DeepSeek-V4-Pro & +0.4 & $[-14.4, +13.7]$ & $[-5.5, +5.7]$ & $55.6\,/\,22.2\,/\,22.2$ \\
MiniMax-M3 $-$ GPT-5.5 & -10.8 & $[-20.3, -2.2]$ & $[-13.5, -6.9]$ & $0.0\,/\,0.0\,/\,100.0$ \\
MiniMax-M3 $-$ GPT-5.6-Luna & +2.9 & $[-5.6, +9.4]$ & $[+0.4, +4.5]$ & $100.0\,/\,0.0\,/\,0.0$ \\
MiniMax-M3 $-$ GPT-5.6-Terra & +0.7 & $[-11.3, +9.6]$ & $[-2.5, +5.3]$ & $66.7\,/\,0.0\,/\,33.3$ \\
MiniMax-M3 $-$ GPT-5.6-Sol & -11.5 & $[-23.6, +1.3]$ & $[-16.0, -7.8]$ & $0.0\,/\,0.0\,/\,100.0$ \\
MiniMax-M3 $-$ GPT-6-Astra & -17.6 & $[-26.4, -10.5]$ & $[-19.5, -14.7]$ & $0.0\,/\,0.0\,/\,100.0$ \\
\midrule
\multicolumn{5}{@{}l}{\textit{Frameworks with MiniMax-M3}} \\
Codex $-$ Claude Code & +2.5 & $[-9.6, +12.6]$ & $[-2.9, +4.5]$ & $77.8\,/\,0.0\,/\,22.2$ \\
Codex $-$ OpenCode & 0.0 & $[-8.8, +8.1]$ & $[-5.1, +1.8]$ & $66.7\,/\,11.1\,/\,22.2$ \\
Codex $-$ Ref-ReAct & -2.8 & $[-10.9, +2.7]$ & $[-5.4, -0.6]$ & $0.0\,/\,0.0\,/\,100.0$ \\
Claude Code $-$ OpenCode & -2.5 & $[-9.6, +4.0]$ & $[-4.2, +0.6]$ & $11.1\,/\,0.0\,/\,88.9$ \\
Claude Code $-$ Ref-ReAct & -5.2 & $[-11.3, +2.6]$ & $[-8.0, -2.5]$ & $0.0\,/\,0.0\,/\,100.0$ \\
OpenCode $-$ Ref-ReAct & -2.8 & $[-8.5, +3.3]$ & $[-4.2, -0.4]$ & $0.0\,/\,0.0\,/\,100.0$ \\
\bottomrule
\end{tabular*}
\par\vspace{3pt}
\begin{minipage}{\linewidth}
\footnotesize\raggedright
$\Delta=100\bigl(\sum K_A/\sum R-\sum K_B/\sum R\bigr)$ is the pooled citation-coverage difference in percentage points.
Bootstrap percentile intervals use 20,000 paired source-incident-group draws.
Leave-one-group-out ranges recompute $\Delta$ after excluding each group in turn. The final column gives the percentages of positive, tied, and negative outcomes.
Both analyses hold the saved executions and recorded settings fixed and assess incident-composition sensitivity.
\end{minipage}\par
\end{table}

MiniMax-M3 outperforms the four earlier MiniMax models under both
checks, while GPT-5.5 and GPT-6-Astra outperform MiniMax-M3.
GPT-5.6-Sol also remains ahead after every single-group omission,
but its bootstrap interval includes zero.
The near-tie between MiniMax-M3 and DeepSeek-V4-Pro can reverse
when the incident mix changes.
Among frameworks, Ref-ReAct leads after every single-group omission,
yet all framework-pair bootstrap intervals include zero.
These results support selected model separations more strongly
than a complete ranking.

\FloatBarrier
\section{Robustness across telemetry conditions}
\label{app:telemetry_robustness}
This appendix first reports each condition's results, then compares the same recoverable actions across conditions and checks stricter evidence-preserving pairs. The final controls distinguish removal of target support from changes in the remaining investigation. All comparisons use the frozen run-selection ledger and fixed support rules.

\subsection{Results under each observation condition}
\label{app:observation_details}
Tables~\ref{tab:restored_views_models} and~\ref{tab:restored_views_frameworks} report acquisition and citation coverage under all seven conditions. Table~\ref{tab:condition_populations} specifies each panel's cases and action counts. Systems share the same available records and recoverable actions within a condition, but these populations can change between conditions. The per-view rates are therefore descriptive, not paired robustness estimates.

\begin{table}[t]
\centering
\footnotesize
\setlength{\tabcolsep}{1.5pt}
\renewcommand{\arraystretch}{1.12}
\caption{All models across seven observation conditions under the current support rules, with shared cases within each condition.}
\label{tab:restored_views_models}
\label{tab:restored_views_gpt}
\begin{tabular*}{\linewidth}{@{}l@{\hspace{5pt}\extracolsep{\fill}}*{14}{c}@{}}
\toprule
Model
& \multicolumn{2}{c}{Full}
& \multicolumn{2}{c}{Endpoint}
& \multicolumn{2}{c}{Network}
& \multicolumn{2}{c}{Retention}
& \multicolumn{2}{c}{Sampling}
& \multicolumn{2}{c}{Targeted}
& \multicolumn{2}{c}{Matched} \\
\cmidrule(lr){2-3}\cmidrule(lr){4-5}\cmidrule(lr){6-7}
\cmidrule(lr){8-9}\cmidrule(lr){10-11}\cmidrule(lr){12-13}
\cmidrule(l){14-15}
& $O/R$ & $K/R$
& $O/R$ & $K/R$
& $O/R$ & $K/R$
& $O/R$ & $K/R$
& $O/R$ & $K/R$
& $O/R$ & $K/R$
& $O/R$ & $K/R$ \\
\midrule
MiniMax-M2 & 21.3 & 4.9 & 15.9 & 2.3 & 16.7 & 5.6 & 16.2 & 2.7 & 16.1 & 3.2 & 20.8 & 8.3 & 17.9 & 2.6 \\
MiniMax-M2.1-HS & 21.3 & 4.9 & 25.0 & 2.3 & 16.7 & 0.0 & 21.6 & 8.1 & 16.1 & 3.2 & 14.6 & 4.2 & 17.9 & 2.6 \\
MiniMax-M2.5-HS & 16.4 & 6.6 & 27.3 & 6.8 & 22.2 & 5.6 & 40.5 & 18.9 & 19.4 & 3.2 & 14.6 & 6.3 & 20.5 & 0.0 \\
MiniMax-M2.7-HS & 42.6 & 19.7 & 25.0 & 9.1 & 27.8 & 0.0 & 27.0 & 13.5 & 22.6 & 6.5 & 20.8 & 4.2 & 30.8 & 5.1 \\
MiniMax-M3 & 59.0 & 45.9 & 34.1 & 22.7 & 50.0 & 33.3 & 43.2 & 21.6 & 38.7 & 25.8 & 43.8 & 25.0 & 46.2 & 33.3 \\
DeepSeek-V4-Pro & 50.8 & 37.7 & 34.1 & 22.7 & 72.2 & 55.6 & 45.9 & 27.0 & 29.0 & 12.9 & 62.5 & 45.8 & 38.5 & 12.8 \\
GPT-5.5 & 60.7 & 45.9 & 63.6 & 45.5 & 55.6 & 55.6 & 51.4 & 32.4 & 48.4 & 29.0 & 64.6 & 50.0 & 53.8 & 30.8 \\
GPT-5.6-Luna & 44.3 & 21.3 & 68.2 & 43.2 & 61.1 & 50.0 & 32.4 & 13.5 & 35.5 & 19.4 & 54.2 & 35.4 & 53.8 & 20.5 \\
GPT-5.6-Terra & 63.9 & 41.0 & 59.1 & 22.7 & 55.6 & 50.0 & 48.6 & 27.0 & 35.5 & 19.4 & 52.1 & 31.3 & 43.6 & 20.5 \\
GPT-5.6-Sol & 65.6 & 45.9 & 72.7 & 43.2 & 83.3 & 66.7 & 64.9 & 27.0 & 61.3 & 35.5 & 77.1 & 50.0 & 71.8 & 33.3 \\
GPT-6-Astra & 80.3 & 50.8 & 79.5 & 56.8 & 77.8 & 55.6 & 62.2 & 37.8 & 74.2 & 38.7 & 79.2 & 56.3 & 71.8 & 38.5 \\
\bottomrule
\end{tabular*}
\end{table}

\begin{table}[htbp]
\caption{Acquisition and citation coverage across seven observation conditions for the four frameworks.}
\label{tab:restored_views_frameworks}
\centering
\small
\setlength{\tabcolsep}{3pt}
\renewcommand{\arraystretch}{1.12}
\begin{tabular*}{\linewidth}{@{}l@{\extracolsep{\fill}}*{8}{c}@{}}
\toprule
View
& \multicolumn{2}{c}{Codex}
& \multicolumn{2}{c}{Claude Code}
& \multicolumn{2}{c}{OpenCode}
& \multicolumn{2}{c}{Ref-ReAct} \\
\cmidrule(lr){2-3}\cmidrule(lr){4-5}
\cmidrule(lr){6-7}\cmidrule(l){8-9}
& $O/R$ & $K/R$
& $O/R$ & $K/R$
& $O/R$ & $K/R$
& $O/R$ & $K/R$ \\
\midrule
$\Omega_0$ Full & 55.2 & 40.3 & 34.3 & 26.9 & 46.3 & 32.8 & 58.2 & 52.2 \\
$\Omega_1$ Endpoint & 25.5 & 16.4 & 23.6 & 14.5 & 25.5 & 14.5 & 18.2 & 14.5 \\
$\Omega_2$ Network & 47.1 & 35.3 & 47.1 & 41.2 & 70.6 & 47.1 & 47.1 & 47.1 \\
$\Omega_3$ Retention & 38.5 & 20.5 & 38.5 & 20.5 & 25.6 & 20.5 & 28.2 & 20.5 \\
$\Omega_4$ Sampling & 44.1 & 32.4 & 33.8 & 20.6 & 45.6 & 35.3 & 33.8 & 32.4 \\
$\Omega_5$ Targeted & 37.9 & 19.0 & 48.3 & 43.1 & 48.3 & 43.1 & 39.7 & 36.2 \\
$\Omega_6$ Matched & 50.0 & 41.4 & 48.3 & 31.0 & 31.0 & 20.7 & 31.0 & 25.9 \\
\bottomrule
\end{tabular*}
\end{table}

GPT-5.6-Sol, for example, sufficiently cites 12 of 18 recoverable action--view pairs under network-only observation (66.7\%) and 28 of 61 under Full (45.9\%). Neither the action set nor the scenario mix is held fixed in this contrast. The comparisons below instead retain actions recoverable in both views.

\FloatBarrier
\subsection{Pairing, eligibility, and input identity}
Each comparison starts from the frozen run-selection ledger. The primary analysis requires a report accepted for scoring, a usable evidence audit, and the recorded runtime-audit selection flag. We compare Full with each changed condition. We first select scenarios with qualifying runs under both conditions for every system in the panel, then retain actions recoverable under both conditions. These pairs reproduce the earlier paired analysis but differ from the 47-case and 56-case sets used for the main rankings. Table~\ref{tab:telemetry-qualification} reports the analysis sets.


\begin{table}[htbp]
\centering\footnotesize
\setlength{\tabcolsep}{3pt}
\caption{Cases and actions included in the paired telemetry comparisons. Scenarios (Sc.), source-incident groups (Gr.), and unique actions (Act.) contribute at least one action recoverable under both conditions after selecting cases shared by all systems in the panel. $M$ counts system--action pairs. Own counts each system's primary paired runs, including pairs with no common recoverable action. Accepted also includes accepted runs with usable audits outside the primary filter. Failed-run occurrences and primary-filter exclusions are counted before selecting shared cases.}
\label{tab:telemetry-qualification}
\begin{tabular*}{\linewidth}{@{}l@{\extracolsep{\fill}}rrrrrrrr@{}}
\toprule
Condition & Sc. & Gr. & Act. & $M$ & Own & Accepted & Failed & Excluded \\
\midrule
\multicolumn{9}{l}{\textit{Eleven models}} \\
$\Omega_1$ Endpoint & 6 & 6 & 41 & 451 & 104 & 107 & 3 & 4 \\
$\Omega_2$ Network & 4 & 4 & 14 & 154 & 101 & 104 & 6 & 3 \\
$\Omega_3$ Retention & 7 & 7 & 33 & 363 & 105 & 106 & 4 & 2 \\
$\Omega_4$ Sampling & 3 & 3 & 19 & 209 & 100 & 106 & 4 & 6 \\
$\Omega_5$ Suppression & 6 & 6 & 38 & 418 & 104 & 106 & 4 & 2 \\
$\Omega_6$ Matched removal & 5 & 5 & 35 & 385 & 102 & 107 & 3 & 5 \\
\midrule
\multicolumn{9}{l}{\textit{Four frameworks}} \\
$\Omega_1$ Endpoint & 8 & 7 & 55 & 220 & 38 & 38 & 3 & 0 \\
$\Omega_2$ Network & 5 & 5 & 17 & 68 & 32 & 32 & 8 & 0 \\
$\Omega_3$ Retention & 7 & 7 & 35 & 140 & 37 & 38 & 2 & 1 \\
$\Omega_4$ Sampling & 8 & 7 & 62 & 248 & 38 & 38 & 2 & 0 \\
$\Omega_5$ Suppression & 8 & 7 & 53 & 212 & 38 & 38 & 2 & 0 \\
$\Omega_6$ Matched removal & 8 & 7 & 58 & 232 & 38 & 38 & 3 & 0 \\
\bottomrule
\end{tabular*}
\end{table}

Each analysis row identifies a system, source-incident group, scenario, action, condition pair, and both selected runs. It also records the rule, input hashes, and execution settings. A shared client name does not imply identical client versions across models. Each pair retains its recorded configuration. Measured token use and configured limits are reported separately. Historical stopping-policy runs remain separate.

Failed or unavailable runs are excluded rather than scored as empty reports. Accepted empty reports remain eligible and receive zero citation coverage. For every included action, the audit establishes whether support was acquired and whether formal-stage citations preserve it. These states locate a coverage loss before or after acquisition, and their totals satisfy $K\leq O\leq R$. Without a usable audit, acquisition would remain unknown even if citation coverage could be scored.

\FloatBarrier
\subsection{Action transitions and their denominators}
For acquisition and citation coverage separately, four counts identify support in both runs, only Full, only the changed condition, or neither. Each system--action pair is counted once per condition contrast, including pairs missed in both runs. Figure~\ref{fig:telemetry-paired-summary} presents the citation transitions. The tables below supply exact counts and distinguish loss before and after acquisition. Records from different systems are never combined to construct a witness.

The complete state matrix uses U when queries do not acquire sufficient support, A when they acquire sufficient support but formal-stage citations do not preserve it, and C when formal-stage citations provide sufficient support. In Table~\ref{tab:telemetry-state-matrix}, the first letter is the Full-run state and the second the changed-view state. All nine combinations are possible because the columns compare separate investigations. They are not a temporal sequence within one run.


\begin{table}[htbp]
\centering\footnotesize
\setlength{\tabcolsep}{3pt}
\caption{Three-state transitions for the shared primary pairs. U denotes insufficient query acquisition, A sufficient acquisition without sufficient formal-stage citations, and C sufficient formal-stage citations. Each two-letter code gives the Full state first and the changed-condition state second. The states come from separate runs. All nine counts sum to $M$.}
\label{tab:telemetry-state-matrix}
\begin{tabular*}{\linewidth}{@{}l@{\extracolsep{\fill}}rrrrrrrrrr@{}}
\toprule
Condition & $M$ & UU & UA & UC & AU & AA & AC & CU & CA & CC \\
\midrule
\multicolumn{11}{l}{\textit{Eleven models}} \\
$\Omega_1$ Endpoint & 451 & 193 & 26 & 19 & 16 & 55 & 18 & 32 & 12 & 80 \\
$\Omega_2$ Network & 154 & 61 & 7 & 7 & 3 & 14 & 9 & 1 & 6 & 46 \\
$\Omega_3$ Retention & 363 & 189 & 19 & 9 & 18 & 36 & 10 & 12 & 15 & 55 \\
$\Omega_4$ Sampling & 209 & 120 & 6 & 0 & 6 & 28 & 6 & 2 & 3 & 38 \\
$\Omega_5$ Suppression & 418 & 186 & 7 & 7 & 16 & 47 & 10 & 15 & 15 & 115 \\
$\Omega_6$ Matched removal & 385 & 198 & 15 & 3 & 16 & 65 & 7 & 5 & 16 & 60 \\
\midrule
\multicolumn{11}{l}{\textit{Four frameworks}} \\
$\Omega_1$ Endpoint & 220 & 103 & 2 & 3 & 13 & 9 & 2 & 53 & 7 & 28 \\
$\Omega_2$ Network & 68 & 30 & 5 & 1 & 2 & 2 & 11 & 0 & 0 & 17 \\
$\Omega_3$ Retention & 140 & 80 & 5 & 3 & 3 & 5 & 5 & 7 & 8 & 24 \\
$\Omega_4$ Sampling & 248 & 112 & 3 & 10 & 13 & 7 & 5 & 26 & 11 & 61 \\
$\Omega_5$ Suppression & 212 & 93 & 3 & 17 & 5 & 5 & 8 & 21 & 6 & 54 \\
$\Omega_6$ Matched removal & 232 & 106 & 8 & 12 & 7 & 6 & 5 & 26 & 10 & 52 \\
\bottomrule
\end{tabular*}
\end{table}

Lost citation coverage equals CU plus CA, while new coverage equals UC plus AC. The matrix therefore also identifies whether lost coverage occurs before or after acquisition. Among actions supported by Full-run citations, $\mathrm{CC}/(\mathrm{CC}+\mathrm{CU}+\mathrm{CA})$ is the fraction still supported by citations under the changed condition. This cross-run quantity differs from citation retention $K/O$, which compares acquired and formally cited support within one run. Unchanged failures such as UU do not establish successful investigation. Table~\ref{tab:telemetry-system-transitions} gives every system's counts, including acquisition under each condition.

For the model Full-to-Endpoint comparison, 80 of 451 system--action pairs retain citation support, 44 lose it, and 37 gain it. The net change is therefore $100(37-44)/451\approx-1.6$ percentage points, while only $80/(80+44)\approx64.5\%$ of Full-supported pairs remain covered. Of the 44 losses, 32 lack acquired support in the endpoint run and twelve retain an acquired witness but lack sufficient formal-stage citations. The small net change thus coexists with losses at both boundaries.


\begingroup\footnotesize
\setlength{\tabcolsep}{3pt}
\setlength{\LTpre}{0pt}\setlength{\LTpost}{0pt}
\setlength{\LTleft}{0pt}\setlength{\LTright}{0pt}
\setlength{\LTcapwidth}{\linewidth}
\begin{longtable}{@{}
>{\raggedright\arraybackslash}p{\dimexpr.22\linewidth-\tabcolsep\relax}
>{\raggedright\arraybackslash}p{\dimexpr.18\linewidth-2\tabcolsep\relax}
>{\raggedleft\arraybackslash}p{\dimexpr.05\linewidth-2\tabcolsep\relax}
>{\raggedleft\arraybackslash}p{\dimexpr.09\linewidth-2\tabcolsep\relax}
>{\raggedleft\arraybackslash}p{\dimexpr.065\linewidth-2\tabcolsep\relax}
>{\raggedleft\arraybackslash}p{\dimexpr.06\linewidth-2\tabcolsep\relax}
>{\raggedleft\arraybackslash}p{\dimexpr.08\linewidth-2\tabcolsep\relax}
>{\raggedleft\arraybackslash}p{\dimexpr.085\linewidth-2\tabcolsep\relax}
>{\raggedleft\arraybackslash}p{\dimexpr.085\linewidth-2\tabcolsep\relax}
>{\raggedleft\arraybackslash}p{\dimexpr.085\linewidth-\tabcolsep\relax}@{}}
\caption{Per-system changes in citation coverage for actions recoverable under both conditions. Each comparison uses primary paired scenarios shared by all systems in its panel. The four transition counts sum to $M$. Lost coverage is split into support not acquired (U) and support acquired but insufficiently cited (A). $O_0$ and $O_v$ count acquired actions under Full and the changed condition.}\label{tab:telemetry-system-transitions}\\
\toprule
System & View & $M$ & $O_0/O_v$ & Both & Lost & Gained & Neither & Lost U & Lost A \\
\midrule
\endfirsthead
\toprule System & View & $M$ & $O_0/O_v$ & Both & Lost & Gained & Neither & Lost U & Lost A \\ \midrule
\endhead
\bottomrule\endfoot
\multicolumn{10}{@{}l}{\textit{Eleven models, Codex fixed}} \\
MiniMax-M2 & $\Omega_1$ Endpoint & 41 & 8/7 & 0 & 0 & 1 & 40 & 0 & 0 \\
MiniMax-M2 & $\Omega_2$ Network & 14 & 4/3 & 0 & 1 & 1 & 12 & 0 & 1 \\
MiniMax-M2 & $\Omega_3$ Retention & 33 & 5/6 & 0 & 0 & 1 & 32 & 0 & 0 \\
MiniMax-M2 & $\Omega_4$ Sampling & 19 & 5/5 & 0 & 0 & 1 & 18 & 0 & 0 \\
MiniMax-M2 & $\Omega_5$ Suppression & 38 & 8/8 & 2 & 1 & 2 & 33 & 0 & 1 \\
MiniMax-M2 & $\Omega_6$ Matched & 35 & 7/7 & 0 & 1 & 1 & 33 & 0 & 1 \\
\addlinespace
MiniMax-M2.1-HS & $\Omega_1$ Endpoint & 41 & 8/11 & 0 & 1 & 1 & 39 & 1 & 0 \\
MiniMax-M2.1-HS & $\Omega_2$ Network & 14 & 4/3 & 0 & 1 & 0 & 13 & 0 & 1 \\
MiniMax-M2.1-HS & $\Omega_3$ Retention & 33 & 5/7 & 0 & 1 & 2 & 30 & 1 & 0 \\
MiniMax-M2.1-HS & $\Omega_4$ Sampling & 19 & 5/5 & 1 & 0 & 0 & 18 & 0 & 0 \\
MiniMax-M2.1-HS & $\Omega_5$ Suppression & 38 & 8/7 & 2 & 1 & 0 & 35 & 0 & 1 \\
MiniMax-M2.1-HS & $\Omega_6$ Matched & 35 & 7/7 & 0 & 0 & 1 & 34 & 0 & 0 \\
\addlinespace
MiniMax-M2.5-HS & $\Omega_1$ Endpoint & 41 & 5/12 & 0 & 1 & 3 & 37 & 0 & 1 \\
MiniMax-M2.5-HS & $\Omega_2$ Network & 14 & 4/4 & 0 & 1 & 1 & 12 & 0 & 1 \\
MiniMax-M2.5-HS & $\Omega_3$ Retention & 33 & 5/14 & 0 & 2 & 7 & 24 & 1 & 1 \\
MiniMax-M2.5-HS & $\Omega_4$ Sampling & 19 & 3/5 & 0 & 0 & 1 & 18 & 0 & 0 \\
MiniMax-M2.5-HS & $\Omega_5$ Suppression & 38 & 7/7 & 3 & 1 & 0 & 34 & 0 & 1 \\
MiniMax-M2.5-HS & $\Omega_6$ Matched & 35 & 5/8 & 0 & 1 & 0 & 34 & 0 & 1 \\
\addlinespace
MiniMax-M2.7-HS & $\Omega_1$ Endpoint & 41 & 21/11 & 2 & 7 & 2 & 30 & 7 & 0 \\
MiniMax-M2.7-HS & $\Omega_2$ Network & 14 & 4/5 & 0 & 1 & 0 & 13 & 0 & 1 \\
MiniMax-M2.7-HS & $\Omega_3$ Retention & 33 & 10/9 & 1 & 1 & 4 & 27 & 0 & 1 \\
MiniMax-M2.7-HS & $\Omega_4$ Sampling & 19 & 5/5 & 0 & 0 & 2 & 17 & 0 & 0 \\
MiniMax-M2.7-HS & $\Omega_5$ Suppression & 38 & 19/10 & 2 & 8 & 0 & 28 & 6 & 2 \\
MiniMax-M2.7-HS & $\Omega_6$ Matched & 35 & 12/12 & 2 & 0 & 0 & 33 & 0 & 0 \\
\addlinespace
MiniMax-M3 & $\Omega_1$ Endpoint & 41 & 23/14 & 9 & 10 & 0 & 22 & 10 & 0 \\
MiniMax-M3 & $\Omega_2$ Network & 14 & 10/9 & 4 & 2 & 2 & 6 & 1 & 1 \\
MiniMax-M3 & $\Omega_3$ Retention & 33 & 16/12 & 5 & 6 & 0 & 22 & 2 & 4 \\
MiniMax-M3 & $\Omega_4$ Sampling & 19 & 8/9 & 5 & 1 & 1 & 12 & 0 & 1 \\
MiniMax-M3 & $\Omega_5$ Suppression & 38 & 25/19 & 11 & 8 & 0 & 19 & 4 & 4 \\
MiniMax-M3 & $\Omega_6$ Matched & 35 & 18/16 & 9 & 6 & 2 & 18 & 1 & 5 \\
\addlinespace
DeepSeek-V4-Pro & $\Omega_1$ Endpoint & 41 & 21/15 & 5 & 9 & 5 & 22 & 8 & 1 \\
DeepSeek-V4-Pro & $\Omega_2$ Network & 14 & 9/13 & 8 & 0 & 2 & 4 & 0 & 0 \\
DeepSeek-V4-Pro & $\Omega_3$ Retention & 33 & 15/15 & 9 & 4 & 0 & 20 & 1 & 3 \\
DeepSeek-V4-Pro & $\Omega_4$ Sampling & 19 & 6/6 & 2 & 0 & 0 & 17 & 0 & 0 \\
DeepSeek-V4-Pro & $\Omega_5$ Suppression & 38 & 23/27 & 19 & 1 & 1 & 17 & 0 & 1 \\
DeepSeek-V4-Pro & $\Omega_6$ Matched & 35 & 13/13 & 4 & 3 & 0 & 28 & 0 & 3 \\
\addlinespace
GPT-5.5 & $\Omega_1$ Endpoint & 41 & 25/27 & 16 & 1 & 3 & 21 & 1 & 0 \\
GPT-5.5 & $\Omega_2$ Network & 14 & 9/10 & 8 & 0 & 2 & 4 & 0 & 0 \\
GPT-5.5 & $\Omega_3$ Retention & 33 & 16/16 & 9 & 2 & 1 & 21 & 1 & 1 \\
GPT-5.5 & $\Omega_4$ Sampling & 19 & 9/11 & 7 & 0 & 1 & 11 & 0 & 0 \\
GPT-5.5 & $\Omega_5$ Suppression & 38 & 25/24 & 16 & 3 & 3 & 16 & 3 & 0 \\
GPT-5.5 & $\Omega_6$ Matched & 35 & 17/19 & 10 & 1 & 0 & 24 & 1 & 0 \\
\addlinespace
GPT-5.6-Luna & $\Omega_1$ Endpoint & 41 & 17/28 & 5 & 5 & 13 & 18 & 1 & 4 \\
GPT-5.6-Luna & $\Omega_2$ Network & 14 & 6/9 & 2 & 0 & 5 & 7 & 0 & 0 \\
GPT-5.6-Luna & $\Omega_3$ Retention & 33 & 15/10 & 3 & 3 & 1 & 26 & 2 & 1 \\
GPT-5.6-Luna & $\Omega_4$ Sampling & 19 & 10/7 & 5 & 2 & 0 & 12 & 1 & 1 \\
GPT-5.6-Luna & $\Omega_5$ Suppression & 38 & 16/19 & 5 & 2 & 9 & 22 & 1 & 1 \\
GPT-5.6-Luna & $\Omega_6$ Matched & 35 & 21/20 & 7 & 3 & 1 & 24 & 0 & 3 \\
\addlinespace
GPT-5.6-Terra & $\Omega_1$ Endpoint & 41 & 25/24 & 9 & 6 & 1 & 25 & 2 & 4 \\
GPT-5.6-Terra & $\Omega_2$ Network & 14 & 10/10 & 8 & 0 & 1 & 5 & 0 & 0 \\
GPT-5.6-Terra & $\Omega_3$ Retention & 33 & 17/16 & 8 & 3 & 1 & 21 & 1 & 2 \\
GPT-5.6-Terra & $\Omega_4$ Sampling & 19 & 10/7 & 5 & 2 & 0 & 12 & 1 & 1 \\
GPT-5.6-Terra & $\Omega_5$ Suppression & 38 & 26/23 & 14 & 3 & 1 & 20 & 1 & 2 \\
GPT-5.6-Terra & $\Omega_6$ Matched & 35 & 20/16 & 5 & 5 & 2 & 23 & 3 & 2 \\
\addlinespace
GPT-5.6-Sol & $\Omega_1$ Endpoint & 41 & 27/29 & 15 & 3 & 3 & 20 & 1 & 2 \\
GPT-5.6-Sol & $\Omega_2$ Network & 14 & 9/11 & 9 & 0 & 1 & 4 & 0 & 0 \\
GPT-5.6-Sol & $\Omega_3$ Retention & 33 & 18/20 & 8 & 3 & 1 & 21 & 1 & 2 \\
GPT-5.6-Sol & $\Omega_4$ Sampling & 19 & 9/8 & 6 & 0 & 0 & 13 & 0 & 0 \\
GPT-5.6-Sol & $\Omega_5$ Suppression & 38 & 29/28 & 19 & 1 & 0 & 18 & 0 & 1 \\
GPT-5.6-Sol & $\Omega_6$ Matched & 35 & 21/24 & 10 & 0 & 2 & 23 & 0 & 0 \\
\addlinespace
GPT-6-Astra & $\Omega_1$ Endpoint & 41 & 33/32 & 19 & 1 & 5 & 16 & 1 & 0 \\
GPT-6-Astra & $\Omega_2$ Network & 14 & 10/12 & 7 & 1 & 1 & 5 & 0 & 1 \\
GPT-6-Astra & $\Omega_3$ Retention & 33 & 24/19 & 12 & 2 & 1 & 18 & 2 & 0 \\
GPT-6-Astra & $\Omega_4$ Sampling & 19 & 13/13 & 7 & 0 & 0 & 12 & 0 & 0 \\
GPT-6-Astra & $\Omega_5$ Suppression & 38 & 32/29 & 22 & 1 & 1 & 14 & 0 & 1 \\
GPT-6-Astra & $\Omega_6$ Matched & 35 & 28/24 & 13 & 1 & 1 & 20 & 0 & 1 \\
\addlinespace
\multicolumn{10}{@{}l}{\textit{Four frameworks, MiniMax-M3 fixed}} \\
Codex & $\Omega_1$ Endpoint & 55 & 32/14 & 9 & 15 & 0 & 31 & 15 & 0 \\
Codex & $\Omega_2$ Network & 17 & 9/8 & 4 & 0 & 2 & 11 & 0 & 0 \\
Codex & $\Omega_3$ Retention & 35 & 15/14 & 6 & 3 & 2 & 24 & 0 & 3 \\
Codex & $\Omega_4$ Sampling & 62 & 35/28 & 17 & 10 & 3 & 32 & 5 & 5 \\
Codex & $\Omega_5$ Suppression & 53 & 28/18 & 10 & 11 & 0 & 32 & 7 & 4 \\
Codex & $\Omega_6$ Matched & 58 & 31/29 & 18 & 5 & 6 & 29 & 1 & 4 \\
\addlinespace
Claude Code & $\Omega_1$ Endpoint & 55 & 20/13 & 5 & 10 & 3 & 37 & 6 & 4 \\
Claude Code & $\Omega_2$ Network & 17 & 7/8 & 4 & 0 & 3 & 10 & 0 & 0 \\
Claude Code & $\Omega_3$ Retention & 35 & 12/15 & 7 & 3 & 1 & 24 & 1 & 2 \\
Claude Code & $\Omega_4$ Sampling & 62 & 22/20 & 8 & 9 & 5 & 40 & 5 & 4 \\
Claude Code & $\Omega_5$ Suppression & 53 & 17/27 & 11 & 3 & 13 & 26 & 2 & 1 \\
Claude Code & $\Omega_6$ Matched & 58 & 17/28 & 9 & 6 & 9 & 34 & 3 & 3 \\
\addlinespace
OpenCode & $\Omega_1$ Endpoint & 55 & 27/14 & 6 & 12 & 2 & 35 & 10 & 2 \\
OpenCode & $\Omega_2$ Network & 17 & 7/12 & 2 & 0 & 6 & 9 & 0 & 0 \\
OpenCode & $\Omega_3$ Retention & 35 & 12/10 & 6 & 3 & 2 & 24 & 2 & 1 \\
OpenCode & $\Omega_4$ Sampling & 62 & 30/29 & 18 & 3 & 6 & 35 & 1 & 2 \\
OpenCode & $\Omega_5$ Suppression & 53 & 24/27 & 15 & 2 & 10 & 26 & 2 & 0 \\
OpenCode & $\Omega_6$ Matched & 58 & 25/18 & 10 & 11 & 2 & 35 & 9 & 2 \\
\addlinespace
Ref-ReAct & $\Omega_1$ Endpoint & 55 & 33/10 & 8 & 23 & 0 & 24 & 22 & 1 \\
Ref-ReAct & $\Omega_2$ Network & 17 & 9/8 & 7 & 0 & 1 & 9 & 0 & 0 \\
Ref-ReAct & $\Omega_3$ Retention & 35 & 13/11 & 5 & 6 & 3 & 21 & 4 & 2 \\
Ref-ReAct & $\Omega_4$ Sampling & 62 & 36/20 & 18 & 15 & 1 & 28 & 15 & 0 \\
Ref-ReAct & $\Omega_5$ Suppression & 53 & 30/21 & 18 & 11 & 2 & 22 & 10 & 1 \\
Ref-ReAct & $\Omega_6$ Matched & 58 & 33/18 & 15 & 14 & 0 & 29 & 13 & 1 \\
\addlinespace
\end{longtable}\endgroup

\subsection{Incident composition and selection sensitivity}
\begin{table}[htbp]
\centering
\small
\setlength{\tabcolsep}{2pt}
\renewcommand{\arraystretch}{1.12}
\caption{Sensitivity of citation-coverage changes to weighting and case selection.}
\label{tab:telemetry-sensitivity}
\begin{tabular*}{\linewidth}{@{}l@{\hspace{5pt}\extracolsep{\fill}}*{10}{c}@{}}
\toprule
Condition
& \multicolumn{2}{c}{Primary}
& \multicolumn{2}{c}{Source}
& \multicolumn{2}{c}{Scenario}
& \multicolumn{2}{c}{Own}
& \multicolumn{2}{c}{Accepted} \\
\cmidrule(lr){2-3}\cmidrule(lr){4-5}\cmidrule(lr){6-7}
\cmidrule(lr){8-9}\cmidrule(l){10-11}
& M & F & M & F & M & F & M & F & M & F \\
\midrule
$\Omega_1$ Endpoint & $-1.6$ & $-25.0$ & $-2.0$ & $-21.1$ & $-2.0$ & $-22.2$ & $-5.3$ & $-22.7$ & $-6.0$ & $-22.7$ \\
$\Omega_2$ Network & $+5.8$ & $+17.6$ & $+6.5$ & $+9.6$ & $+6.5$ & $+9.6$ & $+3.4$ & $+7.2$ & $+4.0$ & $+7.2$ \\
$\Omega_3$ Retention & $-2.2$ & $-5.0$ & $-2.7$ & $-6.5$ & $-2.7$ & $-6.5$ & $-1.6$ & $-10.4$ & $-1.5$ & $-8.4$ \\
$\Omega_4$ Sampling & $+0.5$ & $-8.9$ & $+3.3$ & $-9.1$ & $+3.3$ & $-10.0$ & $-3.9$ & $-7.6$ & $-4.2$ & $-7.6$ \\
$\Omega_5$ Suppression & $-3.1$ & $-0.9$ & $-2.2$ & $+1.0$ & $-2.2$ & $-3.3$ & $-3.0$ & $-2.4$ & $-3.0$ & $-2.4$ \\
$\Omega_6$ Matched removal & $-2.9$ & $-8.2$ & $-2.3$ & $-6.9$ & $-2.3$ & $-8.6$ & $-3.3$ & $-5.4$ & $-2.3$ & $-5.4$ \\
\bottomrule
\end{tabular*}
\par\vspace{3pt}
\begin{minipage}{\linewidth}
\footnotesize\raggedright
M: eleven models; F: four frameworks.
Values are percentage-point changes from Full to the indicated condition.
Primary pools the same recoverable actions across systems.
Source and Scenario give equal weight to source-incident groups and scenarios, respectively.
Own uses each system's primary paired runs.
Accepted also includes accepted runs with usable audits outside the primary filter.
\end{minipage}\par
\end{table}


\paragraph{Weighting.}
\label{app:deep_paired_conditions}
The primary differences use the same system--action pairs as the two coverage rates, so gained minus lost exactly reproduces the change. The Source and Scenario columns give equal weight to source-incident groups and scenarios, respectively.

\paragraph{Bootstrap intervals.}
Each of 20,000 bootstrap draws resamples whole source-incident groups, preserving their systems, actions, and related reconstructions. The saved summary also reports the result after omitting each source-incident group. These analyses measure sensitivity to the incident mix, not variability across repeated runs of the same task. Some comparisons contain few groups, notably three for the model sampling comparison, so intervals are exploratory.

\paragraph{Run-selection sensitivity.}
The Own analysis uses every qualifying primary pair for each system separately. The Accepted analysis also includes scoreable runs excluded by the primary runtime filter. These summaries can use different scenarios across systems, unlike the shared-case primary comparison.

\FloatBarrier
\subsection{Qualification for unchanged-witness comparisons}
\label{app:telemetry_witnesses}
The strict audit examines all 75 eligible actions before reading their investigation outcomes. An action qualifies when Full and endpoint-only views contain identical candidate record IDs and identical registered minimal-witness sets, with at least one witness in both views. Threshold-based rules are evaluated with the frozen scorer rather than reduced to record counts. Forty-five actions across nine reconstructions and eight source-incident groups meet these conditions. The panel-specific paired runs reduce this to 32 actions from five source-incident groups for the model comparison and 37 actions from six groups for the framework comparison (Figure~\ref{fig:telemetry-witness-all-qualified}). Each system's rates use its panel's fixed action denominator.

\begin{wraptable}{r}{0.51\textwidth}
\centering\small
\caption{Full-to-Endpoint evidence screen over all 75 eligible actions, before inspecting investigation outcomes.}
\label{tab:telemetry-witness-qualification}
\setlength{\tabcolsep}{3pt}
\begin{tabularx}{\linewidth}{@{}Xr@{}}
\toprule
Screening outcome & Actions \\
\midrule
All candidates and minimal witnesses retained & 45 \\
Support survives, evidence sets change & 19 \\
Full supported, no Endpoint witness & 9 \\
No Full-view witness & 2 \\
\midrule
Total & 75 \\
\bottomrule
\end{tabularx}
\end{wraptable}

The qualification archive lists every action, its rules, both record sets and minimal witnesses, and all applicable rejection reasons. These reasons can overlap: losing a candidate record can change the witness set and remove complete support. Table~\ref{tab:telemetry-witness-qualification} instead groups actions into mutually exclusive outcomes. The audit excludes the same unresolved compound action as the main evaluator and does not select actions based on their investigation results.

A read-only check of the saved case databases also compares every stored column of the target records, including raw record content. Across twenty saved views and 713 candidate-row occurrences, the qualifying actions retain identical record values and schemas. This verifies equality of stored target evidence. The investigator's surrounding context and tool responses may still differ between views.


Figure~\ref{fig:telemetry-witness-all-qualified} shows that outcomes
vary across systems despite unchanged registered evidence.
Under endpoint-only telemetry, all five GPT models acquire more
support than under Full, while Claude Code's acquisition is unchanged.
Citation coverage nevertheless decreases for all four frameworks.
Results after omitting source-incident groups are retained for each
qualified system subset.

\paragraph{Independent replay checks.}
A separate set-based verifier reconstructs pair eligibility and transitions without calling the analysis functions. The strict audit independently replays availability and citation support for 2,071 action states and acquisition for 1,869 states. Twenty-eight additional rows use hash-authenticated acquisition values from saved replay because local returned-ID projections are unavailable. These checks use existing investigations without changing rules, selecting new attempts, or rerunning systems.

\subsection{Interpreting targeted evidence removal}
\label{app:paired}
Targeted suppression removes registered support for one selected action. Its matched control deletes the same number of background records from the same source format and instance. Time and byte size are matched approximately. Median and maximum time discrepancies are 5.16 and 92.31 seconds, and absolute byte discrepancies are 2.93\% and 37.46\%. We separately examine the environment transformation, coverage of remaining actions, and assertions about the target.

\paragraph{Was the target's support removed?}
The revised available-record check yields target support states $(0,1)$ in the suppression and control arms for nine of the ten archived scenario pairs. Donot instead yields $(0,0)$ after correction of its original control witness. Its unresolved compound target is excluded from current comparisons. The 90\% figure describes this ten-pair environment audit. This check establishes the support contrast independently of report content.

\paragraph{What happens to the remaining investigation?}
\begin{table}[htbp]
\centering
\footnotesize
\setlength{\tabcolsep}{4pt}
\caption{Citation coverage after targeted suppression versus matched benign removal, using the same remaining recoverable actions and current support rules.}
\label{tab:direct_support_pair}
\begin{tabular*}{\linewidth}{@{}l@{\extracolsep{\fill}}rrr@{}}
\toprule
System & Supp. $K/R$ & Matched $K/R$ & Difference [95\% interval] \\
\midrule
\multicolumn{4}{l}{\textit{Eleven models}} \\
MiniMax-M2 & 6.2 & 3.1 & +3.1 [-3.1, +9.6] \\
MiniMax-M2.1-HS & 3.1 & 6.2 & -3.1 [-7.3, -0.0] \\
MiniMax-M2.5-HS & 5.4 & 3.6 & +1.8 [-0.0, +7.0] \\
MiniMax-M2.7-HS & 3.4 & 10.3 & -6.9 [-14.6, -0.0] \\
MiniMax-M3 & 20.3 & 32.8 & -12.5 [-26.9, +3.8] \\
DeepSeek-V4-Pro & 35.9 & 17.2 & +18.8 [+5.1, +36.3] \\
GPT-5.5 & 45.6 & 45.6 & -0.0 [-6.9, +9.6] \\
GPT-5.6-Luna & 33.3 & 28.2 & +5.1 [-5.0, +15.0] \\
GPT-5.6-Terra & 25.0 & 32.8 & -7.8 [-12.7, -2.0] \\
GPT-5.6-Sol & 48.4 & 46.9 & +1.6 [-9.1, +8.9] \\
GPT-6-Astra & 53.1 & 53.1 & -0.0 [-4.7, +4.2] \\
\midrule
\multicolumn{4}{l}{\textit{Four frameworks}} \\
Codex & 20.3 & 32.8 & -12.5 [-26.9, +3.8] \\
Claude Code & 43.8 & 37.5 & +6.2 [-6.9, +24.1] \\
OpenCode & 47.2 & 26.4 & +20.8 [+2.0, +42.2] \\
Ref-ReAct & 32.1 & 28.6 & +3.6 [-8.1, +20.9] \\
\bottomrule
\end{tabular*}
\end{table}

\begin{wraptable}{r}{0.54\textwidth}
\centering
\footnotesize
\setlength{\tabcolsep}{3pt}
\caption{Target recovery under Full and matched benign removal (\%).}
\label{tab:deep-conditions-floor}
\begin{tabular*}{\linewidth}{@{}l@{\extracolsep{\fill}}rrrr@{}}
\toprule
& \multicolumn{2}{c}{Full} & \multicolumn{2}{c}{Matched} \\
\cmidrule(lr){2-3}\cmidrule(l){4-5}
Panel & $O/R$ & $K/R$ & $O/R$ & $K/R$ \\
\midrule
Eleven models & 39.4 & 24.2 & 36.4 & 21.2 \\
Four frameworks & 29.2 & 12.5 & 29.2 & 16.7 \\
\bottomrule
\end{tabular*}
\end{wraptable}

Table~\ref{tab:direct_support_pair} measures citation coverage of the remaining actions recoverable under both conditions. Each configuration uses its paired accepted runs under the current support rules, with the target excluded from both arms. Positive and negative changes show that the separately executed investigations can cover different shares of these actions.

\paragraph{Was the target covered before removal?}
Table~\ref{tab:deep-conditions-floor} includes all eligible targets in each panel's shared intersection of Full, Suppression, and Matched-removal runs, including targets not covered in Full. It contains three targets for the model panel and six for the framework panel.

Under full telemetry, reports cover 24.2\% of model-panel system--target pairs and 12.5\% of framework-panel pairs. Target coverage is already uncommon before suppression. After suppression, the selected target has no registered sufficient support and cannot contribute to citation coverage. Its target-specific coverage ratio is undefined, although other actions can remain recoverable.

\paragraph{How does the report describe the target?}
Appendix~\ref{app:target_claim_audit} separately checks whether audited reports omit the target, express uncertainty, or continue to assert it, and examines their surviving citations. Its unit is a report or assertion. Keeping these analyses separate prevents an environment-support check from being mistaken for evidence of more cautious reporting.

\FloatBarrier
\section{Acquired evidence and citation retention}
\label{app:r1_synthesis}
This appendix examines whether reports preserve support already acquired during investigation. Citation retention, $K/O$, is the fraction of acquired actions whose sufficient support remains in formal-stage citations. We first distinguish acquisition gaps from citation omissions, then examine lead-query returns, shared acquired support, citation selection, and unmet witness requirements. These controls use the current rules and each panel's shared cases. They complement the cross-telemetry comparisons.

\subsection{Acquisition gaps and citation omissions}
\label{app:coverage_boundaries}
\label{app:observation_path_audit}
\begin{figure}[ht]
\centering 
\includegraphics[width=0.95\textwidth]{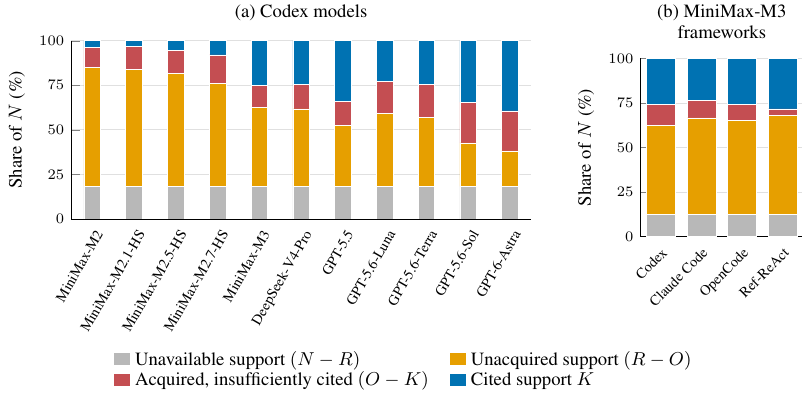} 
\caption{Where support is lost across availability, acquisition, and formal-stage citations. Each bar sums to 100\% of $N$, the eligible reference-action opportunities. Blue is $K/N$, while headline coverage uses $K/R$ and counts only recoverable opportunities in its denominator.}
\label{fig:boundary_decomposition}
\end{figure}

Figure~\ref{fig:boundary_decomposition} partitions all eligible action opportunities, $N$, into supported citations and three gaps: unavailable support, available support not acquired, and acquired support absent from formal-stage citations. Its denominator differs from headline citation coverage, $K/R$, which excludes actions without sufficient available evidence.

Across the models' shared cases, GPT-6-Astra leaves 20.0\% of eligible opportunities with available but unacquired support and 22.4\% with acquired support not preserved in formal-stage citations. These percentages describe the full analysis population.

\paragraph{Zero-coverage investigations.}
\label{app:returned_witness_complexity}
Within the framework comparison, Codex with MiniMax-M3 acquires at least one witness in 13 of its 22 zero-coverage cases (59.1\%), compared with one of 24 for Ref-ReAct (4.2\%). Ref-ReAct returns no source-labelled incident record in 21 of these 24 cases (87.5\%). Each percentage uses that system's own zero-coverage cases. Thus, the same final score can reflect either missing acquisition or failure to preserve acquired support. The OpenCode Spyder case separately verifies that complete witness content appeared in saved tool output (Appendix~\ref{app:examples}).

\subsection{New support and omissions after lead queries}
\label{app:public_lead_expansion}
Table~\ref{tab:lead_reference_guide} distinguishes the fixed lead-query reference, shared first-page returns, and run-specific lead returns. Here, $B_i$ contains actions supported by the exact supplied-lead queries executed in run $i$. Queries must match the supplied pattern, source format, instance, and time interval. Continuation pages count only when consumed at the preceding page's next offset, even if other queries intervene. Unrelated searches and fresh offset-zero reruns are excluded. This classifier applies to all eleven Codex models. Ref-ReAct transforms the queries, so its exact-lead comparison is N/A.

Let $F_i$ contain actions supported by run $i$'s formal-stage citations. Net growth separates into additions and omissions:
\begin{equation}
|F_i|-|B_i|=|F_i\setminus B_i|-|B_i\setminus F_i|.
\label{eq:gain_loss}
\end{equation}
An addition can reuse lead records but requires evidence beyond them to complete a witness. An omission has sufficient lead-query support but no witness in the formal-stage citations. Table~\ref{tab:r1_gain_loss} uses lead-supported opportunities for omission and preservation rates, and all recoverable opportunities for addition rates.

\begin{table}[htbp]
\centering
\footnotesize
\setlength{\tabcolsep}{4pt}
\caption{Support added and omitted in formal-stage citations relative to supplied-lead retrieval. All eleven Codex models use the same shared analysis cases and current support rules.}
\label{tab:r1_gain_loss}
\begin{tabular*}{\linewidth}{@{}l@{\extracolsep{\fill}}rrrrr@{}}
\toprule
Model & Lead $|B|/R$ & Final $K/R$ & Added / $R$ & Omitted / $|B|$ & Net (pp) \\
\midrule
MiniMax-M2 & 14.4 & 4.3 & 0.7 & 75.0 & -10.1 \\
MiniMax-M2.1-HS & 13.7 & 4.0 & 1.1 & 78.9 & -9.7 \\
MiniMax-M2.5-HS & 13.7 & 6.8 & 4.3 & 81.6 & -6.8 \\
MiniMax-M2.7-HS & 14.4 & 9.7 & 7.2 & 82.5 & -4.7 \\
MiniMax-M3 & 14.4 & 30.6 & 20.5 & 30.0 & +16.2 \\
DeepSeek-V4-Pro & 14.4 & 30.2 & 21.6 & 40.0 & +15.8 \\
GPT-5.5 & 14.4 & 41.4 & 29.9 & 20.0 & +27.0 \\
GPT-5.6-Luna & 14.4 & 27.7 & 17.6 & 30.0 & +13.3 \\
GPT-5.6-Terra & 14.4 & 29.9 & 19.8 & 30.0 & +15.5 \\
GPT-5.6-Sol & 14.4 & 42.1 & 30.9 & 22.5 & +27.7 \\
GPT-6-Astra & 14.4 & 48.2 & 37.1 & 22.5 & +33.8 \\
\bottomrule
\end{tabular*}
\end{table}

\begin{figure}[t]
\centering 
\includegraphics[width=0.95\textwidth]{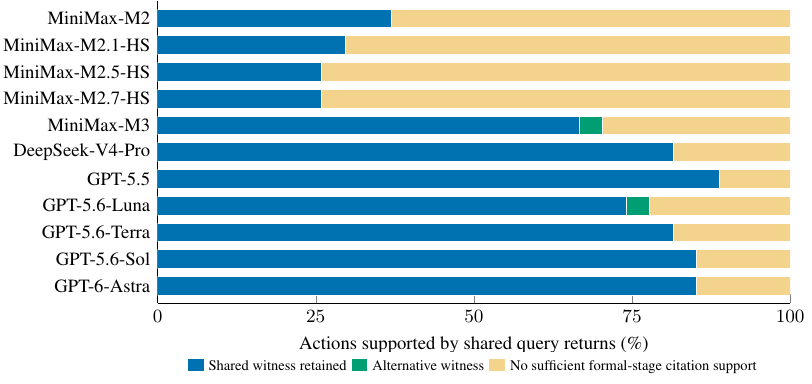} 
\caption{Support returned to every model may still be missing from formal-stage citations. Blue preserves a shared witness, green supplies an alternative witness, and orange lacks sufficient formal-stage citation support.}
\label{fig:r1_shared_support}
\end{figure}

For MiniMax-M3 and DeepSeek-V4-Pro, five of 47 investigations (10.6\%) combine positive net growth with a lead-supported omission. MiniMax-M3's cases occur in Dragonfly and Donot, while DeepSeek-V4-Pro's all come from Dragonfly. These are within-run outcomes with different source concentration across systems.

In MiniMax-M3's Donot sampled-view run, the first page supports 10.0\% of recoverable actions and consumed continuation pages raise this to 30.0\%. Acquisition reaches 60.0\% and citation coverage 40.0\%, but the report preserves only two-thirds of the lead-supported actions. Because continuation pages may arrive later, $B_i$ identifies query provenance rather than a common initial checkpoint.

\subsection{Comparisons with shared or matched acquired support}
\paragraph{Shared first-page records.}
For each case, this control intersects record IDs returned on the first page of supplied-lead queries across all eleven models. The shared records support 27 action--view opportunities spanning seven actions and seven source-incident groups. Figure~\ref{fig:r1_shared_support} distinguishes preservation of a shared witness, use of an alternative witness, and omission of sufficient support.

A complete alternative earns the same action credit as the shared witness. GPT-5.6-Luna preserves a shared witness on 74.1\% of opportunities and an alternative on another 3.7\%. The control fixes shared returned IDs.

\paragraph{Common acquired actions.}
\label{app:matched_acquisition}
This control instead matches action opportunities acquired by every system in a panel, without requiring identical returned records. The model set contains 38 opportunities for eleven actions, and the framework set contains 77 opportunities for 29 actions. Each spans eight source-incident groups. Table~\ref{tab:readability_acquisition_controls} also groups each system's own opportunities by its smallest acquired witness, denoted $w_D$.

\begin{table}[htb]
\centering\small
\setlength{\tabcolsep}{4pt}
\renewcommand{\arraystretch}{1.06}
\caption{Retention on shared acquired actions and two-record witness subsets.}
\label{tab:readability_acquisition_controls}
\begin{tabular*}{\linewidth}{@{}l@{\extracolsep{\fill}}rrr@{}}
\toprule
Investigator & \shortstack{Common acquired\\$K/O$ (\%)} & \shortstack{Own $w_D=2$\\opportunities} & \shortstack{Own $w_D=2$\\$K/O$ (\%)} \\
\midrule
\multicolumn{4}{@{}l}{\textit{Models with Codex}} \\
MiniMax-M2 & 31.6 & 9 & 0.0 \\
MiniMax-M2.1-HS & 23.7 & 10 & 10.0 \\
MiniMax-M2.5-HS & 23.7 & 12 & 8.3 \\
MiniMax-M2.7-HS & 23.7 & 20 & 15.0 \\
MiniMax-M3 & 68.4 & 38 & 55.3 \\
DeepSeek-V4-Pro & 68.4 & 38 & 42.1 \\
GPT-5.5 & 78.9 & 61 & 52.5 \\
GPT-5.6-Luna & 65.8 & 53 & 43.4 \\
GPT-5.6-Terra & 73.7 & 45 & 33.3 \\
GPT-5.6-Sol & 76.3 & 76 & 39.5 \\
GPT-6-Astra & 76.3 & 85 & 50.6 \\
\midrule
\multicolumn{4}{@{}l}{\textit{Frameworks with MiniMax-M3}} \\
Codex & 75.3 & 52 & 53.8 \\
Claude Code & 75.3 & 41 & 61.0 \\
OpenCode & 75.3 & 48 & 64.6 \\
Ref-ReAct & 90.9 & 40 & 77.5 \\
\bottomrule
\end{tabular*}
\end{table}

On the common model set, MiniMax-M3 and DeepSeek-V4-Pro both retain 68.4\%, compared with 78.9\% for GPT-5.5 and 76.3\% for GPT-5.6-Sol and GPT-6-Astra. In source-attested Execution, MiniMax-M3 retains 85.7\% of 21 opportunities acquired by both it and MiniMax-M2.7-HS, compared with 23.8\% for MiniMax-M2.7-HS. These opportunities span five source-incident groups. Removing each group in turn leaves a positive difference of 46.2--68.8 percentage points.

Matching both the opportunity and $w_D=2$ leaves 23 opportunities from six source-incident groups for MiniMax-M3 and DeepSeek-V4-Pro, with retention of 47.8\% and 43.5\%.

\paragraph{Different retained actions.}
\label{app:post_acquisition_complementarity}
Table~\ref{tab:readability_complementarity} compares citation support when both systems acquired a witness. Among jointly acquired opportunities, MiniMax-M3 alone covers 15.7\% and DeepSeek-V4-Pro alone covers 14.7\%. For GPT-5.6-Sol and GPT-6-Astra, the corresponding shares are 7.1\% and 12.5\%. Thus, matching acquisition need not produce the same supported actions in the reports. Their exact records and context may still differ.

\begin{table}[htbp]
\centering\footnotesize
\setlength{\tabcolsep}{3pt}
\caption{Overlap in citation coverage when both systems acquired sufficient support.}
\label{tab:readability_complementarity}
\begin{tabular*}{\linewidth}{@{}L{0.39\linewidth}@{\extracolsep{\fill}}rrrrr@{}}
\toprule
Left / right & Both & Left only & Right only & Neither & Share$^\dagger$ \\
\midrule
MiniMax-M3 / MiniMax-M2.7-HS & 30.6 & 43.1 & 4.2 & 22.2 & 50.0 \\
MiniMax-M3 / DeepSeek-V4-Pro & 52.9 & 15.7 & 14.7 & 16.7 & 50.8 \\
GPT-5.6-Sol / GPT-6-Astra & 56.5 & 7.1 & 12.5 & 23.9 & 83.7 \\
Codex / Ref-ReAct & 73.6 & 2.8 & 16.0 & 7.5 & 29.4 \\
\bottomrule
\end{tabular*}
\par\vspace{3pt}\begin{minipage}{\linewidth}\footnotesize $^\dagger$Share instead conditions on opportunities covered by exactly one member.\end{minipage}
\end{table}

\subsection{Citation selection with saved outputs fixed}
\label{app:citation_selection}
\label{app:current_citation_controls}
For each saved investigation, $D$ contains returned record IDs and $C$ contains formal-stage citation IDs. On the shared framework cases, three offline controls uniformly sample records without replacement:
\begin{enumerate}\setlength{\itemsep}{1pt}\setlength{\parskip}{0pt}
\item Sample $|C|$ records from that report's $D$, preserving its citation count.
\item Sample the case-wise minimum citation count across frameworks from each $D$.
\item Sample that same common count from each report's existing $C$.
\end{enumerate}
The common count is zero in fourteen of the 56 shared cases (25.0\%), which remain included. Each control uses 2,000 archived draws per report, rescored under the current rules with verified sampling-stream hashes. Table~\ref{tab:restored-citation-controls} combines expected coverage, selection intervals, actual coverage, and expanded-field coverage.

\begin{table}[htbp]
\centering\footnotesize
\setlength{\tabcolsep}{3pt}
\renewcommand{\arraystretch}{1.1}
\caption{Citation coverage under fixed-output controls on the shared four-framework population. Random-selection entries give expected $K/R$ (\%) and 95\% selection intervals from 2,000 draws. Actual and expanded-field scores are deterministic.}
\label{tab:restored-citation-controls}
\label{tab:restored-citation-distributions}
\begin{tabular*}{\linewidth}{@{}l@{\extracolsep{\fill}}rrrrr@{}}
\toprule
Framework & Actual & \shortstack{Random $D$\\own count} & \shortstack{Random $D$\\common count} & \shortstack{Thin $C$\\common count} & \shortstack{Expanded\\fields} \\
\midrule
Codex & 29.6 & \shortstack{1.98\\{[0.83, 3.31]}} & \shortstack{1.16\\{[0.28, 2.21]}} & \shortstack{20.66\\{[19.06, 22.10]}} & 31.5 \\
Claude Code & 27.1 & \shortstack{3.65\\{[1.93, 5.25]}} & \shortstack{2.01\\{[0.83, 3.31]}} & \shortstack{18.80\\{[17.40, 20.17]}} & 28.2 \\
OpenCode & 29.6 & \shortstack{4.05\\{[2.21, 6.08]}} & \shortstack{2.03\\{[0.83, 3.31]}} & \shortstack{18.15\\{[16.57, 19.89]}} & 30.7 \\
Ref-ReAct & 32.3 & \shortstack{3.34\\{[1.66, 5.25]}} & \shortstack{1.00\\{[0.00, 2.21]}} & \shortstack{14.20\\{[12.15, 16.30]}} & 33.1 \\
\bottomrule
\end{tabular*}
\end{table}

Actual citations preserve more support than equally many random returned records. Thinning citations to a common count changes the framework ordering: expected coverage is 20.7\% for Codex, 18.8\% for Claude Code, 18.2\% for OpenCode, and 14.2\% for Ref-ReAct. This retrospective control tests citation selection and volume.

Expanding citation extraction to the five non-hypothesis fields raises coverage by 0.8--1.9 percentage points per framework. Some additional fields contain rejected candidates, so the primary score retains the formal-stage boundary. The selection intervals condition on fixed reports and describe random record subsets. Source-incident resampling instead tests sensitivity to the incident sample.

\paragraph{No candidates versus incomplete support.}
Among acquired opportunities lacking sufficient formal-stage citations, no registered candidate is cited for 71.4\% of MiniMax-M3's 42 opportunities, 67.4\% of DeepSeek-V4-Pro's 46, 43.5\% of GPT-5.5's 46, and 80.3\% of GPT-6-Astra's 76. Each rate uses that model's own acquired-but-insufficiently-cited set. Many omissions therefore lack all candidate evidence, rather than an additional record needed to complete a cited combination.

\subsection{Unmet witness requirements}
\label{app:deep_support_structure}
A witness can remain incomplete because records are missing or because the selected records do not establish the required relationship. Fixed conjunctions require complementary records. Recurring-network rules additionally constrain timing and entity bindings. Table~\ref{tab:deep-structure-current} groups recoverable action--view--run opportunities by their smallest available witness and applies the same rule to returned records $D$ and formal-stage citations $C^\cup$ (abbreviated $C$ in the tables). Each panel uses its own shared analysis cases.

\begin{table}[htbp]
\centering
\small
\setlength{\tabcolsep}{2.5pt}
\renewcommand{\arraystretch}{1.12}
\caption{Acquisition and citation coverage by the structure of the required witness, under the current support rules.}
\label{tab:deep-structure-current}
\begin{tabular*}{\linewidth}{@{}l@{\hspace{5pt}\extracolsep{\fill}}*{8}{c}@{}}
\toprule
Structure ($w_E$)
& \multicolumn{2}{c}{Opportunities}
& \multicolumn{2}{c}{Any in $D$ (\%)}
& \multicolumn{2}{c}{$O/R$ (\%)}
& \multicolumn{2}{c}{$K/R$ (\%)} \\
\cmidrule(lr){2-3}\cmidrule(lr){4-5}
\cmidrule(lr){6-7}\cmidrule(l){8-9}
& M & F & M & F & M & F & M & F \\
\midrule
Single record (1) & 1,265 & 528 & 60.4 & 62.3 & 60.1 & 62.1 & 42.8 & 54.7 \\
Fixed conjunction (2) & 1,573 & 808 & 37.9 & 34.2 & 28.1 & 22.3 & 11.6 & 14.1 \\
Fixed conjunction (3) & 33 & 20 & 48.5 & 45.0 & 48.5 & 45.0 & 15.2 & 35.0 \\
Periodic (3) & 132 & 72 & 66.7 & 52.8 & 62.1 & 45.8 & 24.2 & 23.6 \\
Fixed conjunction (4) & 55 & 20 & 100.0 & 95.0 & 100.0 & 95.0 & 3.6 & 10.0 \\
\bottomrule
\end{tabular*}
\par\vspace{3pt}
\begin{minipage}{\linewidth}
\footnotesize\raggedright
M: eleven Codex models; F: four frameworks with MiniMax-M3.
Opportunities count recoverable action--view--run combinations and supply the denominator for all three percentages in a row. Any in $D$ means that at least one candidate record was acquired. $w_E$ is the size of the smallest visible sufficient witness.
\end{minipage}\par
\end{table}

Equal witness size does not imply equal requirements. The three-record fixed conjunction covers one APT27 side-loading action, whereas the periodic group covers four actions. Their citation coverage is 15.2\% versus 24.2\% in the model panel and 35.0\% versus 23.6\% in the framework panel. These groups differ in both structure and source composition. Dragonfly's watering-hole witness requires four records, and the unresolved Donot compound is excluded. We therefore inspect the periodic group's unmet requirements directly rather than attributing its score to record count alone.

\begin{table}[htbp]
\centering\footnotesize
\setlength{\tabcolsep}{3pt}
\caption{Why acquired records and formal-stage citations fail to support recurring-network actions (\%).}
\label{tab:deep-structure-periodic}
\begin{tabular*}{\linewidth}{@{}l@{\extracolsep{\fill}}rrrr@{}}
\toprule
Panel / set & No candidate & Count gap & Constraint gap & Sufficient \\
\midrule
Eleven models: $D$ & 33.3 & 4.5 & 0.0 & 62.1 \\
Eleven models: $C$ & 49.2 & 20.5 & 6.1 & 24.2 \\
Four frameworks: $D$ & 47.2 & 6.9 & 0.0 & 45.8 \\
Four frameworks: $C$ & 54.2 & 22.2 & 0.0 & 23.6 \\
\bottomrule
\end{tabular*}
\par\vspace{3pt}\begin{minipage}{\linewidth}\footnotesize Each row partitions 132 recoverable action--view--run opportunities in the model panel or 72 in the framework panel. Each panel contains four distinct periodic actions from four source-incident groups. $D$ contains acquired records and $C$ formal-stage citations. No candidate means no candidate record is present. \end{minipage}
\end{table}

In the model panel, queries return periodic witnesses for 62.1\% of recoverable opportunities, while formal-stage citations support 24.2\%. The remaining opportunities fall into three groups, all using the same recoverable denominator: 49.2\% contain no cited candidate, 20.5\% fail an observation-count or binding-count check, and 6.1\% pass those counts but fail compatibility. Every compatibility failure has a witness in $D$. Framework acquisition and citation coverage are 45.8\% and 23.6\% on that panel's periodic opportunities.

These aggregate rates are source-dependent. Dragonfly contributes 62.5\% of covered model-panel periodic opportunities and 82.4\% of covered framework opportunities. Excluding it leaves coverage of 15.6\% and 5.8\%, respectively.

Conditioning instead on model-panel periodic opportunities with acquired support but insufficient formal-stage citations, 34.0\% cite no candidate, 50.0\% fail counts, and 16.0\% fail compatibility. A separately archived DeepSeek-V4-Pro Zardoor report under tiered retention illustrates the last case. Its three cited records have successive intervals of about 8,945 and 2,184 seconds, whose ratio of 4.10 exceeds the registered maximum of two. An acquired witness has intervals of about 4,457 and 3,569 seconds and a ratio of 1.25. Both triples contain three records, but only the latter meets the temporal requirement.

Applying the same rule to $D$ and $C^\cup$ distinguishes support missing from query returns from support already returned but not preserved as a witness in formal-stage citations. The saved per-action decisions, exact witnesses, and source memberships make this distinction checkable.

\subsection{Single-record witnesses}
\label{app:single_record_blindspots}
\label{app:canonical_blindspots}
When one record suffices, a missing witness cannot be explained by combining too few records. Table~\ref{tab:restored_blindspots} reports the share of recoverable views in which at least one system independently acquires or cites a witness. These are panel-wide view counts. Records from different systems are never pooled, and different systems may account for coverage in different views.

\begin{table}[htbp]
\centering\footnotesize
\setlength{\tabcolsep}{3pt}
\caption{Single-record process witnesses acquired or sufficiently cited by at least one system in each included recoverable view.}
\label{tab:restored_blindspots}
\begin{tabularx}{\linewidth}{@{}Xrrrr@{}}
\toprule
& \multicolumn{2}{c}{Models} & \multicolumn{2}{c}{Frameworks} \\
\cmidrule(lr){2-3}\cmidrule(l){4-5}
Reference process action & Views & \shortstack{Any acquired /\\sufficiently cited (\%)} & Views & \shortstack{Any acquired /\\sufficiently cited (\%)} \\
\midrule
MuddyWater: first script process & 5 & 100.0 / 100.0 & 5 & 100.0 / 100.0 \\
MuddyWater: later DarkBit-associated process & 5 & 0.0 / 0.0 & 5 & 0.0 / 0.0 \\
Spyder: later Remcos-associated process & 2 & 50.0 / 0.0 & 3 & 0.0 / 0.0 \\
PurpleFox: rootkit-associated process & 3 & 100.0 / 0.0 & 3 & 66.7 / 0.0 \\
PurpleFox: elevation-associated process & 3 & 100.0 / 66.7 & 3 & 33.3 / 0.0 \\
\bottomrule
\end{tabularx}
\end{table}

At least one model covers MuddyWater's first script-process witness in every included recoverable view, but no model acquires the later DarkBit-associated process witness. For Spyder's later Remcos-associated process, at least one model acquires a witness in half of the recoverable views, yet no model sufficiently cites it in any view. PurpleFox's rootkit-associated process likewise has acquired support in every included model-panel view and no sufficient formal-stage citations. These examples distinguish failure to acquire simple witnesses from failure to preserve them.

The case and action masks determine each panel's included views. Source-byte checks establish process identity and creation for the registered candidates. Malware names identify the source descriptions, not verified encryption, backdoor operation, rootkit loading, or privilege escalation, which require further evidence.

\subsection{Confidence filtering}
\label{sec:confidence_selection}
\label{app:claim_diagnostics}
Table~\ref{tab:unified_confidence_tradeoff} removes formal claims below confidence 0.9, rebuilds the citation union, and applies the current support rules. Cases remain fixed, including reports made empty by filtering. Coverage preservation uses unfiltered $K$ as its denominator.

\begin{table}[htbp]
\centering\footnotesize
\setlength{\tabcolsep}{3pt}
\caption{Citation coverage preserved and background citations removed after filtering formal claims at confidence 0.9. Each panel uses its own shared model or framework cases.}
\label{tab:unified_confidence_tradeoff}
\begin{tabular*}{\linewidth}{@{}l@{\extracolsep{\fill}}rrrr@{}}
\toprule
System & Original $K$ & \shortstack{Original background\\citations} & \shortstack{$K$ retained\\(\%)} & \shortstack{Background removed\\(\%)} \\
\midrule
\multicolumn{5}{@{}l}{\textit{Codex models}} \\
MiniMax-M2 & 12 & 143 & 83.3 & 38.5 \\
MiniMax-M2.1-HS & 11 & 243 & 100.0 & 25.1 \\
MiniMax-M2.5-HS & 19 & 248 & 68.4 & 22.2 \\
MiniMax-M2.7-HS & 27 & 211 & 96.3 & 35.1 \\
MiniMax-M3 & 85 & 203 & 81.2 & 40.9 \\
DeepSeek-V4-Pro & 84 & 207 & 46.4 & 57.0 \\
GPT-5.5 & 115 & 163 & 61.7 & 74.8 \\
GPT-5.6-Luna & 77 & 250 & 89.6 & 14.4 \\
GPT-5.6-Terra & 83 & 56 & 79.5 & 17.9 \\
GPT-5.6-Sol & 117 & 30 & 93.2 & 0.0 \\
GPT-6-Astra & 134 & 0 & 99.3 & NA \\
\multicolumn{5}{@{}l}{\textit{MiniMax-M3 frameworks}} \\
Codex & 107 & 245 & 82.2 & 52.7 \\
Claude Code & 98 & 298 & 69.4 & 72.1 \\
OpenCode & 107 & 306 & 79.4 & 39.5 \\
Ref-ReAct & 117 & 975 & 94.0 & 58.4 \\
\bottomrule
\end{tabular*}
\end{table}

At 0.9, MiniMax-M3 preserves 81.2\% of its original coverage and DeepSeek-V4-Pro preserves 46.4\%. GPT-6-Astra preserves 99.3\%, but its background-removal ratio is undefined because its original report set contains no background citations. A common threshold therefore has different effects across systems. Background records can provide legitimate context, and these labels do not establish claim correctness.

\FloatBarrier
\section{Investigation examples and limits of citation scoring}
\label{app:examples}
The following examples show how agents connect records, work with limited telemetry, and omit support already returned by their queries. Their scores illustrate individual investigations.

\subsection{Selection and verification scope}
We selected the GPT traces for inspectable query paths with substantial acquired and cited support, then checked the saved operations, reports, source records, and validation receipts. Both selected runs have accepted reports and no recorded broker-operation errors. The GPT traces preserve operation logs and returned record IDs, but not complete wire-format responses. Their reconstructed source records establish what the records contain. In contrast, the OpenCode example preserves the response page and verifies complete content for five omitted witnesses.

\subsection{Following an APT27 web lead across sources}
\begin{figure}[ht]
\centering\includegraphics[width=\linewidth]{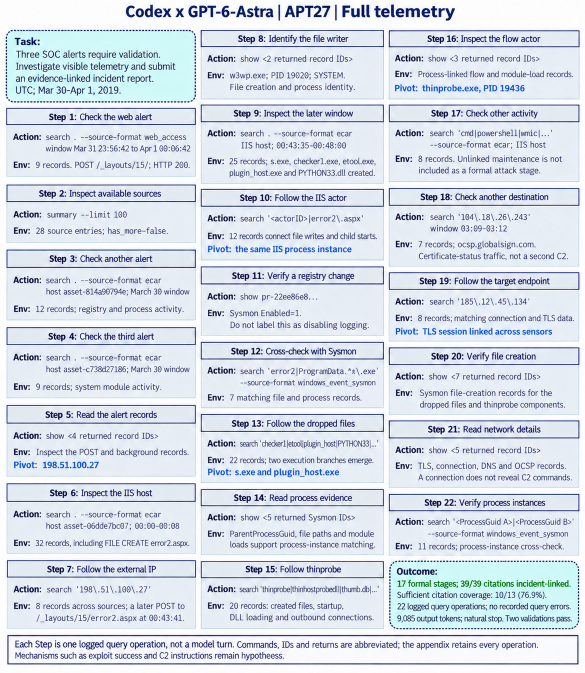}
\caption{Saved Codex--GPT-6-Astra investigation of the 2019 APT27 reconstruction under Full telemetry. All 22 broker operations are shown; a Step is a tool operation, not a model reasoning round. Formal citations support 10 of 13 recoverable actions (76.9\%) under the archived rules. The diagram retains historical execution details and abridged source-record reconstructions; complete original responses are unavailable. Its appendix pointers refer to the archived operation transcripts.}
\label{fig:trajectory_many_requests}
\end{figure}

Codex--GPT-6-Astra investigates the 2019 APT27 reconstruction under Full telemetry (Figure~\ref{fig:trajectory_many_requests}). It follows an unverified web-request lead to a later request for \texttt{error2.aspx}, identifies the \texttt{w3wp.exe} instance that creates the file, and follows process identifiers to \texttt{s.exe}, \texttt{thinprobe.exe}, and a separate \texttt{plugin\_host.exe} branch. A process-associated outbound flow then leads to a matching Zeek TLS session. The report treats nearby certificate-status traffic as background and leaves the exploit mechanism, request-to-file causality, and C2 contents as hypotheses.

The 22 saved operations comprise fourteen searches, seven record lookups, and one source summary. They return 152 distinct record IDs, with no search leaving a continuation page. Under the archived rules, $N=R=O=13$ and $K=10$. The seventeen formal descriptions cite 39 distinct source-labelled incident records. Thus, the report organizes multi-step activity, but its formal-stage citations do not preserve all acquired support.

The three omissions have identifiable causes. A process-start record needed for ASPX placement appears only in an entity description. The \texttt{plugin\_host.exe} claim cites startup and DLL loading but omits the termination required by its compound rule. The IIS worker's later termination is also returned but omitted. All three actions lack a complete witness in the formal-stage citation union.

\subsection{Investigating Spyder with network-only telemetry}
\begin{figure}[ht]
\centering\includegraphics[width=\linewidth]{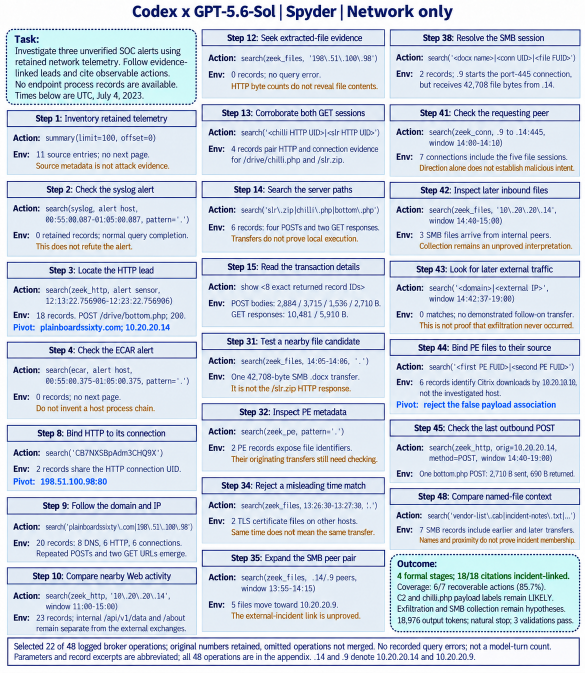}
\caption{Saved Codex--GPT-5.6-Sol investigation of Spyder under network-only telemetry. The 22 selected broker operations retain their original numbers among 48 operations. Formal citations support 6 of 7 recoverable actions (85.7\%) under the archived rules. C2 and payload labels remain qualified claims, while exfiltration and SMB associations remain hypotheses. The diagram retains historical execution details and abridged source-record reconstructions; complete original responses are unavailable. Its appendix pointers refer to the archived operation transcripts.}
\label{fig:trajectory_high_output}
\end{figure}

Codex--GPT-5.6-Sol follows the supplied HTTP lead to \texttt{plainboardssixty.com}, its address \texttt{198.51.100.98}, four POST exchanges, and two GET responses (Figure~\ref{fig:trajectory_high_output}). The original endpoint-alert queries return no retained records. The report does not infer a host-process chain from those missing records and places possible POST-body exfiltration and SMB associations in hypothesis fields.

The 48 saved operations return 615 distinct record IDs, with some continuation pages left unconsumed. Under the archived rules, $N=10$, $R=7$, and $O=K=6$. The four formal descriptions cite eighteen distinct source-labelled incident records and retain support for all six acquired actions. Coverage is therefore 85.7\% of recoverable actions but only 60.0\% of the incident reference set. Network evidence can support the exchanges without establishing execution of the downloaded content or attribution to a particular process. The report's \emph{likely} C2 and payload labels remain formal claims, not independently verified interpretations.

\subsection{Complete returned support omitted from formal citations}
\label{app:complete_content_omissions}
The historical OpenCode content audit matches saved tool output to source records at verified byte locations. Across its 64 cases from nine source-incident groups, 25 of 38 acquired but insufficiently cited action opportunities (65.8\%) have a complete witness verified in the saved response. Of these 25 verified omissions, thirteen cite no registered candidate (52.0\%) and twelve cite only part of a witness (48.0\%). The remaining thirteen omissions lack complete-content certification. This does not establish that their content was absent or truncated. The certified fraction ranges from 60\% to 80\% when each source-incident group is excluded in turn. These results retain the audit's original support rules and population, whose denominator differs from the current framework comparison. They establish complete content in saved tool output, not its inclusion in the provider's final request or the model's attention and understanding.

OpenCode--MiniMax-M3 on Spyder under Full provides a concrete example. Its twelfth record query searches for \texttt{plainboardssixty} and returns 22 records on a completed page. The saved response contains complete witnesses for five actions absent from the report's sufficient citation support: initial payload delivery, hardware-ID exchange, follow-up payload request, archive retrieval, and completion notification. The historical result is $N=10$, $R=9$, $O=6$, and $K=1$. These five omissions occur after complete content appears in the saved response and before formal-stage citation. They cannot be explained by unconsumed pages or shortened previews, although the audit does not establish how the model used the content.

The same report interprets a separate local \texttt{curl} invocation as server-side request forgery. This unsupported interpretation is distinct from the five citation omissions. The following review separates such interpretations from the presence of sufficient cited records.

\subsection{What the cited records establish}
\label{app:target_claim_audit}
Citation coverage checks for a registered witness in the formal-stage citation union. It does not establish that every associated statement about behavior, attribution, timing, or effects is correct. Table~\ref{tab:r1_claim_boundaries} gives three concrete distinctions.

\begin{table}[!htbp]
\centering\small
\caption{Observed facts and stronger interpretations in saved reports. Examples come from the target audit (APT27, Donot) and a separate worked-trace inspection (Spyder). Each judgment concerns the assertion's own citations.}
\label{tab:r1_claim_boundaries}
\begin{tabularx}{\linewidth}{@{}L{0.18\linewidth}XX@{}}
\toprule
Inspected example & Established by the records & Stronger interpretation requiring further support\\
\midrule
Spyder: local command & Local shell invokes \texttt{curl} against a metadata address & Server-side request forgery or membership in the target incident\\
APT27: network activity & Two process-associated connections, followed by process termination & A ten-minute cadence or communication through 14:10:02\\
Donot: Security 4698 & Named task creation and its stored recurrence/action configuration & Subsequent execution, reboot survival, or the initiating script without a separate binding\\
\bottomrule
\end{tabularx}
\end{table}

The historical target review covers 56 reports from four MiniMax-M3 frameworks on seven scenarios from six source-incident groups. Each framework contributes one suppression and one matched-control report per scenario. These 56 reports are distinct from the 56 cases per framework in the current comparison.

\begin{table}[!htbp]
\centering\small
\caption{Historical target review. Panel A includes 28 reports in each arm. Panel B includes only the seventeen reports that mention the target. Entries are counts (percentages) within the stated denominator. The preliminary categories are not an independently adjudicated measure of report correctness.}
\label{tab:target_claim_review}
\setlength{\tabcolsep}{4pt}
\begin{tabularx}{\linewidth}{@{}Xrr@{}}
\toprule
\multicolumn{3}{@{}l}{\textbf{A. Treatment of the selected target} (28 reports per arm)}\\
Target-report state & Suppression & Matched control\\
\midrule
Formal narrower-activity claim; registered witness complete & 0 (0.0\%) & 4 (14.3\%)\\
Formal narrower-activity claim; support outside the registered rule & 4 (14.3\%) & 2 (7.1\%)\\
Formal target claim; registered witness incomplete & 3 (10.7\%) & 4 (14.3\%)\\
Target discussed only as hypothesis or uncertainty & 0 (0.0\%) & 0 (0.0\%)\\
Target not discussed & 21 (75.0\%) & 18 (64.3\%)\\
\bottomrule
\end{tabularx}
\par\medskip
\begin{tabularx}{\linewidth}{@{}Xrrr@{}}
\toprule
\multicolumn{4}{@{}l}{\textbf{B. Interpretation of target mentions} (17 reports)}\\
Support for the narrower observed activity & \shortstack{Unsupported\\detail} & \shortstack{No additional\\mismatch} & Unresolved\\
\midrule
Registered witness complete & 0 (0.0\%) & 1 (5.9\%) & 3 (17.6\%)\\
Support outside the registered rule & 0 (0.0\%) & 3 (17.6\%) & 3 (17.6\%)\\
Registered witness incomplete & 5 (29.4\%) & 0 (0.0\%) & 2 (11.8\%)\\
\midrule
Total & 5 (29.4\%) & 4 (23.5\%) & 8 (47.1\%)\\
\bottomrule
\end{tabularx}
\par\smallskip
\end{table}

Thirty-nine of the 56 reports omit the target. Among the seventeen that mention it, five contain a specific unsupported detail, four have no additional mismatch established, and eight remain unresolved. In thirteen of the 28 framework--scenario pairs, both reports omit the target. All four matched-control reports with a complete registered witness have paired suppression reports that omit it. Omission alone does not show deliberate abstention or behavior that would recur across runs.

\paragraph{Timing and effects require their own evidence.}
In undated APT27, two connections cited by Claude Code are 74 minutes 47.843 seconds apart and do not support the report's claimed ten-minute cadence. Another report describes endpoint and DNS records as periodic Zeek TLS evidence. In APT-C-01, surviving connections establish communication but leave a stronger periodic-malware interpretation unresolved. In Donot, Security Event 4698 establishes task creation and configuration, not subsequent execution or survival after reboot.

\paragraph{Normal context is not incident attribution.}
\label{app:background_semantics}
Claude Code's APT-C-01 endpoint-only report places a routine \texttt{sysstat} event in a confirmed formal stage while describing it as normal activity. Its two citations are source-labelled benign. The description is accurate, but its placement falls outside the formal attack-stage scope. The OpenCode Spyder case instead misinterprets a background \texttt{curl} request to \texttt{169.254.169.254} as SSRF. Those records show an interactive local command, not a server acting on an attacker's behalf. These errors require checking claim meaning as well as citation membership.



\end{document}